\documentclass[
reprint,
superscriptaddress,
amsmath,amssymb,
aps,
floatfix,
]{revtex4-2}

\usepackage{graphicx}
\usepackage{dcolumn}
\usepackage{bm}
\usepackage{hyperref}
\usepackage{siunitx}
\usepackage{chemformula}

\newcommand{\ur}{\textup{r}}
\newcommand{\ui}{\textup{i}}

\newcommand{\ue}{\textup{e}}

\newcommand{\uint}{\textup{int}}
\newcommand{\cpw}{\textup{CPW}}
\newcommand{\ppc}{\textup{PPC}}
\newcommand{\eff}{\textup{eff}}

\begin{document}

\preprint{APS/123-QED}

\title{Low-characteristic-impedance superconducting tadpole resonators in the sub-gigahertz regime}

\author{Miika Rasola}
\email[Corresponding author.\\]{miika.rasola@aalto.fi}
\affiliation{QCD Labs, QTF Centre of Excellence, Department of Applied Physics, Aalto University, P.O. Box 13500, FI-00076 Aalto, Finland}

\author{Samuel Klaver}
\affiliation{QCD Labs, QTF Centre of Excellence, Department of Applied Physics, Aalto University, P.O. Box 13500, FI-00076 Aalto, Finland}

\author{Jian Ma}
\affiliation{QCD Labs, QTF Centre of Excellence, Department of Applied Physics, Aalto University, P.O. Box 13500, FI-00076 Aalto, Finland}

\author{Priyank Singh}
\affiliation{QCD Labs, QTF Centre of Excellence, Department of Applied Physics, Aalto University, P.O. Box 13500, FI-00076 Aalto, Finland}

\author{Tuomas Uusnäkki}
\affiliation{QCD Labs, QTF Centre of Excellence, Department of Applied Physics, Aalto University, P.O. Box 13500, FI-00076 Aalto, Finland}

\author{Heikki Suominen}
\affiliation{QCD Labs, QTF Centre of Excellence, Department of Applied Physics, Aalto University, P.O. Box 13500, FI-00076 Aalto, Finland}

\author{Mikko Möttönen}
\email[Corresponding author.\\]{mikko.mottonen@aalto.fi}
\affiliation{QCD Labs, QTF Centre of Excellence, Department of Applied Physics, Aalto University, P.O. Box 13500, FI-00076 Aalto, Finland}
\affiliation{QTF Centre of Excellence, VTT Technical Research Centre of Finland Ltd., P.O. Box 1000, 02044 VTT, Finland}

\date{\today}

\begin{abstract}

We demonstrate a simple and versatile resonator design based on a short strip of a typical coplanar waveguide shorted at one end to the ground and shunted at the other end with a large parallel-plate capacitor. Due to the shape of the structure, we coin it the tadpole resonator. The design allows tailoring the characteristic impedance of the resonator to especially suit applications requiring low values. We demonstrate characteristic impedances ranging from $Z_\textrm{c}=\qty{2}{\ohm}$ to~$\qty{10}{\ohm}$ and a frequency range from $f_\ur=\qty{290}{MHz}$ to~$\qty{1.1}{GHz}$ while reaching internal quality factors of order $Q_\uint=\num{8.5e3}$ translating into a loss tangent of $\tan(\delta)=\num{1.2e-4}$ for the aluminium oxide used as the dielectric in the parallel plate capacitor. We conclude that these tadpole resonators are well suited for applications requiring low frequency and low characteristic impedance while maintaining a small footprint on chip. The low characteristic impedance of the tadpole resonator renders it a promising candidate for achieving strong inductive coupling to other microwave components.

\end{abstract}

\maketitle


\section{\label{intro}Introduction}

Superconducting quantum circuits have provided unforeseen opportunities to tailor quantum systems for various purposes. This development has led to a new field of quantum microwave engineering, serving a purpose beyond fundamental research with the goal of realizing useful quantum devices~\cite{Wolfgang, Qengineer}. This field has already produced numerous groundbreaking results in quantum computation~\cite{DiCarlo2009, Lucero2012, Zheng2017, Chen2020, Harrigan2021}, communication~\cite{Axline2018, Kurpiers2018, Pogorzalek2019, Kirill2021}, simulation~\cite{Underwood2012, AbdumalikovJr2013, Roushan2017, Kollár2019, Ma2019, Kai2020, Guo2021}, sensing~\cite{Chen2023, Barzanjeh2020, Bienfait2017, Wang2021, Kokkoniemi2019, Kokkoniemi2020, Govenius2016, Gasparinetti2015}, and many more~\cite{Xiang2013, Clerk2020, Tan2017, Timm_thermal, Sundelin2024}.

The coplanar waveguide (CPW) resonator~\cite{Goppl08, Gieres1991, Yoshida1992, Rauch1993} is perhaps the most utilized standard building block in quantum engineering. A CPW resonator is simply a strip of transmission line with shorted or open-circuit boundary conditions at each end. The advantages of CPW resonators include ease of modeling, a wide range of available parameters and design options complemented by simple fabrication~\cite{Goppl08}. Superconducting CPW resonators with frequencies in the gigahertz range and internal quality factors of several hundred thousands can be routinely achieved~\cite{Zikiy2023, Frunzio2005, Barends2007, Goppl08}.

The physical length of a superconducting CPW resonator is of the same order of magnitude as the wavelength of the fundamental microwave field mode in the resonator. In a typical setting regarding superconducting quantum circuits, the CPW resonators are fabricated on a low-loss substrate deposited with a superconducting metal film. The relative permittivity of the substrate largely determines the wavelength of the microwave mode, and therefore physical length of the resonator. For a typically used silicon substrate with $\epsilon_r\approx 11.9$, a resonator at the $\qty{5}{\GHz}$ regime has a length of roughly $\qty{10}{\mm}$. This can be easily fitted on a usual chip size by meandering, but the situation is different at sub-gigahertz frequencies. At $\qty{1}{\GHz}$, the CPW resonator has already a length of about $\qty{60}{\mm}$, exceeding the typical chip dimensions sixfold. This size limits the number of devices per chip considerably and at even lower frequencies may prevent using CPW resonators in practice. More intricate physical geometries, such as twisting the resonator into a spiral~\cite{Zhuravel2012, Maleeva2014, Partanen2016, Rolland2019, Yan2021}, may offer a solution in some cases, but may also raise other concerns with impedance matching, grounding, and parasitic modes~\cite{Wenner2011, Lankwarden2012, Abuwasib2013, Chen2014, Fischer2021}, and render coupling to other devices somewhat challenging.

Apart from problems arising from the large physical size of the low-frequency CPW resonators, another possible issue emerges when coupling these low-frequency resonators with high-frequency components. Theoretical models describing the quantum physics of superconducting quantum circuits typically only consider the fundamental modes of the circuit components. This approximation is fine as long as all the circuit components reside within the same relatively narrow frequency range. However, a low-frequency CPW resonator has a dense spectrum of harmonic modes in contrast to a high-frequency component, which may cause problems upon coupling the two. For example, a $\lambda/2$ CPW resonator with a $\qty{500}{\MHz}$ fundamental mode frequency has ten modes below $\qty{5}{\GHz}$. If these components are to be coupled for their fundamental modes, driving the $\qty{5}{\GHz}$ mode may excite multiple unwanted modes in the $\qty{500}{\MHz}$ resonator. For instance, the detrimental effects of spurious higher harmonics to the qubit lifetime and coherence are a known problem \cite{Houck2008}.

All of the above mentioned issues with CPW resonators are of course well known, and possible alternatives, such as lumped-element devices, have already been studied since the 1990's~\cite{Lancaster1993, Chaloupka2001}. Alternatives to CPW resonators in superconducting circuits have been demonstrated by using interdigital~\cite{Chen, McKenzie2019, Geerlings2012} and parallel-plate~\cite{Zotova2023, Cho2013, Paik2010, Deng2014, Govenius2016, Kouwenhouven2024} capacitors. Although the demonstrated devices solve some issues, most of them also display challenges such as difficulties to align with the fabrication process of a superconducting circuit, complex structures, higher harmonics, excessive losses, or large size. Furthermore, most of the earlier research focuses on devices in the few gigahertz frequency range, leaving uncharted territories in the sub-gigahertz regime. 

\begin{figure}[ht]
\centering
\includegraphics[width=\linewidth]{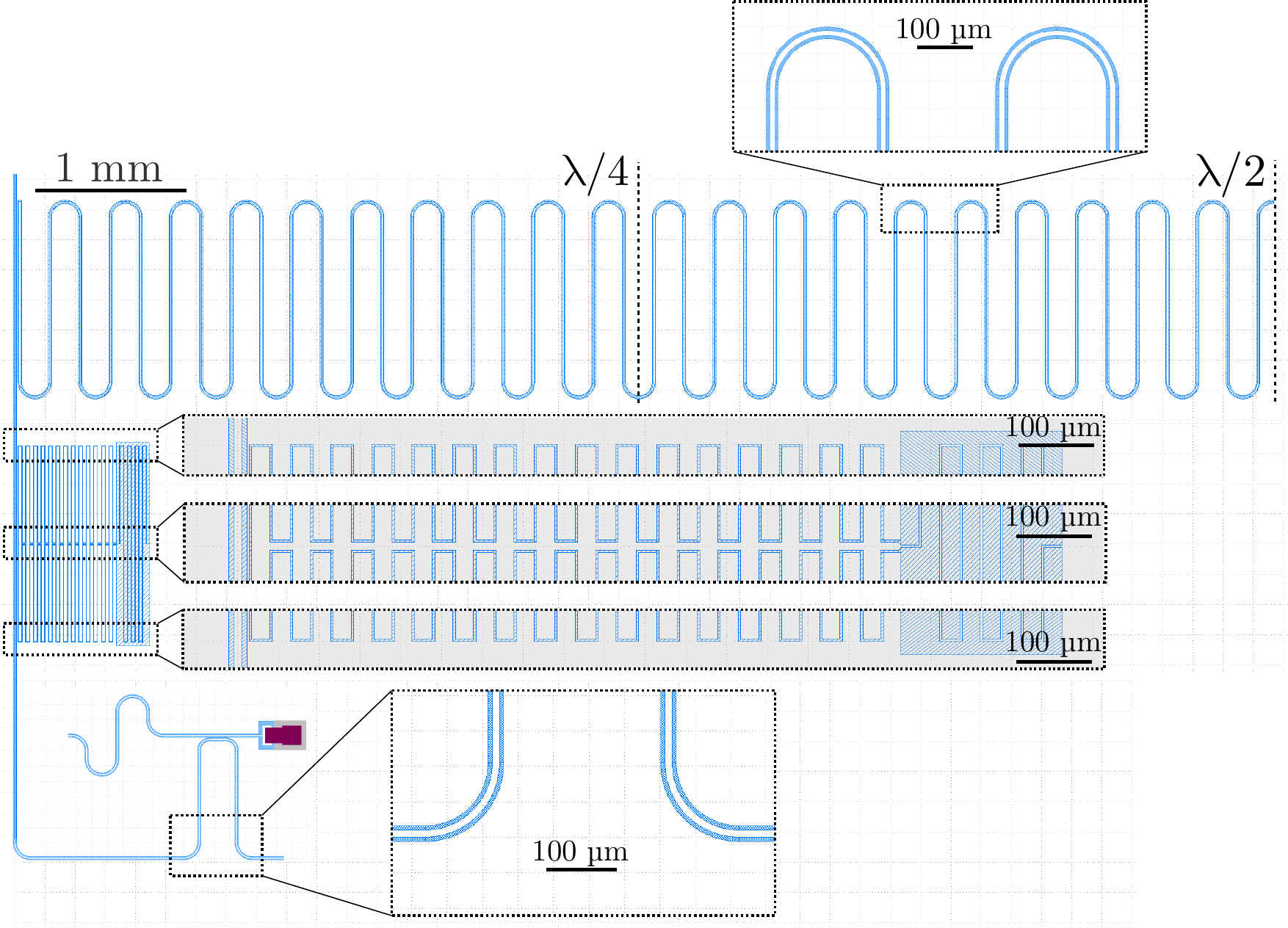}
\caption{Size comparison of three resonators with $\qty{1}{\giga\hertz}$ fundamental-mode frequency. A typical $\qty{50}{\ohm}$ coplanar waveguide resonator (top), a compact inductor-capacitor resonator~\cite{Chen} (middle), and a tadpole resonator (bottom). The coplanar waveguide resonator needs a considerably larger surface area as compared with the two latter options. The white areas represent the superconducting metal thin film whereas the blue indicates the structures where metal has been etched away exposing the substrate. The transmission line used to probe the resonators runs down on the left edge of the chip and makes a curve at the bottom to couple to the CPW strip of the tadpole resonator. The insets show details of the coplanar waveguides and the compact inductor-capacitor resonator. A more detailed schematic of the tadpole resonator can be found in Fig.~\ref{fig:schematic2}. The $\qty{1}{\mm}$ global scale bar at the top left corner applies to the whole figure excluding the insets. Each inset has its own $\qty{100}{\micro\meter}$ scale bar.}
\label{fig:schematic1}
\end{figure}

In this article, we demonstrate an extremely simple and versatile lumped-element resonator design based on a strip of a traditional CPW transmission line shunted with a parallel-plate capacitor (PPC), thus forming a structure reminiscent of a tadpole. Combining the best of both worlds, our design is applicable in a wide range of frequencies, especially in the low-frequency end of the spectrum, while maintaining a relatively small on-chip footprint, retaining the ease of fabrication and implementation, and boasting an extremely robust structure. Importantly, the present design allows for the tuning of the inductance to capacitance ratio, i.e., the characteristic impedance of the resonator, in a wide range below the typical value of $Z_c\sim\qty{50}{\ohm}$, reaching values of the order of $Z_c\sim\qty{1}{\ohm}$. This has the benefit of confining the magnetic field of the resonator mode into a small spatial volume facilitating strong inductive coupling to the resonator. This is potentially beneficial for realizing certain types of superconducting circuits, such as the devices proposed in the Refs.~\cite{Johansson, Kim, Hardal2017, Rasola2024}.

The physical size of the tadpole resonator is compared to two other resonator designs in Fig.~\ref{fig:schematic1}. Note that the tail of the tadpole may be more heavily meandered or even replaced by a meandering wire similar to that in the compact inductor-capacitor resonators, thus rendering the tadpole resonator the smallest of the considered designs.

\section{Low-characteristic-impedance tadpole resonators} 
\subsection{Design and analysis}

\begin{figure}[htb]
\centering
\includegraphics[width=\linewidth]{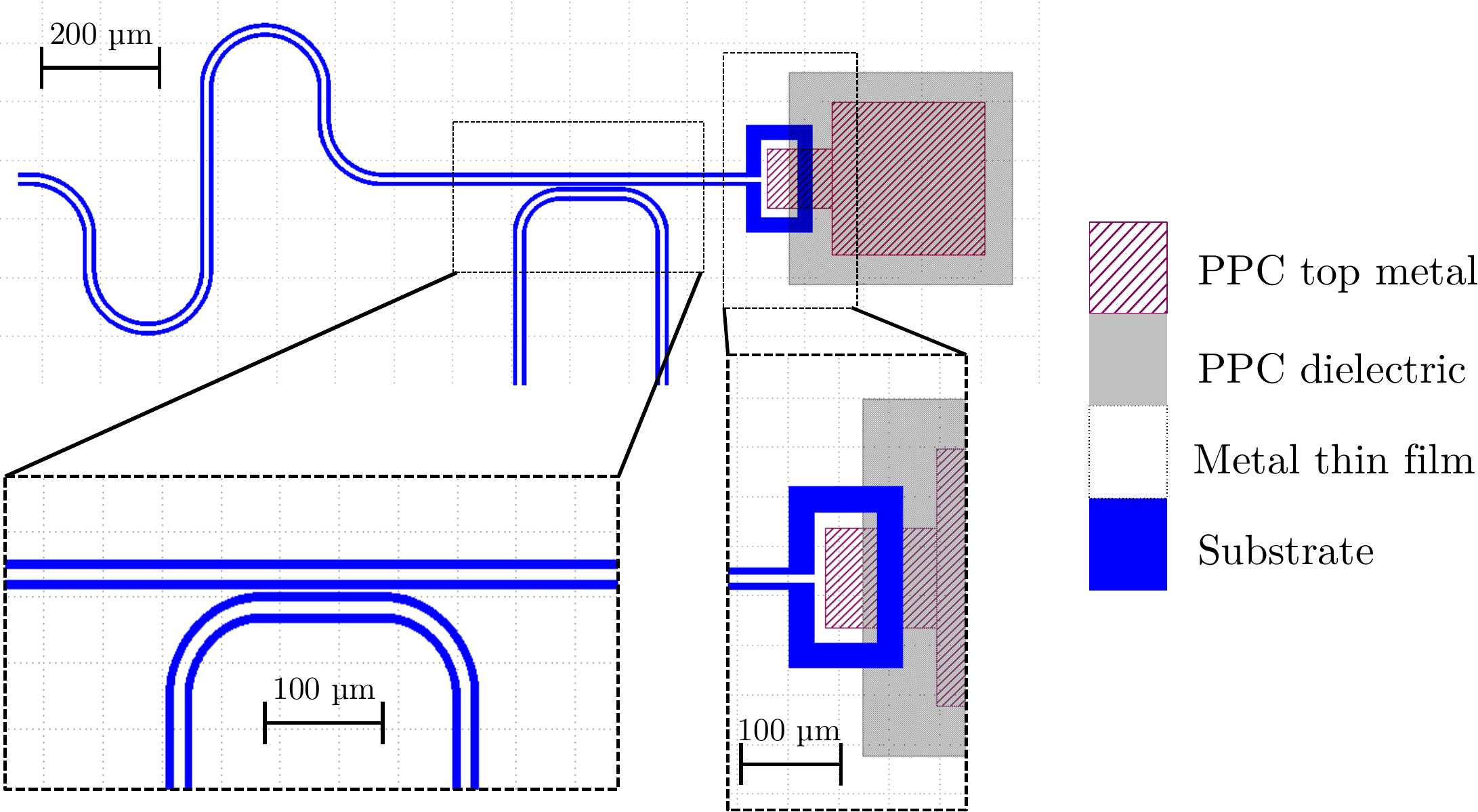}
\caption{Schematic figure of a tadpole resonator viewed from the top and a color-coded layer map indicating the order of different layers from top to bottom. The white areas represent superconducting-metal thin films, the blue color shows where the metal has been etched away thus exposing the substrate, the semitransparent grey indicates the dielectric of the parallel-plate capacitor (PPC), and the semitransparent purple area denotes the superconducting top metal of the PPC. m left inset shows details of the resonator coupling to the feed line which is depicted as the inverted U-shaped structure at the bottom. Note the small strip of the ground plane in between the transmission line and the resonator structure. The bottom right inset shows the PPC shunt, where the PPC top metal is galvanically connected to the center conductor of the resonator.}
\label{fig:schematic2}
\end{figure}

Let us begin by discussing our design on a general level. A detailed schematic of the resonator design can be found in Fig.~\ref{fig:schematic2} for reference. Our resonator design consists of a strip of typical CPW line with one end shorted to the ground and the other shunted to the ground with a large PPC. In the limit of a large PPC and short CPW strip, the the CPW strip essentially provides the inductance and PCC the capacitance, i.e. $C_\ppc\gg C_\cpw$, where $C_\ppc$ and $C_\cpw$ are the capacitances of the PPC and CPW, respectively. Consequently, the physical size of this structure can be much smaller in all dimensions than the wavelength of the resonator mode it houses. This structure can be accurately modelled as a lumped-element resonator where the magnetic field is localized in the short CPW strip and the electric field resides within the PPC. Importantly, this scheme allows for a strong inductive coupling via the CPW strip with a relatively low inductance of the coupler since the total inductance of the resonator is low.

One advantage of the current design is that it is straightforward to estimate the resonance frequency of our resonator, even analytically, given that the effective permittivity of the CPW, $\epsilon_\eff$, and the capacitance per unit area, $c_0$, of the PPC are known. Although analytical expressions for $\epsilon_\eff$ exist~\cite{Gevorgian95}, it can be reliably found, either by finite-element electromagnetic simulations, or by measurement. Capacitance per unit area, on the other hand, is typically known for established fabrication recipes with corresponding characterization data, but can also be evaluated analytically based on the relative permittivity of the chosen dielectric. Once these values are known, one can estimate the capacitance and the inductance of an arbitrary-length CPW strip by the well-known results~\cite{Goppl08} obtained by conformal mapping methods~\cite{Watanabe94, Gevorgian95}. Furthermore, as long as the dielectric of the PPC is thick enough, so that there is no inductive shunt to ground through the dielectric, the capacitance of the PPC can be evaluated as $C_\ppc=c_0A$, where  surface area of the PPC. The frequency of the resonator is thus given by 
\begin{align}
f_\ur=\left[2\pi\sqrt{L(C_\ppc+C_\cpw)}\right]^{-1},
\label{eq:freq}
\end{align}
where $L$ is the inductance of the CPW strip, i.e. the total inductance of the tadpole resonator, and $C_\cpw$ is the small capacitance contribution of the CPW strip. We ignore other possible sources of stray capacitance as they will be tiny as compared to $C_\ppc$. Also note that we neglect the kinetic inductance contribution to the total inductance here, which is typically found to be an accurate approximation~\cite{Goppl08, Frunzio2005, Zotova2023}. To verify this in our case, we follow the methods of Ref.~\cite{Maxfield1965} and find that the estimated kinetic inductance is only about $1\,\%$ of the geometric inductance, thus justifying the use of the approximation. In addition to the resonator frequency, we define the characteristic impedance of the resonator as $Z_\mathrm{c}=\sqrt{L/(C_\ppc+C_\cpw)}$.

We couple our resonators to a feed line by a notch-type coupling at the CPW strip, therefore achieving a mixed coupling with inductive and capacitive nature rather than the typical purely capacitive coupling. We leave a small strip of ground metal in between the resonator CPW strip and the feed line in order to not break the ground plane and hence to reduce the possibility of encountering spurious slotline modes. Although we limit ourselves to a relatively weak coupling here, the design naturally supports strong inductive coupling since the magnetic field is confined to a small volume~\cite{McKenzie2019}. Such strong inductive coupling may be achieved, for instance, via a SQUID~\cite{Johansson, Kim}.

Let us briefly discuss the internal loss sources of the tadpole resonator. Currently, internal quality factors of several hundreds of thousands in CPW resonators can be achieved routinely~\cite{Frunzio2005, Barends2007, Goppl08}. The PPC, on the other hand, may introduce significant additional losses since it adds material prone to quantum two-level systems (TLSs) and losses thereof, such as additional dielectrics, metal--dielectric interfaces and metal--vacuum interfaces~\cite{Sage2011, OConnell2008, Martinis2005, Lindstrom2009, Wang2009}. Thus it is reasonable to model the tadpole as a lumped-element resonator where the losses are dominated by the dielectric losses of the PPC, and an accurate estimate of the loss tangent of the PPC dielectric is provided by the inverse of internal quality factor: $\tan\delta=1/Q_\ui$.

\subsection{Fabrication and experimental methods}

Our CPW strip has a typical cross-sectional geometry with the width of the center conductor $w=\qty{10}{\um}$ and the gap between the center conductor and the ground plane $s=\qty{6}{\um}$. For the sake of simplicity, we equip all our resonators with identical CPW strips of length $\qty{2000}{\um}$, but the length can be, of course, chosen differently to tune the characteristic impedance of the resonator. Furthermore, we fix the thickness of the dielectric layer in the PPC to $d=\qty{42}{\nm}$, and only vary the surface area of the PPC to tune the frequency of the resonators. Since we use aluminium oxide as the dielectric in the PPC, the 42-nm thickness of the layer is more than enough to suppress any leakage through the oxide and their inductive contributions to the PCC~\cite{Zotova2023}. We couple six tadpole resonators to one transmission line for multiplexed readout using a notch-type configuration~\cite{Chen2022}. We design two sets of six resonators: one set with designed frequencies ranging from \SIrange{280}{450}{\MHz} and another with frequencies ranging from \qty{450}{\MHz} to \qty{1.1}{\GHz}.

We fabricate all of the twelve tadpole resonators on a single chip of $\qtyproduct{10 x 10}{\mm}$ high-purity, high-resistivity ($\rho > \qty{10}{\kilo\ohm\cm}$) silicon substrate with a thermally grown $\qty{300}{\nm}$ layer of silicon oxide. The total substrate thickness is $\qty{675}{\um}$. A \ch{Nb} thin film of $\qty{200}{\nm}$ is sputtered on top of the oxide layer. The niobium structures are defined onto AZ5214E photoresist using a Heidelberg Instruments MLA150 maskless aligner followed by a reactive ion etching process in order to remove the niobium from the defined areas.

For the PPC, we first use atomic-layer deposition (ALD) to grow the $\qty{42}{\nm}$ dielectric layer of $\mathrm{AlO}_x$ on the chip with 455 cycles in a \ch{H2O}/TMA (trimethylaluminum) process at \qty{200}{\celsius} in a Beneq TFS-500 system. Next, we protect the dielectric layer at the desired capacitor regions with AZ5214E resist and wet-etch the rest of the aluminium oxide away with a mixture of ammonium fluoride and hydrofluoric acid. Before depositing the top metal of the PPC, we first use argon milling to remove the intrinsic oxide from the niobium contact pad in order to ensure a proper galvanic contact. As the last step before lift-off, dicing, and bonding, we deposit a $\qty{30}{\nm}$ aluminium layer in an electron beam evaporator for the PPC top metal. The PPC fabrication recipe is described in detail in Ref.~\cite{Lake2017}.

For characterization purposes, we cool down the sample in a commercial cryogen-free dilution refrigerator with a base temperature of about $\qty{25}{\milli\K}$. We measure the $S_{21}$ transmission coefficient of each resonator with a Rohde~\&~Schwarz ZNB40 vector network analyzer. The sample is mounted to a sample holder and connected to the electronics via aluminium bond wires. The room temperature control electronics are connected to the sample via coaxial cables with a total attenuation of $80~\si{dB}$ across the levels of the cryostat. For more information about the experimental setup, please refer to the appendix~\ref{app:sample}.

\begin{table*}
\centering
\caption{\label{tab:table1}The most important properties and characterization results of the tadpole resonators. Here $A_\ppc$ is the parallel-plate-capacitor surface area, and $l_\mathrm{tot}/\lambda_0$ is the ratio of the total length of the tadpole resonator to the wavelength of the fundamental mode, as described in the main text. The length of the coplanar-waveguide strip is the same for all resonators here, and the values presented here are determined at the base temperature of the cryostat.}
\begin{ruledtabular}
\begin{tabular}{ccccccccccccc}
Resonator & A & B & C & D & E & F & G & H & I  & J  & K & L \\ 
\hline
Measured $f_\ur$ (MHz) & 290.5 & 315.9 & 346.8 & 377.9 & 413.8 & 450.7 & 467.7 & 559.8 & 695.5 & 790.5 & 944.7 & 1099.1 \\
Predicted $f_\ur$ (MHz) & 286.8 & 311.9 & 342.5 & 373.4 & 410.0 & 445.8 & 486.1 & 574.8 & 686.6 & 817.9 & 933.2 & 1086.6 \\
Relative $f_\ur$ error (\%) & 1.28 & 1.29 & 1.24 & 1.19 & 1.17 & 1.09 & 3.94 & 2.68 & 1.27 & 3.47 & 1.22 & 1.14\\
$Z_\mathrm{c}$ (\unit{\ohm}) & 1.9 & 2.1 & 2.3 & 2.5 & 2.8 & 3.0 & 3.1 & 3.7 & 4.7 & 5.3 & 6.4 & 7.4 \\
$l_\mathrm{tot}/\lambda_0$ & $0.0063$ & $0.0068$ & $0.0073$ & $0.0079$ & $0.0085$ & $0.0091$ & $0.0094$ & $0.011$ & $0.013$ & $0.015$ & $0.018$ & $0.020$ \\
$A_\ppc$ ($\unit{\um^2}$) & 206721 & 174790 & 144846 & 121870 & 101565 & 85446 & 71821 & 51310 & 35905 & 25246 & 19350 & 14221 \\
$C_\cpw/C_\ppc$ (\%) & 0.09 & 0.11 & 0.13 & 0.16 & 0.19 & 0.23 & 0.27 & 0.38 & 0.54 & 0.76 & 0.99 & 1.36 \\
Single-photon power (dBm) & $-147.6$ & $-143.7$ & $-148.9$ & $-138.0$ & $-146.1$ & $-145.7$ & $-143.8$ & $-141.5$ & $-139.1$ & $-144.2$ & $-138.8$ & $-133.1$ \\
\end{tabular}
\end{ruledtabular}
\end{table*}

\section{Results}

\begin{figure}[htb]
\centering
\includegraphics[width=\linewidth]{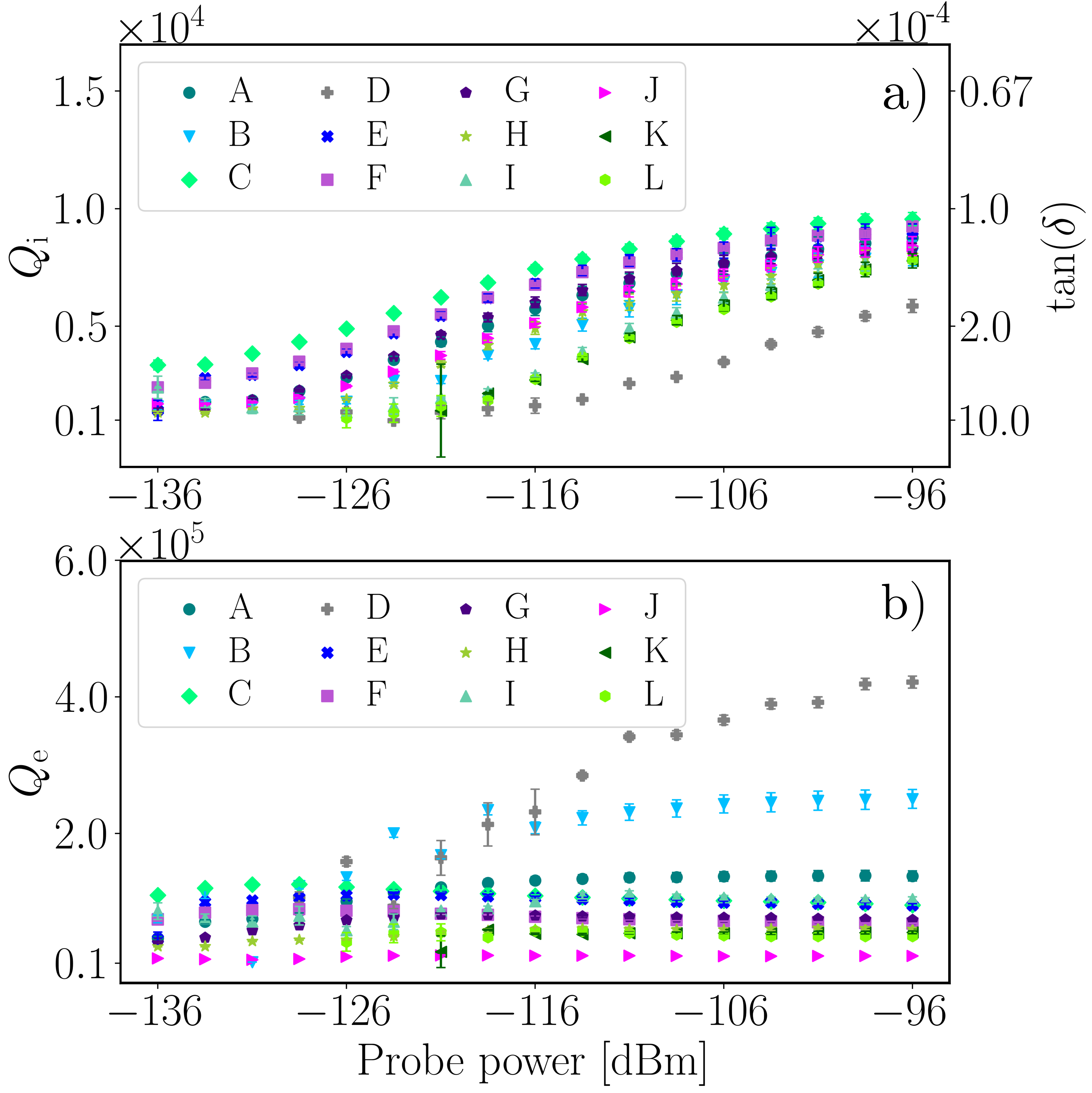}
\caption{Extracted (a) internal and (b) external quality factors (markers) and their $1\sigma$ uncertainties (error bars where larger than marker size) as functions of the probe power for resonators A--L as indicated. The quality factors are determined as an average of 20 measurements each fitted as mentioned in the main text. The vertical axis on the right in panel (a) shows the dielectric loss tangent as defined in the main text. The data is collected as described in the main text with the cryostat at its base temperature.}
\label{fig:q_factors}
\end{figure}

We extract the fundamental resonance frequencies and the quality factors of the tadpole resonators from the measured microwave transmission coefficients through the feed line  with the help of a circular fit in the in-phase--quadrature-phase (IQ) plane, thus utilizing the complete data offered by the transmission coefficient as a function of frequency~\cite{Chen2022, Chen2023, Probst2015}. The Resonator Tools package is used for the fitting and extraction of the parameters~\cite{Probst2015}. More details about the fit can be found in Appendix~\ref{app:fitting}. The extracted internal and external quality factors, $Q_\ui$ and $Q_\ue$, respectively, are presented in Fig.~\ref{fig:q_factors} as a function of the probe power through the feed line. The fundamental resonance frequencies, $f_\ur$, and characteristic impedances of the resonators, $Z_c$, along with other relevant parameters can be found in Table~\ref{tab:table1}.

From the values provided in Table~\ref{tab:table1}, we observe that we can indeed reach very low characteristic impedances for the tadpole resonator. The tunability of the impedance is not limited to this range, however. By varying the length of the CPW strip one can reach values even lower than presented here, while obtaining higher values is also naturally possible.

Another key beneficial characteristic property of the tadpole is its small footprint. Table~\ref{tab:table1} also provides the calculated ratios $l_\mathrm{tot}/\lambda_0$, where $l_\mathrm{tot}$ is the total length of the tadpole resonator including the added length due to the PPC and $\lambda_0=c/(f_\ur\sqrt{\epsilon_\mathrm{eff}})$ is the wave length of the fundamental mode as determined by the speed of light in vacuum, $c$, and the effective relative permittivity, $\epsilon_\mathrm{eff}$, discussed above. Note, that for a PPC size approaching zero, this ratio tends to $1/4$, i.e., the resonator approaches a regular quarter wavelength CPW resonator. From the values given in the table we note that the tadpole resonator reaches significantly smaller values than those of the quarter wavelength resonator, highlighting the small characteristic size of the tadpole.

Let us next discuss the quality factors. Figure~\ref{fig:q_factors}(a) shows that, on average, we can reach internal quality factors of the order of several thousands at reasonably low probe powers. We find an average internal quality factor of $\num{5.2e3}$ at the intermediate probe power of $\qty{-116}{\unit{dBm}}$. The internal quality factor increases linearly with probe power in the intermediate power range and displays saturation at both, low and high power limits. Note that the higher frequency resonators tend to exhibit saturation at higher power at the low-power regime, which is expected as per the higher photon energy. We find that in the low-power limit, the average internal quality factor saturates at $Q_\mathrm{i}=\num{2.3e3}$ and in the high power limit at $Q_\mathrm{i}=\num{8.5e3}$. The single-photon probe power, i.e. the power required to keep an average of one photon in the resonator, is given in Table~\ref{tab:table1}. The single-photon probe power is given by the Resonator Tools package, but for completeness, let us denote that it can be solved from the equation for the average number of photons in the resonator, which is given as the average energy in the resonator divided by the single photon energy~\cite{Sage2011, Probst2015} as
\begin{align}
\langle n_\mathrm{ph}\rangle=\frac{\langle E_\mathrm{r}\rangle}{hf_\ur}=\frac{Q_\mathrm{L}^2}{\pi h f_\ur^2 Q_\ue}P_\mathrm{in},
\label{eq:photon_n}
\end{align}
expressed in terms of the probe power in the input line at the sample, $P_\mathrm{in}$, the loaded quality factor, $Q_\mathrm{L}=\left(1/Q_\ui+1/Q_\ue\right)^{-1}$, and the Planck constant, $h$.

The external quality factors presented in Fig.~\ref{fig:q_factors}(b) are about an order of magnitude higher than the internal quality factors, confirming the rather weak coupling to the feed line. As expected, the external quality factors exhibit no dependence on power. Here, we note that resonators B and D seem to be somewhat of an outliers in the sense that they stand out as having higher external quality factor than the other resonators, especially at high probe power, even though the designed coupling is identical for all resonators. It may be that these two resonators have a weaker coupling to the transmission line due to some fabrication inconsistency resulting in poor signal to noise ratio, and thus, appear as outliers in the data.

The above-mentioned linear behaviour with saturation at both ends of the power spectrum is expected and supports the TLS loss model of the PPC dielectric~\cite{Sage2011, Zotova2023}.
Furthermore, we measured a control sample of CPW resonators without PPCs, fabricated with the same process, and found the quality factor of such resonators to be of order $\num{3e5}$. This supports the assumption that most of the losses arise from the PPC dielectric.

\begin{figure}[htb]
\centering
\includegraphics[width=\linewidth]{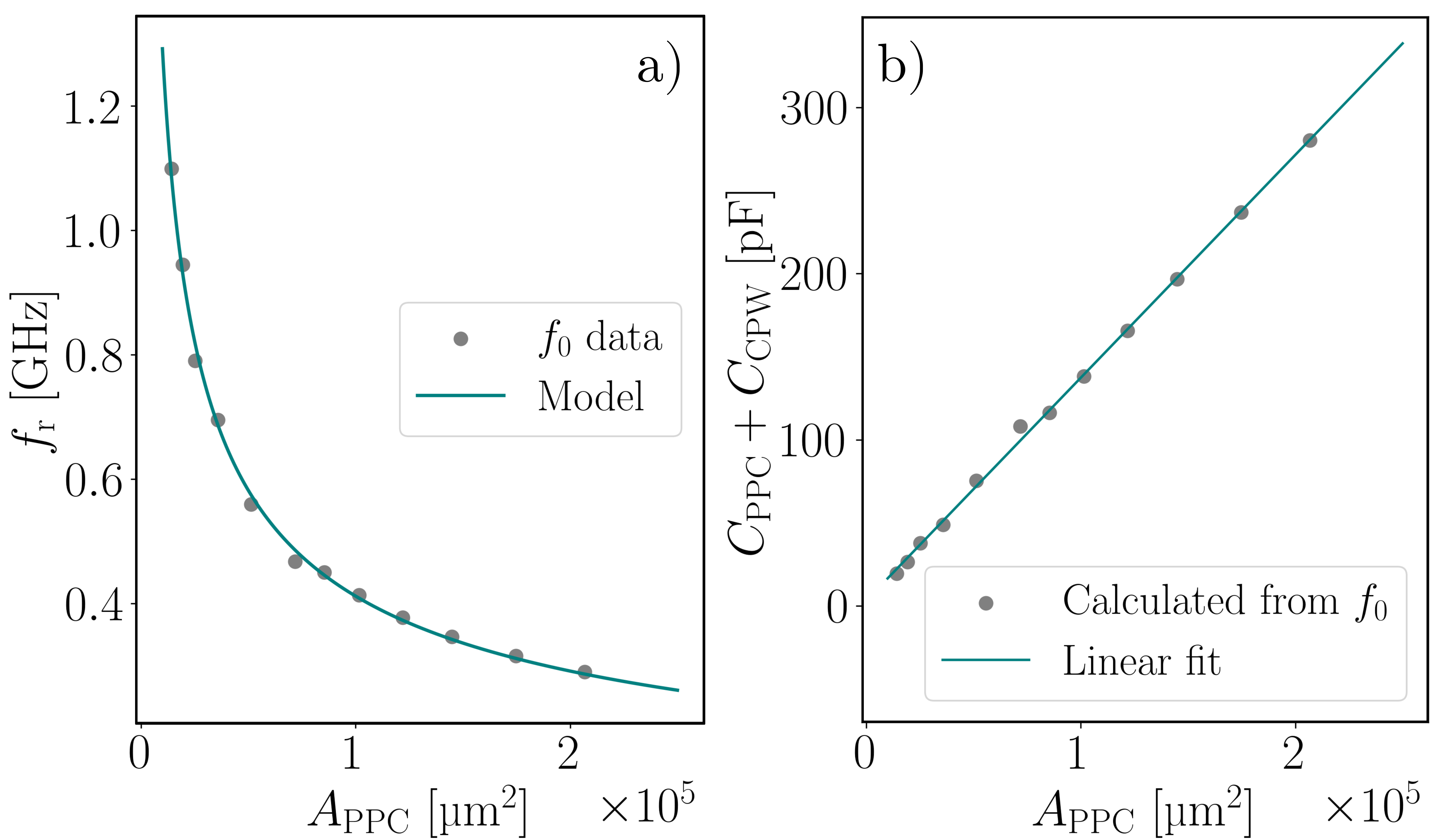}
\caption{Measured (a) resonance frequencies and (b) total capacitances of the resonators A--L as functions of the parallel-plate-capacitor surface area. The frequency data is fitted with the lumped-element-resonator model described in the main text. The total capacitances are computed based on these fits, as described in the main text, and fitted with a straight line. The data was collected as described in the main text at the base temperature of the cryostat.}
\label{fig:fits}
\end{figure}

Let us next compare our lumped-element model for the resonator frequency against the measured data. In Fig.~\ref{fig:fits}(a), we fit the model, defined by equation~\ref{eq:freq}, to the measured resonator frequency data as a function of the PPC surface area, $A_\ppc$. The only fitting parameter is the capacitance per unit area, as we do not have a recent verification for that for the used fabrication recipe. The effective permittivity of the CPW, $\epsilon_{\mathrm{eff}}$, we find by finite-element electromagnetic simulation and estimate the capacitance and inductance per unit length of the CPW analytically~\cite{Goppl08}. From this fit we find the capacitance per unit area to be $c_0=\qty{1.39}{fF/\micro m^2}$, which matches very well to the value, $c_0=\qty{1.4}{fF/\micro m^2}$, reported for the same recipe in reference~\cite{Lake2017}. From table~\ref{tab:table1} we see that, on average, the fitted model can estimate the resonance frequency within 1.7\,\% of the measured value. In Fig.~\ref{fig:fits}(b), we show the total capacitance, $C_\ppc+C_\cpw$, as a function of PPC surface area, $A_\ppc$. The total capacitance values are obtained by solving equation~\ref{eq:freq} for the total capacitance and plugging in the measured resonance frequencies and the determined $c_0$. The data is fitted with a simple straight line to highlight the linear dependence of total capacitance to PPC area. Based on these results, the tadpole resonator can be described by the lumped-element model to a remarkable precision.

\begin{figure}[htb]
\centering
\includegraphics[width=\linewidth]{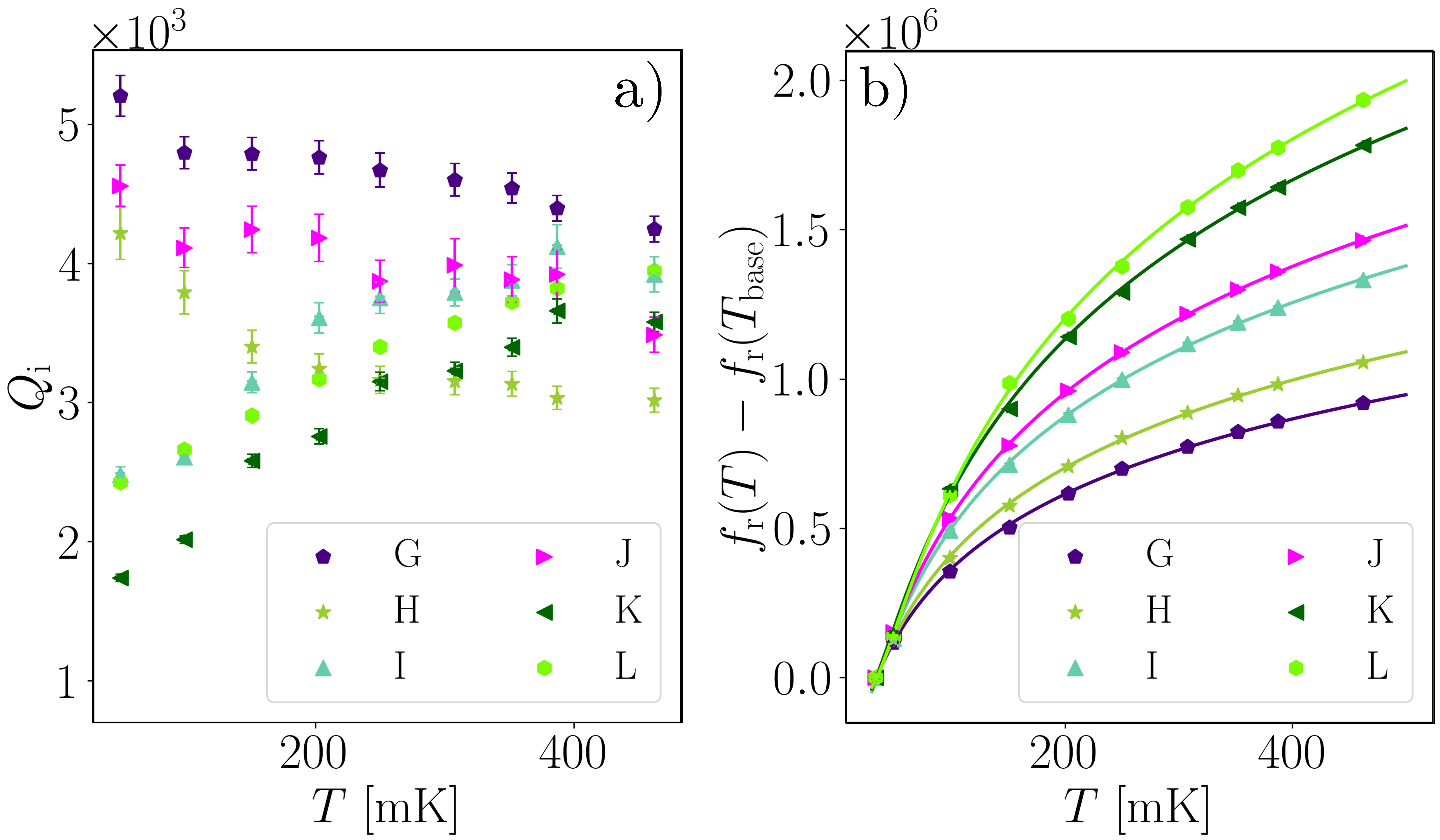}
\caption{Extracted (a) internal quality factors and (b) resonance frequencies of the tadpole resonators G--L (markers) and their $1\sigma$ uncertainties as functions of temperature. The resonance frequencies are expressed as the difference to the base-temperature value $f_\ur(T_\mathrm{base})$ found in Table~\ref{tab:table1}. The solid lines in panel (b) depict fits of the temperature dependence of the resonance frequency according to the two-level-system model~\cite{Sage2011,  Gao2008, Pappas2011} as described in the main text.}
\label{fig:temp}
\end{figure}

In addition to the above characterization of the tadpole resonators at the base temperature of the cryostat, we present the temperature dependence of the internal quality factors and frequencies of resonators A--L in Fig.~\ref{fig:temp}. We find that the quality factors remain relatively constant or slightly increase with increasing temperature, and that the resonance frequencies increase with temperature. Both observations are expected in a system where the intrinsic dissipation is dominated by TLS losses~\cite{Pappas2011, Sage2011, Gao2008}.

\clearpage
Let us utilize the more consistent frequency data to analyze the temperature dependence slightly further. We fit the resonance frequency data with the TLS model for the temperature dependence of the resonance frequency~\cite{Sage2011, Gao2008, Pappas2011} given by
\begin{widetext}
\begin{align}
f_\ur(T)=f_0\left[1+\frac{F\delta_\mathrm{TLS}^0}{\pi}\left\{\mathrm{Re}\left[\Psi\left(\frac{1}{2}+\frac{hf_0}{2\ui\pi k_\mathrm{B}T}\right)\right]-\ln\left(\frac{hf_\ur}{k_\mathrm{B}T}\right)\right\}\right],
\label{eq:freq_temp}
\end{align}
\end{widetext}
where $f_0$ is the resonator frequency at zero temperature, $\delta_\mathrm{TLS}^0$ is the loss tangent at zero temperature, $F$ is the filling factor, defined as the fraction of the total electrical energy of the resonator within the TLS supporting material, $\Psi$ is the digamma function, and $k_\mathrm{B}$ is the Boltzmann constant.

In Eq.~\eqref{eq:freq_temp}, $\delta_\mathrm{TLS}^0$ and $f_0$ are the fitting parameters. For devices with large PPCs, $F$ can be set to unity, since much of the electric field is contained within the PPC dielectric that is also the source of the TLSs~\cite{McRae2020}. Based on Fig.~\ref{fig:temp}(b) we find a good agreement between the model and the experimental data. The fit places $f_0$ very close to the measured base temperature values of $f_\ur$, found in Table~\ref{tab:table1}, with an average relative difference below $0.1\,\%$. In order to compare $\delta_\mathrm{TLS}^0$ to the measured loss tangent, one needs take into account the temperature dependence of the loss tangent approximately given by a hyberbolic tangent as $\tan(\delta)\approx\delta_\mathrm{TLS}^0\tanh[hf_\ur/(2k_\mathrm{B}T)]$, for low probe power~\cite{Gao2008, McRae2020}. We find that this yields an estimate well in line with the measured values presented in Fig.~\ref{fig:q_factors}(a). These findings strongly support the assumption that the TLSs are the the main loss channel of the tadpole resonators.

As shown above, most of the internal losses of the tadpole resonator can be attributed to the TLS losses in the PPC dielectric. Thus, assuming the lumped-element model for the tadpole resonator, one can extract the loss tangent of the $\mathrm{AlO}_x$ used as the dielectric in the PPC. Based on the extracted internal quality factors, we find the average loss tangent to vary between $\tan(\delta)=\num{1.2e-4}$ and $\tan(\delta)=\num{4.3e-4}$ as a function of probe power, as shown in Fig.~\ref{fig:q_factors}(a).

\section{Conclusions}

We demonstrated a simple and versatile low-characteristic-impedance lumped-element resonator design based on a strip of a conventional CPW transmission line shunted with a parallel-plate capacitor resulting in a structure shaped like a tadpole. We fabricated twelve tadpole resonators in the sub-$\qty{10}{\ohm}$ range and characterize the them at subkelvin temperatures reaching internal quality factors of the order of $Q_\ui=\num{1e4}$. We demonstrate characteristic impedances ranging from $Z_\textrm{c}=\qty{2}{\ohm}$ to~$\qty{10}{\ohm}$ and a frequency range from $f_\ur=\qty{290}{MHz}$ to~$\qty{1.1}{GHz}$. We further demonstrated that the resonator can, indeed, be considered as a lumped-element resonator where most of the losses arise from the dielectric losses of the PPC dielectric, translating into a loss tangent of order $\tan(\delta)=1/Q_\ui=\num{1.0e-4}$ for the used $\mathrm{AlO}_x$. 

We further showed that the frequency of the tadpole resonator can be modelled by a simple analytical model to a high accuracy. Typically, the values of the physical quantities needed for the model are known for well-established recipes, but the values can also be found by finite-element electromagnetic simulation or even by analytical expressions and tabulate values for a rough approximation of the resonance frequency.

Since the internal quality factors of the current tadpole resonators are much lower than what is routinely achieved in superconducting CPW resonators~\cite{Zikiy2023, Frunzio2005, Goppl08}, the natural question arises whether this could be improved. As we have showed, the main source of losses are the TLSs in the PPC dielectric. It is thus interesting to search for improvements by trying to reduce the dielectric loss tangent. The two main ways of improving the loss tangent are enhancing the deposition method or changing the dielectric material altogether. In ALD aluminum oxide with similar thickness as used here, loss tangents of the order of $10^{-5}$ have been reported~\cite{Beldi2019}, suggesting that our fabrication process may be subject to improvement. With Josephson junction oxidation and annealing process, loss tangents even as low as $5\times 10^{-7}$ have been reported~\cite{Mamin2021}, but for thin oxide layers of roughly $\qty{2}{\nano\m}$, which could cause the, here unwanted, Josephson inductance becoming significant. For the other option to replace the dielectric entirely, for instance, in $\mathrm{SiO}_x$ loss tangents of order $10^{-5}$ and below have been reported~\cite{Paik2010}. This, however, typically comes with the cost of a low dielectric constant, as is the case for $\mathrm{SiO}_x$, reaching values of roughly half of that of $\mathrm{AlO}_x$.

While the tadpole resonator offered only a moderate internal quality factor, it boasts a number of advantageous properties including versatility, relatively simple fabrication process, small footprint on the chip even at extremely low frequencies, tunable characteristic impedance in fabrication, and inherent capability of strong inductive coupling using low coupling inductance. This renders the tadpole resonator a viable candidate for multiple novel low-frequency applications where reaching top-notch internal quality factors is not of great importance. For instance, the SQUID-mediated inductive coupling proposed in Refs.~\cite{Johansson, Kim} would benefit significantly from the concentrated magnetic field of the tadpole resonator resulting in strong coupling.

\begin{acknowledgments}
We acknowledge the support from the members of the QCD and PICO groups at Aalto University. Especially, we thank Jukka Pekola, Bayan Karimi, Christoforus Satrya, Qiming Chen, and Suman Kundu for fruitful scientific discourse and other help. 

This work was funded by the Academy of Finland Centre of Excellence program (project Nos.~352925, and 336810) and Academy of Finland grant Nos.~316619 and 349594 (THEPOW). We also acknowledge funding from the European Research Council under Advanced Grant No.~101053801 (ConceptQ), and Business Finland under the Quantum Technologies Industrial project (Grant no.~2118781).
\end{acknowledgments}

\appendix

\section{Circle fit}
\label{app:fitting}

\begin{figure}[htb]
\centering
\includegraphics[width=\linewidth]{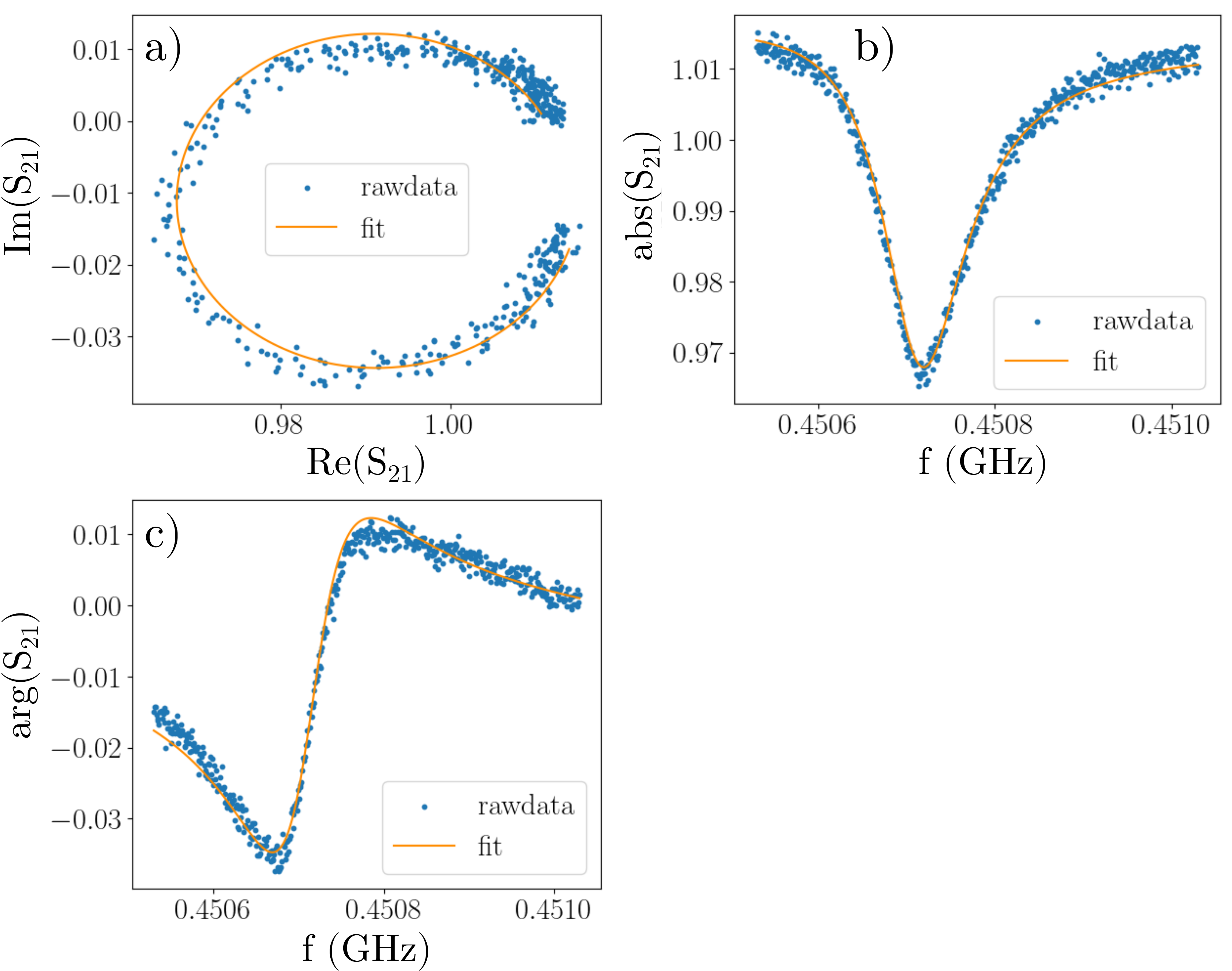}
\caption{Example fit for the transmission coefficient $S_{21}$ produced by the Resonator Tools package (a) in the complex plane and for (b) the magnitude and (c) phase as function of the probe frequency.}
\label{fig:resonator_tools}
\end{figure}

As mentioned in the main text, we employ the Resonator Tools package~\cite{Probst2015} for the purpose of fitting and analyzing the data. Resonator Tools does a circular fit in the in-phase–quadrature-phase (IQ) plane in order to extract the desired parameters from the measured complex-valued microwave transmission coefficient. It uses the following complex-valued function for the fit:
\begin{align}
S_{21}(f)=a\ue^{\ui\alpha}\ue^{-2\pi\ui f\tau}\left[1-\frac{Q_\mathrm{L}/|Q_\ue|\ue^{\ui\phi}}{1+2\ui Q_\mathrm{L}(f/f_\ur-1)}\right],
\label{eq:s21}
\end{align}
where $f$ is the probe frequency, the amplitude $a$, phase shift $\alpha$, and the electric delay $\tau$ take into account the effects caused by the electromagnetic environment of the resonator, such as the connecting cables, and the parameter $\phi$ takes into account the impedance mismatch. All these parameters are varied in the fitting procedure, although only some of them provide interesting information about the system being studied. Fig.~\ref{fig:resonator_tools} presents an example fit produced by the Resonator Tools package.

\section{Finite-element modeling}
\label{app:current}
\begin{figure}[htb]
\centering
\includegraphics[width=0.9\linewidth]{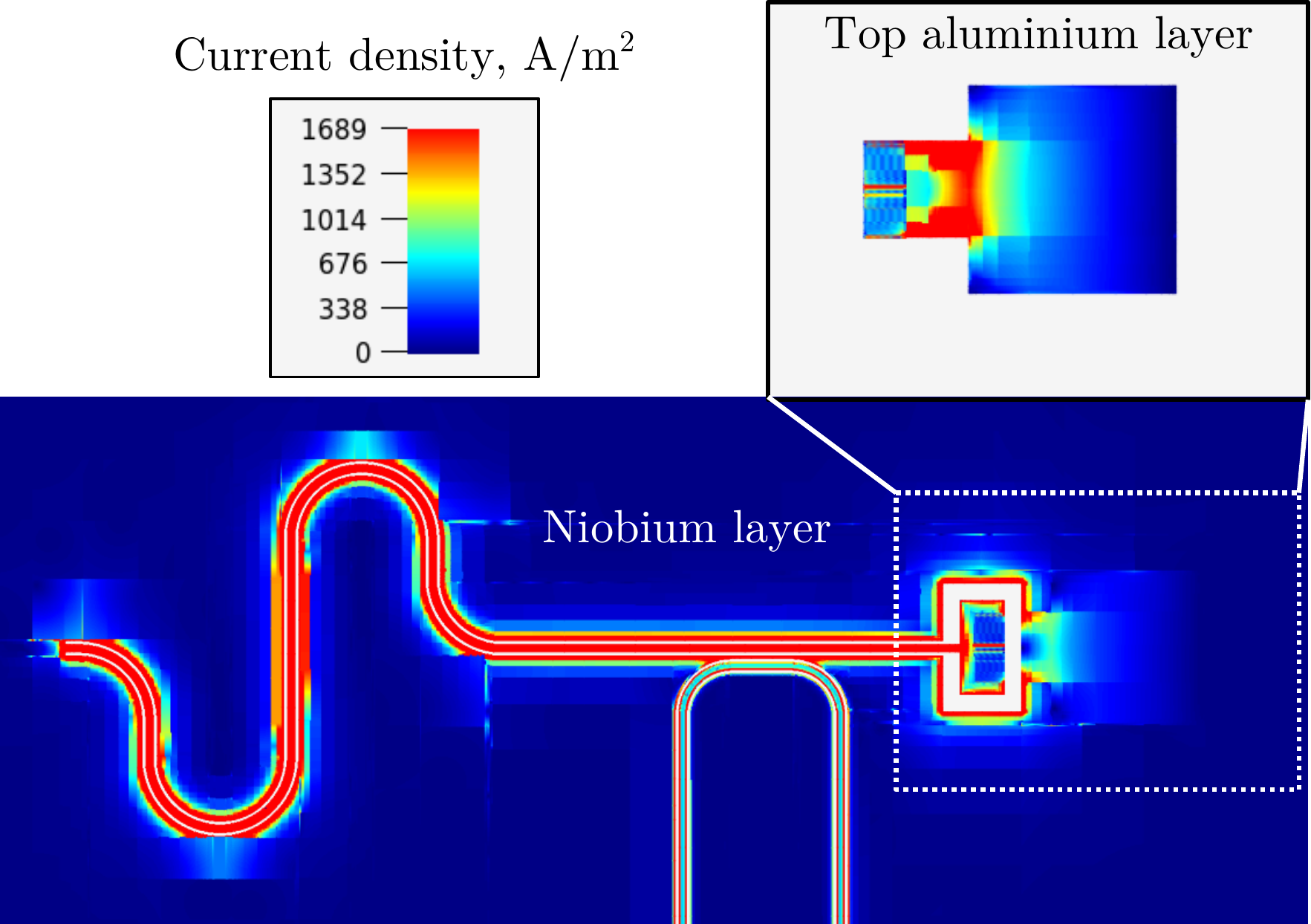}
\caption{Current density in the tadpole resonator at its resonance frequency. The data is produced by a commercial electromagnetic finite-element modeling software Sonnet~\cite{sonnet}. The inset shows the area highlighted with the white dashed rectangle in the main panel.}
\label{fig:current}
\end{figure}

In designing the tadpole resonators, we also carry out electromagnetic finite-element modeling of the resonator. In Fig.~\ref{fig:current}, we present the current density obtained as a result of such a simulation. We observe that the current density is essentially constant over the length of the CPW strip. This suggests that the tadpole acts as a lumped-element resonator with an almost constant field amplitude within the CPW strip as opposed to a traditional CPW resonator.

\section{Cryogenic setup}
\label{app:sample}

\begin{figure}[htb]
\centering
\includegraphics[width=0.9\linewidth]{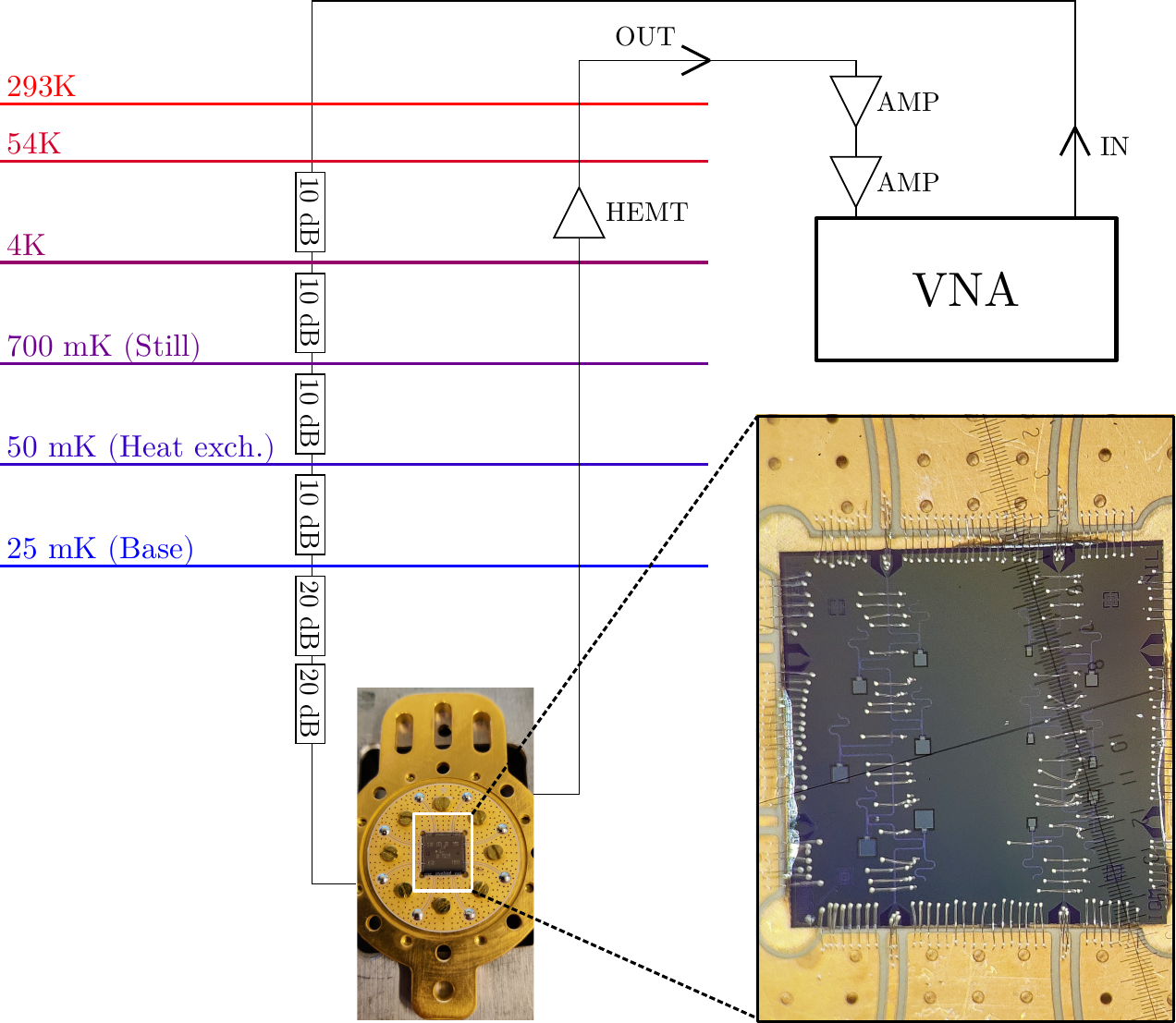}
\caption{Schematic of the experimental setup concerning the cryostat and its wiring together with an image of the sample holder at the bottom left and an inset detailing the sample on bottom right. The wires going to the sample holder are connected to the back of the holder by standard non-magnetic SMA connectors. The box called VNA denotes the vector network analyzer, and the triangles labeled HEMT and AMP denote the high-electron-mobility transistor amplifier and other amplifiers, respectively. The the attenuators are shown on the input line and the arrows labeled IN and OUT provide the directions of the signal.}
\label{fig:cryo_setup}
\end{figure}

Figure~\ref{fig:cryo_setup} presents a schematic of the experimental setup.

\bibliography{cites}

\begin{thebibliography}{80}%
\makeatletter
\providecommand \@ifxundefined [1]{%
 \@ifx{#1\undefined}
}%
\providecommand \@ifnum [1]{%
 \ifnum #1\expandafter \@firstoftwo
 \else \expandafter \@secondoftwo
 \fi
}%
\providecommand \@ifx [1]{%
 \ifx #1\expandafter \@firstoftwo
 \else \expandafter \@secondoftwo
 \fi
}%
\providecommand \natexlab [1]{#1}%
\providecommand \enquote  [1]{``#1''}%
\providecommand \bibnamefont  [1]{#1}%
\providecommand \bibfnamefont [1]{#1}%
\providecommand \citenamefont [1]{#1}%
\providecommand \href@noop [0]{\@secondoftwo}%
\providecommand \href [0]{\begingroup \@sanitize@url \@href}%
\providecommand \@href[1]{\@@startlink{#1}\@@href}%
\providecommand \@@href[1]{\endgroup#1\@@endlink}%
\providecommand \@sanitize@url [0]{\catcode `\\12\catcode `\$12\catcode `\&12\catcode `\#12\catcode `\^12\catcode `\_12\catcode `\%12\relax}%
\providecommand \@@startlink[1]{}%
\providecommand \@@endlink[0]{}%
\providecommand \url  [0]{\begingroup\@sanitize@url \@url }%
\providecommand \@url [1]{\endgroup\@href {#1}{\urlprefix }}%
\providecommand \urlprefix  [0]{URL }%
\providecommand \Eprint [0]{\href }%
\providecommand \doibase [0]{https://doi.org/}%
\providecommand \selectlanguage [0]{\@gobble}%
\providecommand \bibinfo  [0]{\@secondoftwo}%
\providecommand \bibfield  [0]{\@secondoftwo}%
\providecommand \translation [1]{[#1]}%
\providecommand \BibitemOpen [0]{}%
\providecommand \bibitemStop [0]{}%
\providecommand \bibitemNoStop [0]{.\EOS\space}%
\providecommand \EOS [0]{\spacefactor3000\relax}%
\providecommand \BibitemShut  [1]{\csname bibitem#1\endcsname}%
\let\auto@bib@innerbib\@empty
\bibitem [{\citenamefont {Schleich}(2016)}]{Wolfgang}%
  \BibitemOpen
  \bibfield  {author} {\bibinfo {author} {\bibfnamefont {W.~P. e.~a.}\ \bibnamefont {Schleich}},\ }\bibfield  {title} {\bibinfo {title} {Quantum technology: from research to application},\ }\href {https://doi.org/10.1007/s00340-016-6353-8} {\bibfield  {journal} {\bibinfo  {journal} {Applied Physics B}\ }\textbf {\bibinfo {volume} {122}},\ \bibinfo {pages} {130} (\bibinfo {year} {2016})}\BibitemShut {NoStop}%
\bibitem [{\citenamefont {Krantz}\ \emph {et~al.}(2019)\citenamefont {Krantz}, \citenamefont {Kjaergaard}, \citenamefont {Yan}, \citenamefont {Orlando}, \citenamefont {Gustavsson},\ and\ \citenamefont {Oliver}}]{Qengineer}%
  \BibitemOpen
  \bibfield  {author} {\bibinfo {author} {\bibfnamefont {P.}~\bibnamefont {Krantz}}, \bibinfo {author} {\bibfnamefont {M.}~\bibnamefont {Kjaergaard}}, \bibinfo {author} {\bibfnamefont {F.}~\bibnamefont {Yan}}, \bibinfo {author} {\bibfnamefont {T.~P.}\ \bibnamefont {Orlando}}, \bibinfo {author} {\bibfnamefont {S.}~\bibnamefont {Gustavsson}},\ and\ \bibinfo {author} {\bibfnamefont {W.~D.}\ \bibnamefont {Oliver}},\ }\bibfield  {title} {\bibinfo {title} {{A quantum engineer's guide to superconducting qubits}},\ }\href {https://doi.org/10.1063/1.5089550} {\bibfield  {journal} {\bibinfo  {journal} {Applied Physics Reviews}\ }\textbf {\bibinfo {volume} {6}},\ \bibinfo {pages} {021318} (\bibinfo {year} {2019})},\ \Eprint {https://arxiv.org/abs/https://pubs.aip.org/aip/apr/article-pdf/doi/10.1063/1.5089550/16667201/021318\_1\_online.pdf} {https://pubs.aip.org/aip/apr/article-pdf/doi/10.1063/1.5089550/16667201/021318\_1\_online.pdf} \BibitemShut {NoStop}%
\bibitem [{\citenamefont {DiCarlo}\ \emph {et~al.}(2009)\citenamefont {DiCarlo}, \citenamefont {Chow}, \citenamefont {Gambetta}, \citenamefont {Bishop}, \citenamefont {Johnson}, \citenamefont {Schuster}, \citenamefont {Majer}, \citenamefont {Blais}, \citenamefont {Frunzio}, \citenamefont {Girvin},\ and\ \citenamefont {Schoelkopf}}]{DiCarlo2009}%
  \BibitemOpen
  \bibfield  {author} {\bibinfo {author} {\bibfnamefont {L.}~\bibnamefont {DiCarlo}}, \bibinfo {author} {\bibfnamefont {J.~M.}\ \bibnamefont {Chow}}, \bibinfo {author} {\bibfnamefont {J.~M.}\ \bibnamefont {Gambetta}}, \bibinfo {author} {\bibfnamefont {L.~S.}\ \bibnamefont {Bishop}}, \bibinfo {author} {\bibfnamefont {B.~R.}\ \bibnamefont {Johnson}}, \bibinfo {author} {\bibfnamefont {D.~I.}\ \bibnamefont {Schuster}}, \bibinfo {author} {\bibfnamefont {J.}~\bibnamefont {Majer}}, \bibinfo {author} {\bibfnamefont {A.}~\bibnamefont {Blais}}, \bibinfo {author} {\bibfnamefont {L.}~\bibnamefont {Frunzio}}, \bibinfo {author} {\bibfnamefont {S.~M.}\ \bibnamefont {Girvin}},\ and\ \bibinfo {author} {\bibfnamefont {R.~J.}\ \bibnamefont {Schoelkopf}},\ }\bibfield  {title} {\bibinfo {title} {Demonstration of two-qubit algorithms with a superconducting quantum processor},\ }\href {https://doi.org/10.1038/nature08121} {\bibfield  {journal} {\bibinfo  {journal} {Nature}\ }\textbf {\bibinfo {volume} {460}},\ \bibinfo {pages}
  {240} (\bibinfo {year} {2009})}\BibitemShut {NoStop}%
\bibitem [{\citenamefont {Lucero}\ \emph {et~al.}(2012)\citenamefont {Lucero}, \citenamefont {Barends}, \citenamefont {Chen}, \citenamefont {Kelly}, \citenamefont {Mariantoni}, \citenamefont {Megrant}, \citenamefont {O'Malley}, \citenamefont {Sank}, \citenamefont {Vainsencher}, \citenamefont {Wenner}, \citenamefont {White}, \citenamefont {Yin}, \citenamefont {Cleland},\ and\ \citenamefont {Martinis}}]{Lucero2012}%
  \BibitemOpen
  \bibfield  {author} {\bibinfo {author} {\bibfnamefont {E.}~\bibnamefont {Lucero}}, \bibinfo {author} {\bibfnamefont {R.}~\bibnamefont {Barends}}, \bibinfo {author} {\bibfnamefont {Y.}~\bibnamefont {Chen}}, \bibinfo {author} {\bibfnamefont {J.}~\bibnamefont {Kelly}}, \bibinfo {author} {\bibfnamefont {M.}~\bibnamefont {Mariantoni}}, \bibinfo {author} {\bibfnamefont {A.}~\bibnamefont {Megrant}}, \bibinfo {author} {\bibfnamefont {P.}~\bibnamefont {O'Malley}}, \bibinfo {author} {\bibfnamefont {D.}~\bibnamefont {Sank}}, \bibinfo {author} {\bibfnamefont {A.}~\bibnamefont {Vainsencher}}, \bibinfo {author} {\bibfnamefont {J.}~\bibnamefont {Wenner}}, \bibinfo {author} {\bibfnamefont {T.}~\bibnamefont {White}}, \bibinfo {author} {\bibfnamefont {Y.}~\bibnamefont {Yin}}, \bibinfo {author} {\bibfnamefont {A.~N.}\ \bibnamefont {Cleland}},\ and\ \bibinfo {author} {\bibfnamefont {J.~M.}\ \bibnamefont {Martinis}},\ }\bibfield  {title} {\bibinfo {title} {Computing prime factors with a josephson phase qubit quantum
  processor},\ }\href {https://doi.org/10.1038/nphys2385} {\bibfield  {journal} {\bibinfo  {journal} {Nature Physics}\ }\textbf {\bibinfo {volume} {8}},\ \bibinfo {pages} {719} (\bibinfo {year} {2012})}\BibitemShut {NoStop}%
\bibitem [{\citenamefont {Zheng}\ \emph {et~al.}(2017)\citenamefont {Zheng}, \citenamefont {Song}, \citenamefont {Chen}, \citenamefont {Xia}, \citenamefont {Liu}, \citenamefont {Guo}, \citenamefont {Zhang}, \citenamefont {Xu}, \citenamefont {Deng}, \citenamefont {Huang}, \citenamefont {Wu}, \citenamefont {Yan}, \citenamefont {Zheng}, \citenamefont {Lu}, \citenamefont {Pan}, \citenamefont {Wang}, \citenamefont {Lu},\ and\ \citenamefont {Zhu}}]{Zheng2017}%
  \BibitemOpen
  \bibfield  {author} {\bibinfo {author} {\bibfnamefont {Y.}~\bibnamefont {Zheng}}, \bibinfo {author} {\bibfnamefont {C.}~\bibnamefont {Song}}, \bibinfo {author} {\bibfnamefont {M.-C.}\ \bibnamefont {Chen}}, \bibinfo {author} {\bibfnamefont {B.}~\bibnamefont {Xia}}, \bibinfo {author} {\bibfnamefont {W.}~\bibnamefont {Liu}}, \bibinfo {author} {\bibfnamefont {Q.}~\bibnamefont {Guo}}, \bibinfo {author} {\bibfnamefont {L.}~\bibnamefont {Zhang}}, \bibinfo {author} {\bibfnamefont {D.}~\bibnamefont {Xu}}, \bibinfo {author} {\bibfnamefont {H.}~\bibnamefont {Deng}}, \bibinfo {author} {\bibfnamefont {K.}~\bibnamefont {Huang}}, \bibinfo {author} {\bibfnamefont {Y.}~\bibnamefont {Wu}}, \bibinfo {author} {\bibfnamefont {Z.}~\bibnamefont {Yan}}, \bibinfo {author} {\bibfnamefont {D.}~\bibnamefont {Zheng}}, \bibinfo {author} {\bibfnamefont {L.}~\bibnamefont {Lu}}, \bibinfo {author} {\bibfnamefont {J.-W.}\ \bibnamefont {Pan}}, \bibinfo {author} {\bibfnamefont {H.}~\bibnamefont {Wang}}, \bibinfo {author} {\bibfnamefont
  {C.-Y.}\ \bibnamefont {Lu}},\ and\ \bibinfo {author} {\bibfnamefont {X.}~\bibnamefont {Zhu}},\ }\bibfield  {title} {\bibinfo {title} {Solving systems of linear equations with a superconducting quantum processor},\ }\href {https://doi.org/10.1103/PhysRevLett.118.210504} {\bibfield  {journal} {\bibinfo  {journal} {Phys. Rev. Lett.}\ }\textbf {\bibinfo {volume} {118}},\ \bibinfo {pages} {210504} (\bibinfo {year} {2017})}\BibitemShut {NoStop}%
\bibitem [{\citenamefont {Chen}\ \emph {et~al.}(2020)\citenamefont {Chen}, \citenamefont {Gong}, \citenamefont {Xu}, \citenamefont {Yuan}, \citenamefont {Wang}, \citenamefont {Wang}, \citenamefont {Ying}, \citenamefont {Lin}, \citenamefont {Xu}, \citenamefont {Wu}, \citenamefont {Wang}, \citenamefont {Deng}, \citenamefont {Liang}, \citenamefont {Peng}, \citenamefont {Benjamin}, \citenamefont {Zhu}, \citenamefont {Lu},\ and\ \citenamefont {Pan}}]{Chen2020}%
  \BibitemOpen
  \bibfield  {author} {\bibinfo {author} {\bibfnamefont {M.-C.}\ \bibnamefont {Chen}}, \bibinfo {author} {\bibfnamefont {M.}~\bibnamefont {Gong}}, \bibinfo {author} {\bibfnamefont {X.}~\bibnamefont {Xu}}, \bibinfo {author} {\bibfnamefont {X.}~\bibnamefont {Yuan}}, \bibinfo {author} {\bibfnamefont {J.-W.}\ \bibnamefont {Wang}}, \bibinfo {author} {\bibfnamefont {C.}~\bibnamefont {Wang}}, \bibinfo {author} {\bibfnamefont {C.}~\bibnamefont {Ying}}, \bibinfo {author} {\bibfnamefont {J.}~\bibnamefont {Lin}}, \bibinfo {author} {\bibfnamefont {Y.}~\bibnamefont {Xu}}, \bibinfo {author} {\bibfnamefont {Y.}~\bibnamefont {Wu}}, \bibinfo {author} {\bibfnamefont {S.}~\bibnamefont {Wang}}, \bibinfo {author} {\bibfnamefont {H.}~\bibnamefont {Deng}}, \bibinfo {author} {\bibfnamefont {F.}~\bibnamefont {Liang}}, \bibinfo {author} {\bibfnamefont {C.-Z.}\ \bibnamefont {Peng}}, \bibinfo {author} {\bibfnamefont {S.~C.}\ \bibnamefont {Benjamin}}, \bibinfo {author} {\bibfnamefont {X.}~\bibnamefont {Zhu}}, \bibinfo {author}
  {\bibfnamefont {C.-Y.}\ \bibnamefont {Lu}},\ and\ \bibinfo {author} {\bibfnamefont {J.-W.}\ \bibnamefont {Pan}},\ }\bibfield  {title} {\bibinfo {title} {Demonstration of adiabatic variational quantum computing with a superconducting quantum coprocessor},\ }\href {https://doi.org/10.1103/PhysRevLett.125.180501} {\bibfield  {journal} {\bibinfo  {journal} {Phys. Rev. Lett.}\ }\textbf {\bibinfo {volume} {125}},\ \bibinfo {pages} {180501} (\bibinfo {year} {2020})}\BibitemShut {NoStop}%
\bibitem [{\citenamefont {Harrigan}\ \emph {et~al.}(2021)\citenamefont {Harrigan}, \citenamefont {Sung}, \citenamefont {Neeley}, \citenamefont {Satzinger}, \citenamefont {Arute}, \citenamefont {Arya}, \citenamefont {Atalaya}, \citenamefont {Bardin}, \citenamefont {Barends}, \citenamefont {Boixo}, \citenamefont {Broughton}, \citenamefont {Buckley}, \citenamefont {Buell}, \citenamefont {Burkett}, \citenamefont {Bushnell}, \citenamefont {Chen}, \citenamefont {Chen}, \citenamefont {Chiaro}, \citenamefont {Collins}, \citenamefont {Courtney}, \citenamefont {Demura}, \citenamefont {Dunsworth}, \citenamefont {Eppens}, \citenamefont {Fowler}, \citenamefont {Foxen}, \citenamefont {Gidney}, \citenamefont {Giustina}, \citenamefont {Graff}, \citenamefont {Habegger}, \citenamefont {Ho}, \citenamefont {Hong}, \citenamefont {Huang}, \citenamefont {Ioffe}, \citenamefont {Isakov}, \citenamefont {Jeffrey}, \citenamefont {Jiang}, \citenamefont {Jones}, \citenamefont {Kafri}, \citenamefont {Kechedzhi}, \citenamefont {Kelly},
  \citenamefont {Kim}, \citenamefont {Klimov}, \citenamefont {Korotkov}, \citenamefont {Kostritsa}, \citenamefont {Landhuis}, \citenamefont {Laptev}, \citenamefont {Lindmark}, \citenamefont {Leib}, \citenamefont {Martin}, \citenamefont {Martinis}, \citenamefont {McClean}, \citenamefont {McEwen}, \citenamefont {Megrant}, \citenamefont {Mi}, \citenamefont {Mohseni}, \citenamefont {Mruczkiewicz}, \citenamefont {Mutus}, \citenamefont {Naaman}, \citenamefont {Neill}, \citenamefont {Neukart}, \citenamefont {Niu}, \citenamefont {O'Brien}, \citenamefont {O'Gorman}, \citenamefont {Ostby}, \citenamefont {Petukhov}, \citenamefont {Putterman}, \citenamefont {Quintana}, \citenamefont {Roushan}, \citenamefont {Rubin}, \citenamefont {Sank}, \citenamefont {Skolik}, \citenamefont {Smelyanskiy}, \citenamefont {Strain}, \citenamefont {Streif}, \citenamefont {Szalay}, \citenamefont {Vainsencher}, \citenamefont {White}, \citenamefont {Yao}, \citenamefont {Yeh}, \citenamefont {Zalcman}, \citenamefont {Zhou}, \citenamefont {Neven},
  \citenamefont {Bacon}, \citenamefont {Lucero}, \citenamefont {Farhi},\ and\ \citenamefont {Babbush}}]{Harrigan2021}%
  \BibitemOpen
  \bibfield  {author} {\bibinfo {author} {\bibfnamefont {M.~P.}\ \bibnamefont {Harrigan}}, \bibinfo {author} {\bibfnamefont {K.~J.}\ \bibnamefont {Sung}}, \bibinfo {author} {\bibfnamefont {M.}~\bibnamefont {Neeley}}, \bibinfo {author} {\bibfnamefont {K.~J.}\ \bibnamefont {Satzinger}}, \bibinfo {author} {\bibfnamefont {F.}~\bibnamefont {Arute}}, \bibinfo {author} {\bibfnamefont {K.}~\bibnamefont {Arya}}, \bibinfo {author} {\bibfnamefont {J.}~\bibnamefont {Atalaya}}, \bibinfo {author} {\bibfnamefont {J.~C.}\ \bibnamefont {Bardin}}, \bibinfo {author} {\bibfnamefont {R.}~\bibnamefont {Barends}}, \bibinfo {author} {\bibfnamefont {S.}~\bibnamefont {Boixo}}, \bibinfo {author} {\bibfnamefont {M.}~\bibnamefont {Broughton}}, \bibinfo {author} {\bibfnamefont {B.~B.}\ \bibnamefont {Buckley}}, \bibinfo {author} {\bibfnamefont {D.~A.}\ \bibnamefont {Buell}}, \bibinfo {author} {\bibfnamefont {B.}~\bibnamefont {Burkett}}, \bibinfo {author} {\bibfnamefont {N.}~\bibnamefont {Bushnell}}, \bibinfo {author} {\bibfnamefont
  {Y.}~\bibnamefont {Chen}}, \bibinfo {author} {\bibfnamefont {Z.}~\bibnamefont {Chen}}, \bibinfo {author} {\bibfnamefont {B.}~\bibnamefont {Chiaro}}, \bibinfo {author} {\bibfnamefont {R.}~\bibnamefont {Collins}}, \bibinfo {author} {\bibfnamefont {W.}~\bibnamefont {Courtney}}, \bibinfo {author} {\bibfnamefont {S.}~\bibnamefont {Demura}}, \bibinfo {author} {\bibfnamefont {A.}~\bibnamefont {Dunsworth}}, \bibinfo {author} {\bibfnamefont {D.}~\bibnamefont {Eppens}}, \bibinfo {author} {\bibfnamefont {A.}~\bibnamefont {Fowler}}, \bibinfo {author} {\bibfnamefont {B.}~\bibnamefont {Foxen}}, \bibinfo {author} {\bibfnamefont {C.}~\bibnamefont {Gidney}}, \bibinfo {author} {\bibfnamefont {M.}~\bibnamefont {Giustina}}, \bibinfo {author} {\bibfnamefont {R.}~\bibnamefont {Graff}}, \bibinfo {author} {\bibfnamefont {S.}~\bibnamefont {Habegger}}, \bibinfo {author} {\bibfnamefont {A.}~\bibnamefont {Ho}}, \bibinfo {author} {\bibfnamefont {S.}~\bibnamefont {Hong}}, \bibinfo {author} {\bibfnamefont {T.}~\bibnamefont {Huang}},
  \bibinfo {author} {\bibfnamefont {L.~B.}\ \bibnamefont {Ioffe}}, \bibinfo {author} {\bibfnamefont {S.~V.}\ \bibnamefont {Isakov}}, \bibinfo {author} {\bibfnamefont {E.}~\bibnamefont {Jeffrey}}, \bibinfo {author} {\bibfnamefont {Z.}~\bibnamefont {Jiang}}, \bibinfo {author} {\bibfnamefont {C.}~\bibnamefont {Jones}}, \bibinfo {author} {\bibfnamefont {D.}~\bibnamefont {Kafri}}, \bibinfo {author} {\bibfnamefont {K.}~\bibnamefont {Kechedzhi}}, \bibinfo {author} {\bibfnamefont {J.}~\bibnamefont {Kelly}}, \bibinfo {author} {\bibfnamefont {S.}~\bibnamefont {Kim}}, \bibinfo {author} {\bibfnamefont {P.~V.}\ \bibnamefont {Klimov}}, \bibinfo {author} {\bibfnamefont {A.~N.}\ \bibnamefont {Korotkov}}, \bibinfo {author} {\bibfnamefont {F.}~\bibnamefont {Kostritsa}}, \bibinfo {author} {\bibfnamefont {D.}~\bibnamefont {Landhuis}}, \bibinfo {author} {\bibfnamefont {P.}~\bibnamefont {Laptev}}, \bibinfo {author} {\bibfnamefont {M.}~\bibnamefont {Lindmark}}, \bibinfo {author} {\bibfnamefont {M.}~\bibnamefont {Leib}}, \bibinfo
  {author} {\bibfnamefont {O.}~\bibnamefont {Martin}}, \bibinfo {author} {\bibfnamefont {J.~M.}\ \bibnamefont {Martinis}}, \bibinfo {author} {\bibfnamefont {J.~R.}\ \bibnamefont {McClean}}, \bibinfo {author} {\bibfnamefont {M.}~\bibnamefont {McEwen}}, \bibinfo {author} {\bibfnamefont {A.}~\bibnamefont {Megrant}}, \bibinfo {author} {\bibfnamefont {X.}~\bibnamefont {Mi}}, \bibinfo {author} {\bibfnamefont {M.}~\bibnamefont {Mohseni}}, \bibinfo {author} {\bibfnamefont {W.}~\bibnamefont {Mruczkiewicz}}, \bibinfo {author} {\bibfnamefont {J.}~\bibnamefont {Mutus}}, \bibinfo {author} {\bibfnamefont {O.}~\bibnamefont {Naaman}}, \bibinfo {author} {\bibfnamefont {C.}~\bibnamefont {Neill}}, \bibinfo {author} {\bibfnamefont {F.}~\bibnamefont {Neukart}}, \bibinfo {author} {\bibfnamefont {M.~Y.}\ \bibnamefont {Niu}}, \bibinfo {author} {\bibfnamefont {T.~E.}\ \bibnamefont {O'Brien}}, \bibinfo {author} {\bibfnamefont {B.}~\bibnamefont {O'Gorman}}, \bibinfo {author} {\bibfnamefont {E.}~\bibnamefont {Ostby}}, \bibinfo {author}
  {\bibfnamefont {A.}~\bibnamefont {Petukhov}}, \bibinfo {author} {\bibfnamefont {H.}~\bibnamefont {Putterman}}, \bibinfo {author} {\bibfnamefont {C.}~\bibnamefont {Quintana}}, \bibinfo {author} {\bibfnamefont {P.}~\bibnamefont {Roushan}}, \bibinfo {author} {\bibfnamefont {N.~C.}\ \bibnamefont {Rubin}}, \bibinfo {author} {\bibfnamefont {D.}~\bibnamefont {Sank}}, \bibinfo {author} {\bibfnamefont {A.}~\bibnamefont {Skolik}}, \bibinfo {author} {\bibfnamefont {V.}~\bibnamefont {Smelyanskiy}}, \bibinfo {author} {\bibfnamefont {D.}~\bibnamefont {Strain}}, \bibinfo {author} {\bibfnamefont {M.}~\bibnamefont {Streif}}, \bibinfo {author} {\bibfnamefont {M.}~\bibnamefont {Szalay}}, \bibinfo {author} {\bibfnamefont {A.}~\bibnamefont {Vainsencher}}, \bibinfo {author} {\bibfnamefont {T.}~\bibnamefont {White}}, \bibinfo {author} {\bibfnamefont {Z.~J.}\ \bibnamefont {Yao}}, \bibinfo {author} {\bibfnamefont {P.}~\bibnamefont {Yeh}}, \bibinfo {author} {\bibfnamefont {A.}~\bibnamefont {Zalcman}}, \bibinfo {author}
  {\bibfnamefont {L.}~\bibnamefont {Zhou}}, \bibinfo {author} {\bibfnamefont {H.}~\bibnamefont {Neven}}, \bibinfo {author} {\bibfnamefont {D.}~\bibnamefont {Bacon}}, \bibinfo {author} {\bibfnamefont {E.}~\bibnamefont {Lucero}}, \bibinfo {author} {\bibfnamefont {E.}~\bibnamefont {Farhi}},\ and\ \bibinfo {author} {\bibfnamefont {R.}~\bibnamefont {Babbush}},\ }\bibfield  {title} {\bibinfo {title} {Quantum approximate optimization of non-planar graph problems on a planar superconducting processor},\ }\href {https://doi.org/10.1038/s41567-020-01105-y} {\bibfield  {journal} {\bibinfo  {journal} {Nature Physics}\ }\textbf {\bibinfo {volume} {17}},\ \bibinfo {pages} {332} (\bibinfo {year} {2021})}\BibitemShut {NoStop}%
\bibitem [{\citenamefont {Axline}\ \emph {et~al.}(2018)\citenamefont {Axline}, \citenamefont {Burkhart}, \citenamefont {Pfaff}, \citenamefont {Zhang}, \citenamefont {Chou}, \citenamefont {Campagne-Ibarcq}, \citenamefont {Reinhold}, \citenamefont {Frunzio}, \citenamefont {Girvin}, \citenamefont {Jiang}, \citenamefont {Devoret},\ and\ \citenamefont {Schoelkopf}}]{Axline2018}%
  \BibitemOpen
  \bibfield  {author} {\bibinfo {author} {\bibfnamefont {C.~J.}\ \bibnamefont {Axline}}, \bibinfo {author} {\bibfnamefont {L.~D.}\ \bibnamefont {Burkhart}}, \bibinfo {author} {\bibfnamefont {W.}~\bibnamefont {Pfaff}}, \bibinfo {author} {\bibfnamefont {M.}~\bibnamefont {Zhang}}, \bibinfo {author} {\bibfnamefont {K.}~\bibnamefont {Chou}}, \bibinfo {author} {\bibfnamefont {P.}~\bibnamefont {Campagne-Ibarcq}}, \bibinfo {author} {\bibfnamefont {P.}~\bibnamefont {Reinhold}}, \bibinfo {author} {\bibfnamefont {L.}~\bibnamefont {Frunzio}}, \bibinfo {author} {\bibfnamefont {S.~M.}\ \bibnamefont {Girvin}}, \bibinfo {author} {\bibfnamefont {L.}~\bibnamefont {Jiang}}, \bibinfo {author} {\bibfnamefont {M.~H.}\ \bibnamefont {Devoret}},\ and\ \bibinfo {author} {\bibfnamefont {R.~J.}\ \bibnamefont {Schoelkopf}},\ }\bibfield  {title} {\bibinfo {title} {On-demand quantum state transfer and entanglement between remote microwave cavity memories},\ }\href {https://doi.org/10.1038/s41567-018-0115-y} {\bibfield  {journal} {\bibinfo
  {journal} {Nature Physics}\ }\textbf {\bibinfo {volume} {14}},\ \bibinfo {pages} {705} (\bibinfo {year} {2018})}\BibitemShut {NoStop}%
\bibitem [{\citenamefont {Kurpiers}\ \emph {et~al.}(2018)\citenamefont {Kurpiers}, \citenamefont {Magnard}, \citenamefont {Walter}, \citenamefont {Royer}, \citenamefont {Pechal}, \citenamefont {Heinsoo}, \citenamefont {Salath{\'e}}, \citenamefont {Akin}, \citenamefont {Storz}, \citenamefont {Besse}, \citenamefont {Gasparinetti}, \citenamefont {Blais},\ and\ \citenamefont {Wallraff}}]{Kurpiers2018}%
  \BibitemOpen
  \bibfield  {author} {\bibinfo {author} {\bibfnamefont {P.}~\bibnamefont {Kurpiers}}, \bibinfo {author} {\bibfnamefont {P.}~\bibnamefont {Magnard}}, \bibinfo {author} {\bibfnamefont {T.}~\bibnamefont {Walter}}, \bibinfo {author} {\bibfnamefont {B.}~\bibnamefont {Royer}}, \bibinfo {author} {\bibfnamefont {M.}~\bibnamefont {Pechal}}, \bibinfo {author} {\bibfnamefont {J.}~\bibnamefont {Heinsoo}}, \bibinfo {author} {\bibfnamefont {Y.}~\bibnamefont {Salath{\'e}}}, \bibinfo {author} {\bibfnamefont {A.}~\bibnamefont {Akin}}, \bibinfo {author} {\bibfnamefont {S.}~\bibnamefont {Storz}}, \bibinfo {author} {\bibfnamefont {J.-C.}\ \bibnamefont {Besse}}, \bibinfo {author} {\bibfnamefont {S.}~\bibnamefont {Gasparinetti}}, \bibinfo {author} {\bibfnamefont {A.}~\bibnamefont {Blais}},\ and\ \bibinfo {author} {\bibfnamefont {A.}~\bibnamefont {Wallraff}},\ }\bibfield  {title} {\bibinfo {title} {Deterministic quantum state transfer and remote entanglement using microwave photons},\ }\href
  {https://doi.org/10.1038/s41586-018-0195-y} {\bibfield  {journal} {\bibinfo  {journal} {Nature}\ }\textbf {\bibinfo {volume} {558}},\ \bibinfo {pages} {264} (\bibinfo {year} {2018})}\BibitemShut {NoStop}%
\bibitem [{\citenamefont {Pogorzalek}\ \emph {et~al.}(2019)\citenamefont {Pogorzalek}, \citenamefont {Fedorov}, \citenamefont {Xu}, \citenamefont {Parra-Rodriguez}, \citenamefont {Sanz}, \citenamefont {Fischer}, \citenamefont {Xie}, \citenamefont {Inomata}, \citenamefont {Nakamura}, \citenamefont {Solano}, \citenamefont {Marx}, \citenamefont {Deppe},\ and\ \citenamefont {Gross}}]{Pogorzalek2019}%
  \BibitemOpen
  \bibfield  {author} {\bibinfo {author} {\bibfnamefont {S.}~\bibnamefont {Pogorzalek}}, \bibinfo {author} {\bibfnamefont {K.~G.}\ \bibnamefont {Fedorov}}, \bibinfo {author} {\bibfnamefont {M.}~\bibnamefont {Xu}}, \bibinfo {author} {\bibfnamefont {A.}~\bibnamefont {Parra-Rodriguez}}, \bibinfo {author} {\bibfnamefont {M.}~\bibnamefont {Sanz}}, \bibinfo {author} {\bibfnamefont {M.}~\bibnamefont {Fischer}}, \bibinfo {author} {\bibfnamefont {E.}~\bibnamefont {Xie}}, \bibinfo {author} {\bibfnamefont {K.}~\bibnamefont {Inomata}}, \bibinfo {author} {\bibfnamefont {Y.}~\bibnamefont {Nakamura}}, \bibinfo {author} {\bibfnamefont {E.}~\bibnamefont {Solano}}, \bibinfo {author} {\bibfnamefont {A.}~\bibnamefont {Marx}}, \bibinfo {author} {\bibfnamefont {F.}~\bibnamefont {Deppe}},\ and\ \bibinfo {author} {\bibfnamefont {R.}~\bibnamefont {Gross}},\ }\bibfield  {title} {\bibinfo {title} {Secure quantum remote state preparation of squeezed microwave states},\ }\href {https://doi.org/10.1038/s41467-019-10727-7} {\bibfield
  {journal} {\bibinfo  {journal} {Nature Communications}\ }\textbf {\bibinfo {volume} {10}},\ \bibinfo {pages} {2604} (\bibinfo {year} {2019})}\BibitemShut {NoStop}%
\bibitem [{\citenamefont {Fedorov}\ \emph {et~al.}(2021)\citenamefont {Fedorov}, \citenamefont {Renger}, \citenamefont {Pogorzalek}, \citenamefont {Candia}, \citenamefont {Chen}, \citenamefont {Nojiri}, \citenamefont {Inomata}, \citenamefont {Nakamura}, \citenamefont {Partanen}, \citenamefont {Marx}, \citenamefont {Gross},\ and\ \citenamefont {Deppe}}]{Kirill2021}%
  \BibitemOpen
  \bibfield  {author} {\bibinfo {author} {\bibfnamefont {K.~G.}\ \bibnamefont {Fedorov}}, \bibinfo {author} {\bibfnamefont {M.}~\bibnamefont {Renger}}, \bibinfo {author} {\bibfnamefont {S.}~\bibnamefont {Pogorzalek}}, \bibinfo {author} {\bibfnamefont {R.~D.}\ \bibnamefont {Candia}}, \bibinfo {author} {\bibfnamefont {Q.}~\bibnamefont {Chen}}, \bibinfo {author} {\bibfnamefont {Y.}~\bibnamefont {Nojiri}}, \bibinfo {author} {\bibfnamefont {K.}~\bibnamefont {Inomata}}, \bibinfo {author} {\bibfnamefont {Y.}~\bibnamefont {Nakamura}}, \bibinfo {author} {\bibfnamefont {M.}~\bibnamefont {Partanen}}, \bibinfo {author} {\bibfnamefont {A.}~\bibnamefont {Marx}}, \bibinfo {author} {\bibfnamefont {R.}~\bibnamefont {Gross}},\ and\ \bibinfo {author} {\bibfnamefont {F.}~\bibnamefont {Deppe}},\ }\bibfield  {title} {\bibinfo {title} {Experimental quantum teleportation of propagating microwaves},\ }\href {https://doi.org/10.1126/sciadv.abk0891} {\bibfield  {journal} {\bibinfo  {journal} {Science Advances}\ }\textbf {\bibinfo
  {volume} {7}},\ \bibinfo {pages} {eabk0891} (\bibinfo {year} {2021})},\ \Eprint {https://arxiv.org/abs/https://www.science.org/doi/pdf/10.1126/sciadv.abk0891} {https://www.science.org/doi/pdf/10.1126/sciadv.abk0891} \BibitemShut {NoStop}%
\bibitem [{\citenamefont {Underwood}\ \emph {et~al.}(2012)\citenamefont {Underwood}, \citenamefont {Shanks}, \citenamefont {Koch},\ and\ \citenamefont {Houck}}]{Underwood2012}%
  \BibitemOpen
  \bibfield  {author} {\bibinfo {author} {\bibfnamefont {D.~L.}\ \bibnamefont {Underwood}}, \bibinfo {author} {\bibfnamefont {W.~E.}\ \bibnamefont {Shanks}}, \bibinfo {author} {\bibfnamefont {J.}~\bibnamefont {Koch}},\ and\ \bibinfo {author} {\bibfnamefont {A.~A.}\ \bibnamefont {Houck}},\ }\bibfield  {title} {\bibinfo {title} {Low-disorder microwave cavity lattices for quantum simulation with photons},\ }\href {https://doi.org/10.1103/PhysRevA.86.023837} {\bibfield  {journal} {\bibinfo  {journal} {Phys. Rev. A}\ }\textbf {\bibinfo {volume} {86}},\ \bibinfo {pages} {023837} (\bibinfo {year} {2012})}\BibitemShut {NoStop}%
\bibitem [{\citenamefont {Abdumalikov~Jr}\ \emph {et~al.}(2013)\citenamefont {Abdumalikov~Jr}, \citenamefont {Fink}, \citenamefont {Juliusson}, \citenamefont {Pechal}, \citenamefont {Berger}, \citenamefont {Wallraff},\ and\ \citenamefont {Filipp}}]{AbdumalikovJr2013}%
  \BibitemOpen
  \bibfield  {author} {\bibinfo {author} {\bibfnamefont {A.~A.}\ \bibnamefont {Abdumalikov~Jr}}, \bibinfo {author} {\bibfnamefont {J.~M.}\ \bibnamefont {Fink}}, \bibinfo {author} {\bibfnamefont {K.}~\bibnamefont {Juliusson}}, \bibinfo {author} {\bibfnamefont {M.}~\bibnamefont {Pechal}}, \bibinfo {author} {\bibfnamefont {S.}~\bibnamefont {Berger}}, \bibinfo {author} {\bibfnamefont {A.}~\bibnamefont {Wallraff}},\ and\ \bibinfo {author} {\bibfnamefont {S.}~\bibnamefont {Filipp}},\ }\bibfield  {title} {\bibinfo {title} {Experimental realization of non-abelian non-adiabatic geometric gates},\ }\href {https://doi.org/10.1038/nature12010} {\bibfield  {journal} {\bibinfo  {journal} {Nature}\ }\textbf {\bibinfo {volume} {496}},\ \bibinfo {pages} {482} (\bibinfo {year} {2013})}\BibitemShut {NoStop}%
\bibitem [{\citenamefont {Roushan}\ \emph {et~al.}(2017)\citenamefont {Roushan}, \citenamefont {Neill}, \citenamefont {Megrant}, \citenamefont {Chen}, \citenamefont {Babbush}, \citenamefont {Barends}, \citenamefont {Campbell}, \citenamefont {Chen}, \citenamefont {Chiaro}, \citenamefont {Dunsworth}, \citenamefont {Fowler}, \citenamefont {Jeffrey}, \citenamefont {Kelly}, \citenamefont {Lucero}, \citenamefont {Mutus}, \citenamefont {O'Malley}, \citenamefont {Neeley}, \citenamefont {Quintana}, \citenamefont {Sank}, \citenamefont {Vainsencher}, \citenamefont {Wenner}, \citenamefont {White}, \citenamefont {Kapit}, \citenamefont {Neven},\ and\ \citenamefont {Martinis}}]{Roushan2017}%
  \BibitemOpen
  \bibfield  {author} {\bibinfo {author} {\bibfnamefont {P.}~\bibnamefont {Roushan}}, \bibinfo {author} {\bibfnamefont {C.}~\bibnamefont {Neill}}, \bibinfo {author} {\bibfnamefont {A.}~\bibnamefont {Megrant}}, \bibinfo {author} {\bibfnamefont {Y.}~\bibnamefont {Chen}}, \bibinfo {author} {\bibfnamefont {R.}~\bibnamefont {Babbush}}, \bibinfo {author} {\bibfnamefont {R.}~\bibnamefont {Barends}}, \bibinfo {author} {\bibfnamefont {B.}~\bibnamefont {Campbell}}, \bibinfo {author} {\bibfnamefont {Z.}~\bibnamefont {Chen}}, \bibinfo {author} {\bibfnamefont {B.}~\bibnamefont {Chiaro}}, \bibinfo {author} {\bibfnamefont {A.}~\bibnamefont {Dunsworth}}, \bibinfo {author} {\bibfnamefont {A.}~\bibnamefont {Fowler}}, \bibinfo {author} {\bibfnamefont {E.}~\bibnamefont {Jeffrey}}, \bibinfo {author} {\bibfnamefont {J.}~\bibnamefont {Kelly}}, \bibinfo {author} {\bibfnamefont {E.}~\bibnamefont {Lucero}}, \bibinfo {author} {\bibfnamefont {J.}~\bibnamefont {Mutus}}, \bibinfo {author} {\bibfnamefont {P.~J.~J.}\ \bibnamefont
  {O'Malley}}, \bibinfo {author} {\bibfnamefont {M.}~\bibnamefont {Neeley}}, \bibinfo {author} {\bibfnamefont {C.}~\bibnamefont {Quintana}}, \bibinfo {author} {\bibfnamefont {D.}~\bibnamefont {Sank}}, \bibinfo {author} {\bibfnamefont {A.}~\bibnamefont {Vainsencher}}, \bibinfo {author} {\bibfnamefont {J.}~\bibnamefont {Wenner}}, \bibinfo {author} {\bibfnamefont {T.}~\bibnamefont {White}}, \bibinfo {author} {\bibfnamefont {E.}~\bibnamefont {Kapit}}, \bibinfo {author} {\bibfnamefont {H.}~\bibnamefont {Neven}},\ and\ \bibinfo {author} {\bibfnamefont {J.}~\bibnamefont {Martinis}},\ }\bibfield  {title} {\bibinfo {title} {Chiral ground-state currents of interacting photons in a synthetic magnetic field},\ }\href {https://doi.org/10.1038/nphys3930} {\bibfield  {journal} {\bibinfo  {journal} {Nature Physics}\ }\textbf {\bibinfo {volume} {13}},\ \bibinfo {pages} {146} (\bibinfo {year} {2017})}\BibitemShut {NoStop}%
\bibitem [{\citenamefont {Koll{\'a}r}\ \emph {et~al.}(2019)\citenamefont {Koll{\'a}r}, \citenamefont {Fitzpatrick},\ and\ \citenamefont {Houck}}]{Kollár2019}%
  \BibitemOpen
  \bibfield  {author} {\bibinfo {author} {\bibfnamefont {A.~J.}\ \bibnamefont {Koll{\'a}r}}, \bibinfo {author} {\bibfnamefont {M.}~\bibnamefont {Fitzpatrick}},\ and\ \bibinfo {author} {\bibfnamefont {A.~A.}\ \bibnamefont {Houck}},\ }\bibfield  {title} {\bibinfo {title} {Hyperbolic lattices in circuit quantum electrodynamics},\ }\href {https://doi.org/10.1038/s41586-019-1348-3} {\bibfield  {journal} {\bibinfo  {journal} {Nature}\ }\textbf {\bibinfo {volume} {571}},\ \bibinfo {pages} {45} (\bibinfo {year} {2019})}\BibitemShut {NoStop}%
\bibitem [{\citenamefont {Ma}\ \emph {et~al.}(2019)\citenamefont {Ma}, \citenamefont {Saxberg}, \citenamefont {Owens}, \citenamefont {Leung}, \citenamefont {Lu}, \citenamefont {Simon},\ and\ \citenamefont {Schuster}}]{Ma2019}%
  \BibitemOpen
  \bibfield  {author} {\bibinfo {author} {\bibfnamefont {R.}~\bibnamefont {Ma}}, \bibinfo {author} {\bibfnamefont {B.}~\bibnamefont {Saxberg}}, \bibinfo {author} {\bibfnamefont {C.}~\bibnamefont {Owens}}, \bibinfo {author} {\bibfnamefont {N.}~\bibnamefont {Leung}}, \bibinfo {author} {\bibfnamefont {Y.}~\bibnamefont {Lu}}, \bibinfo {author} {\bibfnamefont {J.}~\bibnamefont {Simon}},\ and\ \bibinfo {author} {\bibfnamefont {D.~I.}\ \bibnamefont {Schuster}},\ }\bibfield  {title} {\bibinfo {title} {A dissipatively stabilized mott insulator of photons},\ }\href {https://doi.org/10.1038/s41586-019-0897-9} {\bibfield  {journal} {\bibinfo  {journal} {Nature}\ }\textbf {\bibinfo {volume} {566}},\ \bibinfo {pages} {51} (\bibinfo {year} {2019})}\BibitemShut {NoStop}%
\bibitem [{\citenamefont {Xu}\ \emph {et~al.}(2020)\citenamefont {Xu}, \citenamefont {Sun}, \citenamefont {Liu}, \citenamefont {Zhang}, \citenamefont {Li}, \citenamefont {Dong}, \citenamefont {Ren}, \citenamefont {Zhang}, \citenamefont {Nori}, \citenamefont {Zheng}, \citenamefont {Fan},\ and\ \citenamefont {Wang}}]{Kai2020}%
  \BibitemOpen
  \bibfield  {author} {\bibinfo {author} {\bibfnamefont {K.}~\bibnamefont {Xu}}, \bibinfo {author} {\bibfnamefont {Z.-H.}\ \bibnamefont {Sun}}, \bibinfo {author} {\bibfnamefont {W.}~\bibnamefont {Liu}}, \bibinfo {author} {\bibfnamefont {Y.-R.}\ \bibnamefont {Zhang}}, \bibinfo {author} {\bibfnamefont {H.}~\bibnamefont {Li}}, \bibinfo {author} {\bibfnamefont {H.}~\bibnamefont {Dong}}, \bibinfo {author} {\bibfnamefont {W.}~\bibnamefont {Ren}}, \bibinfo {author} {\bibfnamefont {P.}~\bibnamefont {Zhang}}, \bibinfo {author} {\bibfnamefont {F.}~\bibnamefont {Nori}}, \bibinfo {author} {\bibfnamefont {D.}~\bibnamefont {Zheng}}, \bibinfo {author} {\bibfnamefont {H.}~\bibnamefont {Fan}},\ and\ \bibinfo {author} {\bibfnamefont {H.}~\bibnamefont {Wang}},\ }\bibfield  {title} {\bibinfo {title} {Probing dynamical phase transitions with a superconducting quantum simulator},\ }\href {https://doi.org/10.1126/sciadv.aba4935} {\bibfield  {journal} {\bibinfo  {journal} {Science Advances}\ }\textbf {\bibinfo {volume} {6}},\
  \bibinfo {pages} {eaba4935} (\bibinfo {year} {2020})},\ \Eprint {https://arxiv.org/abs/https://www.science.org/doi/pdf/10.1126/sciadv.aba4935} {https://www.science.org/doi/pdf/10.1126/sciadv.aba4935} \BibitemShut {NoStop}%
\bibitem [{\citenamefont {Guo}\ \emph {et~al.}(2021)\citenamefont {Guo}, \citenamefont {Cheng}, \citenamefont {Sun}, \citenamefont {Song}, \citenamefont {Li}, \citenamefont {Wang}, \citenamefont {Ren}, \citenamefont {Dong}, \citenamefont {Zheng}, \citenamefont {Zhang}, \citenamefont {Mondaini}, \citenamefont {Fan},\ and\ \citenamefont {Wang}}]{Guo2021}%
  \BibitemOpen
  \bibfield  {author} {\bibinfo {author} {\bibfnamefont {Q.}~\bibnamefont {Guo}}, \bibinfo {author} {\bibfnamefont {C.}~\bibnamefont {Cheng}}, \bibinfo {author} {\bibfnamefont {Z.-H.}\ \bibnamefont {Sun}}, \bibinfo {author} {\bibfnamefont {Z.}~\bibnamefont {Song}}, \bibinfo {author} {\bibfnamefont {H.}~\bibnamefont {Li}}, \bibinfo {author} {\bibfnamefont {Z.}~\bibnamefont {Wang}}, \bibinfo {author} {\bibfnamefont {W.}~\bibnamefont {Ren}}, \bibinfo {author} {\bibfnamefont {H.}~\bibnamefont {Dong}}, \bibinfo {author} {\bibfnamefont {D.}~\bibnamefont {Zheng}}, \bibinfo {author} {\bibfnamefont {Y.-R.}\ \bibnamefont {Zhang}}, \bibinfo {author} {\bibfnamefont {R.}~\bibnamefont {Mondaini}}, \bibinfo {author} {\bibfnamefont {H.}~\bibnamefont {Fan}},\ and\ \bibinfo {author} {\bibfnamefont {H.}~\bibnamefont {Wang}},\ }\bibfield  {title} {\bibinfo {title} {Observation of energy-resolved many-body localization},\ }\href {https://doi.org/10.1038/s41567-020-1035-1} {\bibfield  {journal} {\bibinfo  {journal} {Nature
  Physics}\ }\textbf {\bibinfo {volume} {17}},\ \bibinfo {pages} {234} (\bibinfo {year} {2021})}\BibitemShut {NoStop}%
\bibitem [{\citenamefont {Chen}\ \emph {et~al.}(2023{\natexlab{a}})\citenamefont {Chen}, \citenamefont {Fischer}, \citenamefont {Nojiri}, \citenamefont {Renger}, \citenamefont {Xie}, \citenamefont {Partanen}, \citenamefont {Pogorzalek}, \citenamefont {Fedorov}, \citenamefont {Marx}, \citenamefont {Deppe},\ and\ \citenamefont {Gross}}]{Chen2023}%
  \BibitemOpen
  \bibfield  {author} {\bibinfo {author} {\bibfnamefont {Q.-M.}\ \bibnamefont {Chen}}, \bibinfo {author} {\bibfnamefont {M.}~\bibnamefont {Fischer}}, \bibinfo {author} {\bibfnamefont {Y.}~\bibnamefont {Nojiri}}, \bibinfo {author} {\bibfnamefont {M.}~\bibnamefont {Renger}}, \bibinfo {author} {\bibfnamefont {E.}~\bibnamefont {Xie}}, \bibinfo {author} {\bibfnamefont {M.}~\bibnamefont {Partanen}}, \bibinfo {author} {\bibfnamefont {S.}~\bibnamefont {Pogorzalek}}, \bibinfo {author} {\bibfnamefont {K.~G.}\ \bibnamefont {Fedorov}}, \bibinfo {author} {\bibfnamefont {A.}~\bibnamefont {Marx}}, \bibinfo {author} {\bibfnamefont {F.}~\bibnamefont {Deppe}},\ and\ \bibinfo {author} {\bibfnamefont {R.}~\bibnamefont {Gross}},\ }\bibfield  {title} {\bibinfo {title} {Quantum behavior of the duffing oscillator at the dissipative phase transition},\ }\href {https://doi.org/10.1038/s41467-023-38217-x} {\bibfield  {journal} {\bibinfo  {journal} {Nature Communications}\ }\textbf {\bibinfo {volume} {14}},\ \bibinfo {pages} {2896}
  (\bibinfo {year} {2023}{\natexlab{a}})}\BibitemShut {NoStop}%
\bibitem [{\citenamefont {Barzanjeh}\ \emph {et~al.}(2020)\citenamefont {Barzanjeh}, \citenamefont {Pirandola}, \citenamefont {Vitali},\ and\ \citenamefont {Fink}}]{Barzanjeh2020}%
  \BibitemOpen
  \bibfield  {author} {\bibinfo {author} {\bibfnamefont {S.}~\bibnamefont {Barzanjeh}}, \bibinfo {author} {\bibfnamefont {S.}~\bibnamefont {Pirandola}}, \bibinfo {author} {\bibfnamefont {D.}~\bibnamefont {Vitali}},\ and\ \bibinfo {author} {\bibfnamefont {J.~M.}\ \bibnamefont {Fink}},\ }\bibfield  {title} {\bibinfo {title} {Microwave quantum illumination using a digital receiver},\ }\href {https://doi.org/10.1126/sciadv.abb0451} {\bibfield  {journal} {\bibinfo  {journal} {Science Advances}\ }\textbf {\bibinfo {volume} {6}},\ \bibinfo {pages} {eabb0451} (\bibinfo {year} {2020})},\ \Eprint {https://arxiv.org/abs/https://www.science.org/doi/pdf/10.1126/sciadv.abb0451} {https://www.science.org/doi/pdf/10.1126/sciadv.abb0451} \BibitemShut {NoStop}%
\bibitem [{\citenamefont {Bienfait}\ \emph {et~al.}(2017)\citenamefont {Bienfait}, \citenamefont {Campagne-Ibarcq}, \citenamefont {Kiilerich}, \citenamefont {Zhou}, \citenamefont {Probst}, \citenamefont {Pla}, \citenamefont {Schenkel}, \citenamefont {Vion}, \citenamefont {Esteve}, \citenamefont {Morton}, \citenamefont {Moelmer},\ and\ \citenamefont {Bertet}}]{Bienfait2017}%
  \BibitemOpen
  \bibfield  {author} {\bibinfo {author} {\bibfnamefont {A.}~\bibnamefont {Bienfait}}, \bibinfo {author} {\bibfnamefont {P.}~\bibnamefont {Campagne-Ibarcq}}, \bibinfo {author} {\bibfnamefont {A.~H.}\ \bibnamefont {Kiilerich}}, \bibinfo {author} {\bibfnamefont {X.}~\bibnamefont {Zhou}}, \bibinfo {author} {\bibfnamefont {S.}~\bibnamefont {Probst}}, \bibinfo {author} {\bibfnamefont {J.~J.}\ \bibnamefont {Pla}}, \bibinfo {author} {\bibfnamefont {T.}~\bibnamefont {Schenkel}}, \bibinfo {author} {\bibfnamefont {D.}~\bibnamefont {Vion}}, \bibinfo {author} {\bibfnamefont {D.}~\bibnamefont {Esteve}}, \bibinfo {author} {\bibfnamefont {J.~J.~L.}\ \bibnamefont {Morton}}, \bibinfo {author} {\bibfnamefont {K.}~\bibnamefont {Moelmer}},\ and\ \bibinfo {author} {\bibfnamefont {P.}~\bibnamefont {Bertet}},\ }\bibfield  {title} {\bibinfo {title} {Magnetic resonance with squeezed microwaves},\ }\href {https://doi.org/10.1103/PhysRevX.7.041011} {\bibfield  {journal} {\bibinfo  {journal} {Phys. Rev. X}\ }\textbf {\bibinfo {volume}
  {7}},\ \bibinfo {pages} {041011} (\bibinfo {year} {2017})}\BibitemShut {NoStop}%
\bibitem [{\citenamefont {Wang}\ \emph {et~al.}(2021)\citenamefont {Wang}, \citenamefont {Xu}, \citenamefont {Han}, \citenamefont {Fu}, \citenamefont {Puri}, \citenamefont {Girvin}, \citenamefont {Tang}, \citenamefont {Shankar},\ and\ \citenamefont {Devoret}}]{Wang2021}%
  \BibitemOpen
  \bibfield  {author} {\bibinfo {author} {\bibfnamefont {Z.}~\bibnamefont {Wang}}, \bibinfo {author} {\bibfnamefont {M.}~\bibnamefont {Xu}}, \bibinfo {author} {\bibfnamefont {X.}~\bibnamefont {Han}}, \bibinfo {author} {\bibfnamefont {W.}~\bibnamefont {Fu}}, \bibinfo {author} {\bibfnamefont {S.}~\bibnamefont {Puri}}, \bibinfo {author} {\bibfnamefont {S.~M.}\ \bibnamefont {Girvin}}, \bibinfo {author} {\bibfnamefont {H.~X.}\ \bibnamefont {Tang}}, \bibinfo {author} {\bibfnamefont {S.}~\bibnamefont {Shankar}},\ and\ \bibinfo {author} {\bibfnamefont {M.~H.}\ \bibnamefont {Devoret}},\ }\bibfield  {title} {\bibinfo {title} {Quantum microwave radiometry with a superconducting qubit},\ }\href {https://doi.org/10.1103/PhysRevLett.126.180501} {\bibfield  {journal} {\bibinfo  {journal} {Phys. Rev. Lett.}\ }\textbf {\bibinfo {volume} {126}},\ \bibinfo {pages} {180501} (\bibinfo {year} {2021})}\BibitemShut {NoStop}%
\bibitem [{\citenamefont {Kokkoniemi}\ \emph {et~al.}(2019)\citenamefont {Kokkoniemi}, \citenamefont {Govenius}, \citenamefont {Vesterinen}, \citenamefont {Lake}, \citenamefont {Gunyh{\'o}}, \citenamefont {Tan}, \citenamefont {Simbierowicz}, \citenamefont {Gr{\"o}nberg}, \citenamefont {Lehtinen}, \citenamefont {Prunnila}, \citenamefont {Hassel}, \citenamefont {Lamminen}, \citenamefont {Saira},\ and\ \citenamefont {M{\"o}tt{\"o}nen}}]{Kokkoniemi2019}%
  \BibitemOpen
  \bibfield  {author} {\bibinfo {author} {\bibfnamefont {R.}~\bibnamefont {Kokkoniemi}}, \bibinfo {author} {\bibfnamefont {J.}~\bibnamefont {Govenius}}, \bibinfo {author} {\bibfnamefont {V.}~\bibnamefont {Vesterinen}}, \bibinfo {author} {\bibfnamefont {R.~E.}\ \bibnamefont {Lake}}, \bibinfo {author} {\bibfnamefont {A.~M.}\ \bibnamefont {Gunyh{\'o}}}, \bibinfo {author} {\bibfnamefont {K.~Y.}\ \bibnamefont {Tan}}, \bibinfo {author} {\bibfnamefont {S.}~\bibnamefont {Simbierowicz}}, \bibinfo {author} {\bibfnamefont {L.}~\bibnamefont {Gr{\"o}nberg}}, \bibinfo {author} {\bibfnamefont {J.}~\bibnamefont {Lehtinen}}, \bibinfo {author} {\bibfnamefont {M.}~\bibnamefont {Prunnila}}, \bibinfo {author} {\bibfnamefont {J.}~\bibnamefont {Hassel}}, \bibinfo {author} {\bibfnamefont {A.}~\bibnamefont {Lamminen}}, \bibinfo {author} {\bibfnamefont {O.-P.}\ \bibnamefont {Saira}},\ and\ \bibinfo {author} {\bibfnamefont {M.}~\bibnamefont {M{\"o}tt{\"o}nen}},\ }\bibfield  {title} {\bibinfo {title} {Nanobolometer with ultralow noise
  equivalent power},\ }\href {https://doi.org/10.1038/s42005-019-0225-6} {\bibfield  {journal} {\bibinfo  {journal} {Communications Physics}\ }\textbf {\bibinfo {volume} {2}},\ \bibinfo {pages} {124} (\bibinfo {year} {2019})}\BibitemShut {NoStop}%
\bibitem [{\citenamefont {Kokkoniemi}\ \emph {et~al.}(2020)\citenamefont {Kokkoniemi}, \citenamefont {Girard}, \citenamefont {Hazra}, \citenamefont {Laitinen}, \citenamefont {Govenius}, \citenamefont {Lake}, \citenamefont {Sallinen}, \citenamefont {Vesterinen}, \citenamefont {Partanen}, \citenamefont {Tan}, \citenamefont {Chan}, \citenamefont {Tan}, \citenamefont {Hakonen},\ and\ \citenamefont {M{\"o}tt{\"o}nen}}]{Kokkoniemi2020}%
  \BibitemOpen
  \bibfield  {author} {\bibinfo {author} {\bibfnamefont {R.}~\bibnamefont {Kokkoniemi}}, \bibinfo {author} {\bibfnamefont {J.-P.}\ \bibnamefont {Girard}}, \bibinfo {author} {\bibfnamefont {D.}~\bibnamefont {Hazra}}, \bibinfo {author} {\bibfnamefont {A.}~\bibnamefont {Laitinen}}, \bibinfo {author} {\bibfnamefont {J.}~\bibnamefont {Govenius}}, \bibinfo {author} {\bibfnamefont {R.~E.}\ \bibnamefont {Lake}}, \bibinfo {author} {\bibfnamefont {I.}~\bibnamefont {Sallinen}}, \bibinfo {author} {\bibfnamefont {V.}~\bibnamefont {Vesterinen}}, \bibinfo {author} {\bibfnamefont {M.}~\bibnamefont {Partanen}}, \bibinfo {author} {\bibfnamefont {J.~Y.}\ \bibnamefont {Tan}}, \bibinfo {author} {\bibfnamefont {K.~W.}\ \bibnamefont {Chan}}, \bibinfo {author} {\bibfnamefont {K.~Y.}\ \bibnamefont {Tan}}, \bibinfo {author} {\bibfnamefont {P.}~\bibnamefont {Hakonen}},\ and\ \bibinfo {author} {\bibfnamefont {M.}~\bibnamefont {M{\"o}tt{\"o}nen}},\ }\bibfield  {title} {\bibinfo {title} {Bolometer operating at the threshold for circuit
  quantum electrodynamics},\ }\href {https://doi.org/10.1038/s41586-020-2753-3} {\bibfield  {journal} {\bibinfo  {journal} {Nature}\ }\textbf {\bibinfo {volume} {586}},\ \bibinfo {pages} {47} (\bibinfo {year} {2020})}\BibitemShut {NoStop}%
\bibitem [{\citenamefont {Govenius}\ \emph {et~al.}(2016)\citenamefont {Govenius}, \citenamefont {Lake}, \citenamefont {Tan},\ and\ \citenamefont {M\"ott\"onen}}]{Govenius2016}%
  \BibitemOpen
  \bibfield  {author} {\bibinfo {author} {\bibfnamefont {J.}~\bibnamefont {Govenius}}, \bibinfo {author} {\bibfnamefont {R.~E.}\ \bibnamefont {Lake}}, \bibinfo {author} {\bibfnamefont {K.~Y.}\ \bibnamefont {Tan}},\ and\ \bibinfo {author} {\bibfnamefont {M.}~\bibnamefont {M\"ott\"onen}},\ }\bibfield  {title} {\bibinfo {title} {Detection of zeptojoule microwave pulses using electrothermal feedback in proximity-induced josephson junctions},\ }\href {https://doi.org/10.1103/PhysRevLett.117.030802} {\bibfield  {journal} {\bibinfo  {journal} {Phys. Rev. Lett.}\ }\textbf {\bibinfo {volume} {117}},\ \bibinfo {pages} {030802} (\bibinfo {year} {2016})}\BibitemShut {NoStop}%
\bibitem [{\citenamefont {Gasparinetti}\ \emph {et~al.}(2015)\citenamefont {Gasparinetti}, \citenamefont {Viisanen}, \citenamefont {Saira}, \citenamefont {Faivre}, \citenamefont {Arzeo}, \citenamefont {Meschke},\ and\ \citenamefont {Pekola}}]{Gasparinetti2015}%
  \BibitemOpen
  \bibfield  {author} {\bibinfo {author} {\bibfnamefont {S.}~\bibnamefont {Gasparinetti}}, \bibinfo {author} {\bibfnamefont {K.~L.}\ \bibnamefont {Viisanen}}, \bibinfo {author} {\bibfnamefont {O.-P.}\ \bibnamefont {Saira}}, \bibinfo {author} {\bibfnamefont {T.}~\bibnamefont {Faivre}}, \bibinfo {author} {\bibfnamefont {M.}~\bibnamefont {Arzeo}}, \bibinfo {author} {\bibfnamefont {M.}~\bibnamefont {Meschke}},\ and\ \bibinfo {author} {\bibfnamefont {J.~P.}\ \bibnamefont {Pekola}},\ }\bibfield  {title} {\bibinfo {title} {Fast electron thermometry for ultrasensitive calorimetric detection},\ }\href {https://doi.org/10.1103/PhysRevApplied.3.014007} {\bibfield  {journal} {\bibinfo  {journal} {Phys. Rev. Appl.}\ }\textbf {\bibinfo {volume} {3}},\ \bibinfo {pages} {014007} (\bibinfo {year} {2015})}\BibitemShut {NoStop}%
\bibitem [{\citenamefont {Xiang}\ \emph {et~al.}(2013)\citenamefont {Xiang}, \citenamefont {Ashhab}, \citenamefont {You},\ and\ \citenamefont {Nori}}]{Xiang2013}%
  \BibitemOpen
  \bibfield  {author} {\bibinfo {author} {\bibfnamefont {Z.-L.}\ \bibnamefont {Xiang}}, \bibinfo {author} {\bibfnamefont {S.}~\bibnamefont {Ashhab}}, \bibinfo {author} {\bibfnamefont {J.~Q.}\ \bibnamefont {You}},\ and\ \bibinfo {author} {\bibfnamefont {F.}~\bibnamefont {Nori}},\ }\bibfield  {title} {\bibinfo {title} {Hybrid quantum circuits: Superconducting circuits interacting with other quantum systems},\ }\href {https://doi.org/10.1103/RevModPhys.85.623} {\bibfield  {journal} {\bibinfo  {journal} {Rev. Mod. Phys.}\ }\textbf {\bibinfo {volume} {85}},\ \bibinfo {pages} {623} (\bibinfo {year} {2013})}\BibitemShut {NoStop}%
\bibitem [{\citenamefont {Clerk}\ \emph {et~al.}(2020)\citenamefont {Clerk}, \citenamefont {Lehnert}, \citenamefont {Bertet}, \citenamefont {Petta},\ and\ \citenamefont {Nakamura}}]{Clerk2020}%
  \BibitemOpen
  \bibfield  {author} {\bibinfo {author} {\bibfnamefont {A.~A.}\ \bibnamefont {Clerk}}, \bibinfo {author} {\bibfnamefont {K.~W.}\ \bibnamefont {Lehnert}}, \bibinfo {author} {\bibfnamefont {P.}~\bibnamefont {Bertet}}, \bibinfo {author} {\bibfnamefont {J.~R.}\ \bibnamefont {Petta}},\ and\ \bibinfo {author} {\bibfnamefont {Y.}~\bibnamefont {Nakamura}},\ }\bibfield  {title} {\bibinfo {title} {Hybrid quantum systems with circuit quantum electrodynamics},\ }\href {https://doi.org/10.1038/s41567-020-0797-9} {\bibfield  {journal} {\bibinfo  {journal} {Nature Physics}\ }\textbf {\bibinfo {volume} {16}},\ \bibinfo {pages} {257} (\bibinfo {year} {2020})}\BibitemShut {NoStop}%
\bibitem [{\citenamefont {Tan}\ \emph {et~al.}(2017)\citenamefont {Tan}, \citenamefont {Partanen}, \citenamefont {Lake}, \citenamefont {Govenius}, \citenamefont {Masuda},\ and\ \citenamefont {M{\"o}tt{\"o}nen}}]{Tan2017}%
  \BibitemOpen
  \bibfield  {author} {\bibinfo {author} {\bibfnamefont {K.~Y.}\ \bibnamefont {Tan}}, \bibinfo {author} {\bibfnamefont {M.}~\bibnamefont {Partanen}}, \bibinfo {author} {\bibfnamefont {R.~E.}\ \bibnamefont {Lake}}, \bibinfo {author} {\bibfnamefont {J.}~\bibnamefont {Govenius}}, \bibinfo {author} {\bibfnamefont {S.}~\bibnamefont {Masuda}},\ and\ \bibinfo {author} {\bibfnamefont {M.}~\bibnamefont {M{\"o}tt{\"o}nen}},\ }\bibfield  {title} {\bibinfo {title} {Quantum-circuit refrigerator},\ }\href {https://doi.org/10.1038/ncomms15189} {\bibfield  {journal} {\bibinfo  {journal} {Nature Communications}\ }\textbf {\bibinfo {volume} {8}},\ \bibinfo {pages} {15189} (\bibinfo {year} {2017})}\BibitemShut {NoStop}%
\bibitem [{\citenamefont {Mörstedt}\ \emph {et~al.}(2024)\citenamefont {Mörstedt}, \citenamefont {Teixeira}, \citenamefont {Viitanen}, \citenamefont {Kivijärvi}, \citenamefont {Tiiri}, \citenamefont {Rasola}, \citenamefont {Gunyho}, \citenamefont {Kundu}, \citenamefont {Lattier}, \citenamefont {Vadimov}, \citenamefont {Catelani}, \citenamefont {Sevriuk}, \citenamefont {Heinsoo}, \citenamefont {Räbinä}, \citenamefont {Ankerhold},\ and\ \citenamefont {Möttönen}}]{Timm_thermal}%
  \BibitemOpen
  \bibfield  {author} {\bibinfo {author} {\bibfnamefont {T.~F.}\ \bibnamefont {Mörstedt}}, \bibinfo {author} {\bibfnamefont {W.~S.}\ \bibnamefont {Teixeira}}, \bibinfo {author} {\bibfnamefont {A.}~\bibnamefont {Viitanen}}, \bibinfo {author} {\bibfnamefont {H.}~\bibnamefont {Kivijärvi}}, \bibinfo {author} {\bibfnamefont {M.}~\bibnamefont {Tiiri}}, \bibinfo {author} {\bibfnamefont {M.}~\bibnamefont {Rasola}}, \bibinfo {author} {\bibfnamefont {A.~M.}\ \bibnamefont {Gunyho}}, \bibinfo {author} {\bibfnamefont {S.}~\bibnamefont {Kundu}}, \bibinfo {author} {\bibfnamefont {L.}~\bibnamefont {Lattier}}, \bibinfo {author} {\bibfnamefont {V.}~\bibnamefont {Vadimov}}, \bibinfo {author} {\bibfnamefont {G.}~\bibnamefont {Catelani}}, \bibinfo {author} {\bibfnamefont {V.}~\bibnamefont {Sevriuk}}, \bibinfo {author} {\bibfnamefont {J.}~\bibnamefont {Heinsoo}}, \bibinfo {author} {\bibfnamefont {J.}~\bibnamefont {Räbinä}}, \bibinfo {author} {\bibfnamefont {J.}~\bibnamefont {Ankerhold}},\ and\ \bibinfo {author} {\bibfnamefont
  {M.}~\bibnamefont {Möttönen}},\ }\href {https://arxiv.org/abs/2402.09594} {\bibinfo {title} {Rapid on-demand generation of thermal states in superconducting quantum circuits}} (\bibinfo {year} {2024}),\ \Eprint {https://arxiv.org/abs/2402.09594} {arXiv:2402.09594 [quant-ph]} \BibitemShut {NoStop}%
\bibitem [{\citenamefont {Sundelin}\ \emph {et~al.}(2024)\citenamefont {Sundelin}, \citenamefont {Aamir}, \citenamefont {Kulkarni}, \citenamefont {Castillo-Moreno},\ and\ \citenamefont {Gasparinetti}}]{Sundelin2024}%
  \BibitemOpen
  \bibfield  {author} {\bibinfo {author} {\bibfnamefont {S.}~\bibnamefont {Sundelin}}, \bibinfo {author} {\bibfnamefont {M.~A.}\ \bibnamefont {Aamir}}, \bibinfo {author} {\bibfnamefont {V.~M.}\ \bibnamefont {Kulkarni}}, \bibinfo {author} {\bibfnamefont {C.}~\bibnamefont {Castillo-Moreno}},\ and\ \bibinfo {author} {\bibfnamefont {S.}~\bibnamefont {Gasparinetti}},\ }\href {https://arxiv.org/abs/2403.03373} {\bibinfo {title} {Quantum refrigeration powered by noise in a superconducting circuit}} (\bibinfo {year} {2024}),\ \Eprint {https://arxiv.org/abs/2403.03373} {arXiv:2403.03373 [quant-ph]} \BibitemShut {NoStop}%
\bibitem [{\citenamefont {Göppl}\ \emph {et~al.}(2008)\citenamefont {Göppl}, \citenamefont {Fragner}, \citenamefont {Baur}, \citenamefont {Bianchetti}, \citenamefont {Filipp}, \citenamefont {Fink}, \citenamefont {Leek}, \citenamefont {Puebla}, \citenamefont {Steffen},\ and\ \citenamefont {Wallraff}}]{Goppl08}%
  \BibitemOpen
  \bibfield  {author} {\bibinfo {author} {\bibfnamefont {M.}~\bibnamefont {Göppl}}, \bibinfo {author} {\bibfnamefont {A.}~\bibnamefont {Fragner}}, \bibinfo {author} {\bibfnamefont {M.}~\bibnamefont {Baur}}, \bibinfo {author} {\bibfnamefont {R.}~\bibnamefont {Bianchetti}}, \bibinfo {author} {\bibfnamefont {S.}~\bibnamefont {Filipp}}, \bibinfo {author} {\bibfnamefont {J.~M.}\ \bibnamefont {Fink}}, \bibinfo {author} {\bibfnamefont {P.~J.}\ \bibnamefont {Leek}}, \bibinfo {author} {\bibfnamefont {G.}~\bibnamefont {Puebla}}, \bibinfo {author} {\bibfnamefont {L.}~\bibnamefont {Steffen}},\ and\ \bibinfo {author} {\bibfnamefont {A.}~\bibnamefont {Wallraff}},\ }\bibfield  {title} {\bibinfo {title} {Coplanar waveguide resonators for circuit quantum electrodynamics},\ }\href {https://doi.org/10.1063/1.3010859} {\bibfield  {journal} {\bibinfo  {journal} {Journal of Applied Physics}\ }\textbf {\bibinfo {volume} {104}},\ \bibinfo {pages} {113904} (\bibinfo {year} {2008})},\ \Eprint
  {https://arxiv.org/abs/https://doi.org/10.1063/1.3010859} {https://doi.org/10.1063/1.3010859} \BibitemShut {NoStop}%
\bibitem [{\citenamefont {Gieres}\ \emph {et~al.}(1991)\citenamefont {Gieres}, \citenamefont {Kessler}, \citenamefont {Kraus}, \citenamefont {Roas}, \citenamefont {Russer}, \citenamefont {Soelkner},\ and\ \citenamefont {Valenzuela}}]{Gieres1991}%
  \BibitemOpen
  \bibfield  {author} {\bibinfo {author} {\bibfnamefont {G.}~\bibnamefont {Gieres}}, \bibinfo {author} {\bibfnamefont {J.}~\bibnamefont {Kessler}}, \bibinfo {author} {\bibfnamefont {J.}~\bibnamefont {Kraus}}, \bibinfo {author} {\bibfnamefont {B.}~\bibnamefont {Roas}}, \bibinfo {author} {\bibfnamefont {P.}~\bibnamefont {Russer}}, \bibinfo {author} {\bibfnamefont {G.}~\bibnamefont {Soelkner}},\ and\ \bibinfo {author} {\bibfnamefont {A.~A.}\ \bibnamefont {Valenzuela}},\ }\bibfield  {title} {\bibinfo {title} {High-frequency characterization of yba2cu3o7-x thin films with coplanar resonators},\ }\href {https://doi.org/10.1088/0953-2048/4/11/020} {\bibfield  {journal} {\bibinfo  {journal} {Supercond. Sci. Technol.}\ }\textbf {\bibinfo {volume} {4}},\ \bibinfo {pages} {629} (\bibinfo {year} {1991})}\BibitemShut {NoStop}%
\bibitem [{\citenamefont {Yoshida}\ \emph {et~al.}(1992)\citenamefont {Yoshida}, \citenamefont {Hossain}, \citenamefont {Kisu}, \citenamefont {Enpuku},\ and\ \citenamefont {Yamafuji}}]{Yoshida1992}%
  \BibitemOpen
  \bibfield  {author} {\bibinfo {author} {\bibfnamefont {K.}~\bibnamefont {Yoshida}}, \bibinfo {author} {\bibfnamefont {M.~S.}\ \bibnamefont {Hossain}}, \bibinfo {author} {\bibfnamefont {T.}~\bibnamefont {Kisu}}, \bibinfo {author} {\bibfnamefont {K.~E.~K.}\ \bibnamefont {Enpuku}},\ and\ \bibinfo {author} {\bibfnamefont {K.~Y.~K.}\ \bibnamefont {Yamafuji}},\ }\bibfield  {title} {\bibinfo {title} {Modeling of kinetic-inductance coplanar striplin with nbn thin films},\ }\href {https://doi.org/10.1143/JJAP.31.3844} {\bibfield  {journal} {\bibinfo  {journal} {Japanese Journal of Applied Physics}\ }\textbf {\bibinfo {volume} {31}},\ \bibinfo {pages} {3844} (\bibinfo {year} {1992})}\BibitemShut {NoStop}%
\bibitem [{\citenamefont {Rauch}\ \emph {et~al.}(1993)\citenamefont {Rauch}, \citenamefont {Gornik}, \citenamefont {Sölkner}, \citenamefont {Valenzuela}, \citenamefont {Fox},\ and\ \citenamefont {Behner}}]{Rauch1993}%
  \BibitemOpen
  \bibfield  {author} {\bibinfo {author} {\bibfnamefont {W.}~\bibnamefont {Rauch}}, \bibinfo {author} {\bibfnamefont {E.}~\bibnamefont {Gornik}}, \bibinfo {author} {\bibfnamefont {G.}~\bibnamefont {Sölkner}}, \bibinfo {author} {\bibfnamefont {A.~A.}\ \bibnamefont {Valenzuela}}, \bibinfo {author} {\bibfnamefont {F.}~\bibnamefont {Fox}},\ and\ \bibinfo {author} {\bibfnamefont {H.}~\bibnamefont {Behner}},\ }\bibfield  {title} {\bibinfo {title} {Microwave properties of yba2cu3o7-x thin films studied with coplanar transmission line resonators},\ }\href {https://doi.org/10.1063/1.353173} {\bibfield  {journal} {\bibinfo  {journal} {Journal of Applied Physics}\ }\textbf {\bibinfo {volume} {73}},\ \bibinfo {pages} {1866} (\bibinfo {year} {1993})},\ \Eprint {https://arxiv.org/abs/https://pubs.aip.org/aip/jap/article-pdf/73/4/1866/18654921/1866\_1\_online.pdf} {https://pubs.aip.org/aip/jap/article-pdf/73/4/1866/18654921/1866\_1\_online.pdf} \BibitemShut {NoStop}%
\bibitem [{\citenamefont {Zikiy}\ \emph {et~al.}(2023)\citenamefont {Zikiy}, \citenamefont {Ivanov}, \citenamefont {Smirnov}, \citenamefont {Moskalev}, \citenamefont {Polozov}, \citenamefont {Matanin}, \citenamefont {Malevannaya}, \citenamefont {Echeistov}, \citenamefont {Konstantinova},\ and\ \citenamefont {Rodionov}}]{Zikiy2023}%
  \BibitemOpen
  \bibfield  {author} {\bibinfo {author} {\bibfnamefont {E.~V.}\ \bibnamefont {Zikiy}}, \bibinfo {author} {\bibfnamefont {A.~I.}\ \bibnamefont {Ivanov}}, \bibinfo {author} {\bibfnamefont {N.~S.}\ \bibnamefont {Smirnov}}, \bibinfo {author} {\bibfnamefont {D.~O.}\ \bibnamefont {Moskalev}}, \bibinfo {author} {\bibfnamefont {V.~I.}\ \bibnamefont {Polozov}}, \bibinfo {author} {\bibfnamefont {A.~R.}\ \bibnamefont {Matanin}}, \bibinfo {author} {\bibfnamefont {E.~I.}\ \bibnamefont {Malevannaya}}, \bibinfo {author} {\bibfnamefont {V.~V.}\ \bibnamefont {Echeistov}}, \bibinfo {author} {\bibfnamefont {T.~G.}\ \bibnamefont {Konstantinova}},\ and\ \bibinfo {author} {\bibfnamefont {I.~A.}\ \bibnamefont {Rodionov}},\ }\bibfield  {title} {\bibinfo {title} {High-q trenched aluminum coplanar resonators with an ultrasonic edge microcutting for superconducting quantum devices},\ }\href {https://doi.org/10.1038/s41598-023-42332-6} {\bibfield  {journal} {\bibinfo  {journal} {Scientific Reports}\ }\textbf {\bibinfo {volume} {13}},\
  \bibinfo {pages} {15536} (\bibinfo {year} {2023})}\BibitemShut {NoStop}%
\bibitem [{\citenamefont {Frunzio}\ \emph {et~al.}(2005)\citenamefont {Frunzio}, \citenamefont {Wallraff}, \citenamefont {Schuster}, \citenamefont {Majer},\ and\ \citenamefont {Schoelkopf}}]{Frunzio2005}%
  \BibitemOpen
  \bibfield  {author} {\bibinfo {author} {\bibfnamefont {L.}~\bibnamefont {Frunzio}}, \bibinfo {author} {\bibfnamefont {A.}~\bibnamefont {Wallraff}}, \bibinfo {author} {\bibfnamefont {D.}~\bibnamefont {Schuster}}, \bibinfo {author} {\bibfnamefont {J.}~\bibnamefont {Majer}},\ and\ \bibinfo {author} {\bibfnamefont {R.}~\bibnamefont {Schoelkopf}},\ }\bibfield  {title} {\bibinfo {title} {Fabrication and characterization of superconducting circuit qed devices for quantum computation},\ }\href {https://doi.org/10.1109/TASC.2005.850084} {\bibfield  {journal} {\bibinfo  {journal} {IEEE Transactions on Applied Superconductivity}\ }\textbf {\bibinfo {volume} {15}},\ \bibinfo {pages} {860} (\bibinfo {year} {2005})}\BibitemShut {NoStop}%
\bibitem [{\citenamefont {Barends}\ \emph {et~al.}(2007)\citenamefont {Barends}, \citenamefont {Baselmans}, \citenamefont {Hovenier}, \citenamefont {Gao}, \citenamefont {Yates}, \citenamefont {Klapwijk},\ and\ \citenamefont {Hoevers}}]{Barends2007}%
  \BibitemOpen
  \bibfield  {author} {\bibinfo {author} {\bibfnamefont {R.}~\bibnamefont {Barends}}, \bibinfo {author} {\bibfnamefont {J.~J.~A.}\ \bibnamefont {Baselmans}}, \bibinfo {author} {\bibfnamefont {J.~N.}\ \bibnamefont {Hovenier}}, \bibinfo {author} {\bibfnamefont {J.~R.}\ \bibnamefont {Gao}}, \bibinfo {author} {\bibfnamefont {S.~J.~C.}\ \bibnamefont {Yates}}, \bibinfo {author} {\bibfnamefont {T.~M.}\ \bibnamefont {Klapwijk}},\ and\ \bibinfo {author} {\bibfnamefont {H.~F.~C.}\ \bibnamefont {Hoevers}},\ }\bibfield  {title} {\bibinfo {title} {Niobium and tantalum high q resonators for photon detectors},\ }\href {https://doi.org/10.1109/TASC.2007.898541} {\bibfield  {journal} {\bibinfo  {journal} {IEEE Transactions on Applied Superconductivity}\ }\textbf {\bibinfo {volume} {17}},\ \bibinfo {pages} {263} (\bibinfo {year} {2007})}\BibitemShut {NoStop}%
\bibitem [{\citenamefont {Zhuravel}\ \emph {et~al.}(2012)\citenamefont {Zhuravel}, \citenamefont {Kurter}, \citenamefont {Ustinov},\ and\ \citenamefont {Anlage}}]{Zhuravel2012}%
  \BibitemOpen
  \bibfield  {author} {\bibinfo {author} {\bibfnamefont {A.~P.}\ \bibnamefont {Zhuravel}}, \bibinfo {author} {\bibfnamefont {C.}~\bibnamefont {Kurter}}, \bibinfo {author} {\bibfnamefont {A.~V.}\ \bibnamefont {Ustinov}},\ and\ \bibinfo {author} {\bibfnamefont {S.~M.}\ \bibnamefont {Anlage}},\ }\bibfield  {title} {\bibinfo {title} {Unconventional rf photoresponse from a superconducting spiral resonator},\ }\href {https://doi.org/10.1103/PhysRevB.85.134535} {\bibfield  {journal} {\bibinfo  {journal} {Phys. Rev. B}\ }\textbf {\bibinfo {volume} {85}},\ \bibinfo {pages} {134535} (\bibinfo {year} {2012})}\BibitemShut {NoStop}%
\bibitem [{\citenamefont {Maleeva}\ \emph {et~al.}(2014)\citenamefont {Maleeva}, \citenamefont {Fistul}, \citenamefont {Karpov}, \citenamefont {Zhuravel}, \citenamefont {Averkin}, \citenamefont {Jung},\ and\ \citenamefont {Ustinov}}]{Maleeva2014}%
  \BibitemOpen
  \bibfield  {author} {\bibinfo {author} {\bibfnamefont {N.}~\bibnamefont {Maleeva}}, \bibinfo {author} {\bibfnamefont {M.~V.}\ \bibnamefont {Fistul}}, \bibinfo {author} {\bibfnamefont {A.}~\bibnamefont {Karpov}}, \bibinfo {author} {\bibfnamefont {A.~P.}\ \bibnamefont {Zhuravel}}, \bibinfo {author} {\bibfnamefont {A.}~\bibnamefont {Averkin}}, \bibinfo {author} {\bibfnamefont {P.}~\bibnamefont {Jung}},\ and\ \bibinfo {author} {\bibfnamefont {A.~V.}\ \bibnamefont {Ustinov}},\ }\bibfield  {title} {\bibinfo {title} {{Electrodynamics of a ring-shaped spiral resonator}},\ }\href {https://doi.org/10.1063/1.4863835} {\bibfield  {journal} {\bibinfo  {journal} {Journal of Applied Physics}\ }\textbf {\bibinfo {volume} {115}},\ \bibinfo {pages} {064910} (\bibinfo {year} {2014})},\ \Eprint {https://arxiv.org/abs/https://pubs.aip.org/aip/jap/article-pdf/doi/10.1063/1.4863835/13380381/064910\_1\_online.pdf} {https://pubs.aip.org/aip/jap/article-pdf/doi/10.1063/1.4863835/13380381/064910\_1\_online.pdf} \BibitemShut {NoStop}%
\bibitem [{\citenamefont {Partanen}\ \emph {et~al.}(2016)\citenamefont {Partanen}, \citenamefont {Tan}, \citenamefont {Govenius}, \citenamefont {Lake}, \citenamefont {M{\"a}kel{\"a}}, \citenamefont {Tanttu},\ and\ \citenamefont {M{\"o}tt{\"o}nen}}]{Partanen2016}%
  \BibitemOpen
  \bibfield  {author} {\bibinfo {author} {\bibfnamefont {M.}~\bibnamefont {Partanen}}, \bibinfo {author} {\bibfnamefont {K.~Y.}\ \bibnamefont {Tan}}, \bibinfo {author} {\bibfnamefont {J.}~\bibnamefont {Govenius}}, \bibinfo {author} {\bibfnamefont {R.~E.}\ \bibnamefont {Lake}}, \bibinfo {author} {\bibfnamefont {M.~K.}\ \bibnamefont {M{\"a}kel{\"a}}}, \bibinfo {author} {\bibfnamefont {T.}~\bibnamefont {Tanttu}},\ and\ \bibinfo {author} {\bibfnamefont {M.}~\bibnamefont {M{\"o}tt{\"o}nen}},\ }\bibfield  {title} {\bibinfo {title} {Quantum-limited heat conduction over macroscopic distances},\ }\href {https://doi.org/10.1038/nphys3642} {\bibfield  {journal} {\bibinfo  {journal} {Nature Physics}\ }\textbf {\bibinfo {volume} {12}},\ \bibinfo {pages} {460} (\bibinfo {year} {2016})}\BibitemShut {NoStop}%
\bibitem [{\citenamefont {Rolland}\ \emph {et~al.}(2019)\citenamefont {Rolland}, \citenamefont {Peugeot}, \citenamefont {Dambach}, \citenamefont {Westig}, \citenamefont {Kubala}, \citenamefont {Mukharsky}, \citenamefont {Altimiras}, \citenamefont {le~Sueur}, \citenamefont {Joyez}, \citenamefont {Vion}, \citenamefont {Roche}, \citenamefont {Esteve}, \citenamefont {Ankerhold},\ and\ \citenamefont {Portier}}]{Rolland2019}%
  \BibitemOpen
  \bibfield  {author} {\bibinfo {author} {\bibfnamefont {C.}~\bibnamefont {Rolland}}, \bibinfo {author} {\bibfnamefont {A.}~\bibnamefont {Peugeot}}, \bibinfo {author} {\bibfnamefont {S.}~\bibnamefont {Dambach}}, \bibinfo {author} {\bibfnamefont {M.}~\bibnamefont {Westig}}, \bibinfo {author} {\bibfnamefont {B.}~\bibnamefont {Kubala}}, \bibinfo {author} {\bibfnamefont {Y.}~\bibnamefont {Mukharsky}}, \bibinfo {author} {\bibfnamefont {C.}~\bibnamefont {Altimiras}}, \bibinfo {author} {\bibfnamefont {H.}~\bibnamefont {le~Sueur}}, \bibinfo {author} {\bibfnamefont {P.}~\bibnamefont {Joyez}}, \bibinfo {author} {\bibfnamefont {D.}~\bibnamefont {Vion}}, \bibinfo {author} {\bibfnamefont {P.}~\bibnamefont {Roche}}, \bibinfo {author} {\bibfnamefont {D.}~\bibnamefont {Esteve}}, \bibinfo {author} {\bibfnamefont {J.}~\bibnamefont {Ankerhold}},\ and\ \bibinfo {author} {\bibfnamefont {F.}~\bibnamefont {Portier}},\ }\bibfield  {title} {\bibinfo {title} {Antibunched photons emitted by a dc-biased josephson junction},\ }\href
  {https://doi.org/10.1103/PhysRevLett.122.186804} {\bibfield  {journal} {\bibinfo  {journal} {Phys. Rev. Lett.}\ }\textbf {\bibinfo {volume} {122}},\ \bibinfo {pages} {186804} (\bibinfo {year} {2019})}\BibitemShut {NoStop}%
\bibitem [{\citenamefont {Yan}\ \emph {et~al.}(2021)\citenamefont {Yan}, \citenamefont {Hassel}, \citenamefont {Vesterinen}, \citenamefont {Zhang}, \citenamefont {Ikonen}, \citenamefont {Gr{\"o}nberg}, \citenamefont {Goetz},\ and\ \citenamefont {M{\"o}tt{\"o}nen}}]{Yan2021}%
  \BibitemOpen
  \bibfield  {author} {\bibinfo {author} {\bibfnamefont {C.}~\bibnamefont {Yan}}, \bibinfo {author} {\bibfnamefont {J.}~\bibnamefont {Hassel}}, \bibinfo {author} {\bibfnamefont {V.}~\bibnamefont {Vesterinen}}, \bibinfo {author} {\bibfnamefont {J.}~\bibnamefont {Zhang}}, \bibinfo {author} {\bibfnamefont {J.}~\bibnamefont {Ikonen}}, \bibinfo {author} {\bibfnamefont {L.}~\bibnamefont {Gr{\"o}nberg}}, \bibinfo {author} {\bibfnamefont {J.}~\bibnamefont {Goetz}},\ and\ \bibinfo {author} {\bibfnamefont {M.}~\bibnamefont {M{\"o}tt{\"o}nen}},\ }\bibfield  {title} {\bibinfo {title} {A low-noise on-chip coherent microwave source},\ }\href {https://doi.org/10.1038/s41928-021-00680-z} {\bibfield  {journal} {\bibinfo  {journal} {Nature Electronics}\ }\textbf {\bibinfo {volume} {4}},\ \bibinfo {pages} {885} (\bibinfo {year} {2021})}\BibitemShut {NoStop}%
\bibitem [{\citenamefont {Wenner}\ \emph {et~al.}(2011)\citenamefont {Wenner}, \citenamefont {Neeley}, \citenamefont {Bialczak}, \citenamefont {Lenander}, \citenamefont {Lucero}, \citenamefont {O’Connell}, \citenamefont {Sank}, \citenamefont {Wang}, \citenamefont {Weides}, \citenamefont {Cleland},\ and\ \citenamefont {Martinis}}]{Wenner2011}%
  \BibitemOpen
  \bibfield  {author} {\bibinfo {author} {\bibfnamefont {J.}~\bibnamefont {Wenner}}, \bibinfo {author} {\bibfnamefont {M.}~\bibnamefont {Neeley}}, \bibinfo {author} {\bibfnamefont {R.~C.}\ \bibnamefont {Bialczak}}, \bibinfo {author} {\bibfnamefont {M.}~\bibnamefont {Lenander}}, \bibinfo {author} {\bibfnamefont {E.}~\bibnamefont {Lucero}}, \bibinfo {author} {\bibfnamefont {A.~D.}\ \bibnamefont {O’Connell}}, \bibinfo {author} {\bibfnamefont {D.}~\bibnamefont {Sank}}, \bibinfo {author} {\bibfnamefont {H.}~\bibnamefont {Wang}}, \bibinfo {author} {\bibfnamefont {M.}~\bibnamefont {Weides}}, \bibinfo {author} {\bibfnamefont {A.~N.}\ \bibnamefont {Cleland}},\ and\ \bibinfo {author} {\bibfnamefont {J.~M.}\ \bibnamefont {Martinis}},\ }\bibfield  {title} {\bibinfo {title} {Wirebond crosstalk and cavity modes in large chip mounts for superconducting qubits},\ }\href {https://doi.org/10.1088/0953-2048/24/6/065001} {\bibfield  {journal} {\bibinfo  {journal} {Superconductor Science and Technology}\ }\textbf {\bibinfo
  {volume} {24}},\ \bibinfo {pages} {065001} (\bibinfo {year} {2011})}\BibitemShut {NoStop}%
\bibitem [{\citenamefont {Lankwarden}\ \emph {et~al.}(2012)\citenamefont {Lankwarden}, \citenamefont {Endo}, \citenamefont {Baselmans},\ and\ \citenamefont {Bruijn}}]{Lankwarden2012}%
  \BibitemOpen
  \bibfield  {author} {\bibinfo {author} {\bibfnamefont {Y.~J.~Y.}\ \bibnamefont {Lankwarden}}, \bibinfo {author} {\bibfnamefont {A.}~\bibnamefont {Endo}}, \bibinfo {author} {\bibfnamefont {J.~J.~A.}\ \bibnamefont {Baselmans}},\ and\ \bibinfo {author} {\bibfnamefont {M.~P.}\ \bibnamefont {Bruijn}},\ }\bibfield  {title} {\bibinfo {title} {Development of nbtin-al direct antenna coupled kinetic inductance detectors},\ }\href {https://doi.org/10.1007/s10909-012-0503-0} {\bibfield  {journal} {\bibinfo  {journal} {Journal of Low Temperature Physics}\ }\textbf {\bibinfo {volume} {167}},\ \bibinfo {pages} {367} (\bibinfo {year} {2012})}\BibitemShut {NoStop}%
\bibitem [{\citenamefont {Abuwasib}\ \emph {et~al.}(2013)\citenamefont {Abuwasib}, \citenamefont {Krantz},\ and\ \citenamefont {Delsing}}]{Abuwasib2013}%
  \BibitemOpen
  \bibfield  {author} {\bibinfo {author} {\bibfnamefont {M.}~\bibnamefont {Abuwasib}}, \bibinfo {author} {\bibfnamefont {P.}~\bibnamefont {Krantz}},\ and\ \bibinfo {author} {\bibfnamefont {P.}~\bibnamefont {Delsing}},\ }\bibfield  {title} {\bibinfo {title} {Fabrication of large dimension aluminum air-bridges for superconducting quantum circuits},\ }\href {https://doi.org/10.1116/1.4798399} {\bibfield  {journal} {\bibinfo  {journal} {Journal of Vacuum Science and Technology B}\ }\textbf {\bibinfo {volume} {31}},\ \bibinfo {pages} {031601} (\bibinfo {year} {2013})},\ \Eprint {https://arxiv.org/abs/https://pubs.aip.org/avs/jvb/article-pdf/doi/10.1116/1.4798399/15916545/031601\_1\_online.pdf} {https://pubs.aip.org/avs/jvb/article-pdf/doi/10.1116/1.4798399/15916545/031601\_1\_online.pdf} \BibitemShut {NoStop}%
\bibitem [{\citenamefont {Chen}\ \emph {et~al.}(2014)\citenamefont {Chen}, \citenamefont {Megrant}, \citenamefont {Kelly}, \citenamefont {Barends}, \citenamefont {Bochmann}, \citenamefont {Chen}, \citenamefont {Chiaro}, \citenamefont {Dunsworth}, \citenamefont {Jeffrey}, \citenamefont {Mutus}, \citenamefont {O'Malley}, \citenamefont {Neill}, \citenamefont {Roushan}, \citenamefont {Sank}, \citenamefont {Vainsencher}, \citenamefont {Wenner}, \citenamefont {White}, \citenamefont {Cleland},\ and\ \citenamefont {Martinis}}]{Chen2014}%
  \BibitemOpen
  \bibfield  {author} {\bibinfo {author} {\bibfnamefont {Z.}~\bibnamefont {Chen}}, \bibinfo {author} {\bibfnamefont {A.}~\bibnamefont {Megrant}}, \bibinfo {author} {\bibfnamefont {J.}~\bibnamefont {Kelly}}, \bibinfo {author} {\bibfnamefont {R.}~\bibnamefont {Barends}}, \bibinfo {author} {\bibfnamefont {J.}~\bibnamefont {Bochmann}}, \bibinfo {author} {\bibfnamefont {Y.}~\bibnamefont {Chen}}, \bibinfo {author} {\bibfnamefont {B.}~\bibnamefont {Chiaro}}, \bibinfo {author} {\bibfnamefont {A.}~\bibnamefont {Dunsworth}}, \bibinfo {author} {\bibfnamefont {E.}~\bibnamefont {Jeffrey}}, \bibinfo {author} {\bibfnamefont {J.~Y.}\ \bibnamefont {Mutus}}, \bibinfo {author} {\bibfnamefont {P.~J.~J.}\ \bibnamefont {O'Malley}}, \bibinfo {author} {\bibfnamefont {C.}~\bibnamefont {Neill}}, \bibinfo {author} {\bibfnamefont {P.}~\bibnamefont {Roushan}}, \bibinfo {author} {\bibfnamefont {D.}~\bibnamefont {Sank}}, \bibinfo {author} {\bibfnamefont {A.}~\bibnamefont {Vainsencher}}, \bibinfo {author} {\bibfnamefont {J.}~\bibnamefont
  {Wenner}}, \bibinfo {author} {\bibfnamefont {T.~C.}\ \bibnamefont {White}}, \bibinfo {author} {\bibfnamefont {A.~N.}\ \bibnamefont {Cleland}},\ and\ \bibinfo {author} {\bibfnamefont {J.~M.}\ \bibnamefont {Martinis}},\ }\bibfield  {title} {\bibinfo {title} {Fabrication and characterization of aluminum airbridges for superconducting microwave circuits},\ }\href {https://doi.org/10.1063/1.4863745} {\bibfield  {journal} {\bibinfo  {journal} {Applied Physics Letters}\ }\textbf {\bibinfo {volume} {104}},\ \bibinfo {pages} {052602} (\bibinfo {year} {2014})},\ \Eprint {https://arxiv.org/abs/https://pubs.aip.org/aip/apl/article-pdf/doi/10.1063/1.4863745/14305812/052602\_1\_online.pdf} {https://pubs.aip.org/aip/apl/article-pdf/doi/10.1063/1.4863745/14305812/052602\_1\_online.pdf} \BibitemShut {NoStop}%
\bibitem [{\citenamefont {Fischer}\ \emph {et~al.}(2021)\citenamefont {Fischer}, \citenamefont {Chen}, \citenamefont {Besson}, \citenamefont {Eder}, \citenamefont {Goetz}, \citenamefont {Pogorzalek}, \citenamefont {Renger}, \citenamefont {Xie}, \citenamefont {Hartmann}, \citenamefont {Fedorov}, \citenamefont {Marx}, \citenamefont {Deppe},\ and\ \citenamefont {Gross}}]{Fischer2021}%
  \BibitemOpen
  \bibfield  {author} {\bibinfo {author} {\bibfnamefont {M.}~\bibnamefont {Fischer}}, \bibinfo {author} {\bibfnamefont {Q.-M.}\ \bibnamefont {Chen}}, \bibinfo {author} {\bibfnamefont {C.}~\bibnamefont {Besson}}, \bibinfo {author} {\bibfnamefont {P.}~\bibnamefont {Eder}}, \bibinfo {author} {\bibfnamefont {J.}~\bibnamefont {Goetz}}, \bibinfo {author} {\bibfnamefont {S.}~\bibnamefont {Pogorzalek}}, \bibinfo {author} {\bibfnamefont {M.}~\bibnamefont {Renger}}, \bibinfo {author} {\bibfnamefont {E.}~\bibnamefont {Xie}}, \bibinfo {author} {\bibfnamefont {M.~J.}\ \bibnamefont {Hartmann}}, \bibinfo {author} {\bibfnamefont {K.~G.}\ \bibnamefont {Fedorov}}, \bibinfo {author} {\bibfnamefont {A.}~\bibnamefont {Marx}}, \bibinfo {author} {\bibfnamefont {F.}~\bibnamefont {Deppe}},\ and\ \bibinfo {author} {\bibfnamefont {R.}~\bibnamefont {Gross}},\ }\bibfield  {title} {\bibinfo {title} {In situ tunable nonlinearity and competing signal paths in coupled superconducting resonators},\ }\href
  {https://doi.org/10.1103/PhysRevB.103.094515} {\bibfield  {journal} {\bibinfo  {journal} {Phys. Rev. B}\ }\textbf {\bibinfo {volume} {103}},\ \bibinfo {pages} {094515} (\bibinfo {year} {2021})}\BibitemShut {NoStop}%
\bibitem [{\citenamefont {Houck}\ \emph {et~al.}(2008)\citenamefont {Houck}, \citenamefont {Schreier}, \citenamefont {Johnson}, \citenamefont {Chow}, \citenamefont {Koch}, \citenamefont {Gambetta}, \citenamefont {Schuster}, \citenamefont {Frunzio}, \citenamefont {Devoret}, \citenamefont {Girvin},\ and\ \citenamefont {Schoelkopf}}]{Houck2008}%
  \BibitemOpen
  \bibfield  {author} {\bibinfo {author} {\bibfnamefont {A.~A.}\ \bibnamefont {Houck}}, \bibinfo {author} {\bibfnamefont {J.~A.}\ \bibnamefont {Schreier}}, \bibinfo {author} {\bibfnamefont {B.~R.}\ \bibnamefont {Johnson}}, \bibinfo {author} {\bibfnamefont {J.~M.}\ \bibnamefont {Chow}}, \bibinfo {author} {\bibfnamefont {J.}~\bibnamefont {Koch}}, \bibinfo {author} {\bibfnamefont {J.~M.}\ \bibnamefont {Gambetta}}, \bibinfo {author} {\bibfnamefont {D.~I.}\ \bibnamefont {Schuster}}, \bibinfo {author} {\bibfnamefont {L.}~\bibnamefont {Frunzio}}, \bibinfo {author} {\bibfnamefont {M.~H.}\ \bibnamefont {Devoret}}, \bibinfo {author} {\bibfnamefont {S.~M.}\ \bibnamefont {Girvin}},\ and\ \bibinfo {author} {\bibfnamefont {R.~J.}\ \bibnamefont {Schoelkopf}},\ }\bibfield  {title} {\bibinfo {title} {Controlling the spontaneous emission of a superconducting transmon qubit},\ }\href {https://doi.org/10.1103/PhysRevLett.101.080502} {\bibfield  {journal} {\bibinfo  {journal} {Phys. Rev. Lett.}\ }\textbf {\bibinfo {volume}
  {101}},\ \bibinfo {pages} {080502} (\bibinfo {year} {2008})}\BibitemShut {NoStop}%
\bibitem [{\citenamefont {Lancaster}\ \emph {et~al.}(1993)\citenamefont {Lancaster}, \citenamefont {Li}, \citenamefont {Porch},\ and\ \citenamefont {Chew}}]{Lancaster1993}%
  \BibitemOpen
  \bibfield  {author} {\bibinfo {author} {\bibfnamefont {M.}~\bibnamefont {Lancaster}}, \bibinfo {author} {\bibfnamefont {J.}~\bibnamefont {Li}}, \bibinfo {author} {\bibfnamefont {A.}~\bibnamefont {Porch}},\ and\ \bibinfo {author} {\bibfnamefont {N.}~\bibnamefont {Chew}},\ }\bibfield  {title} {\bibinfo {title} {High temperature superconductor lumped element resonator},\ }\href {https://doi.org/10.1049/el:19931149} {\bibfield  {journal} {\bibinfo  {journal} {Electronics Letters}\ }\textbf {\bibinfo {volume} {29}},\ \bibinfo {pages} {1728} (\bibinfo {year} {1993})}\BibitemShut {NoStop}%
\bibitem [{\citenamefont {Chaloupka}\ and\ \citenamefont {Kolesov}(2001)}]{Chaloupka2001}%
  \BibitemOpen
  \bibfield  {author} {\bibinfo {author} {\bibfnamefont {H.~J.}\ \bibnamefont {Chaloupka}}\ and\ \bibinfo {author} {\bibfnamefont {S.}~\bibnamefont {Kolesov}},\ }\bibinfo {title} {Design of lumped-element 2d rf devices},\ in\ \href {https://doi.org/10.1007/978-94-010-0450-3_9} {\emph {\bibinfo {booktitle} {Microwave Superconductivity}}},\ \bibinfo {editor} {edited by\ \bibinfo {editor} {\bibfnamefont {H.}~\bibnamefont {Weinstock}}\ and\ \bibinfo {editor} {\bibfnamefont {M.}~\bibnamefont {Nisenoff}}}\ (\bibinfo  {publisher} {Springer Netherlands},\ \bibinfo {address} {Dordrecht},\ \bibinfo {year} {2001})\ pp.\ \bibinfo {pages} {205--238}\BibitemShut {NoStop}%
\bibitem [{\citenamefont {Chen}\ \emph {et~al.}(2023{\natexlab{b}})\citenamefont {Chen}, \citenamefont {Singh}, \citenamefont {Duda}, \citenamefont {Catto}, \citenamefont {Ker\"anen}, \citenamefont {Alizadeh}, \citenamefont {M\"orstedt}, \citenamefont {Sah}, \citenamefont {Gunyh\'o}, \citenamefont {Liu},\ and\ \citenamefont {M\"ott\"onen}}]{Chen}%
  \BibitemOpen
  \bibfield  {author} {\bibinfo {author} {\bibfnamefont {Q.-M.}\ \bibnamefont {Chen}}, \bibinfo {author} {\bibfnamefont {P.}~\bibnamefont {Singh}}, \bibinfo {author} {\bibfnamefont {R.}~\bibnamefont {Duda}}, \bibinfo {author} {\bibfnamefont {G.}~\bibnamefont {Catto}}, \bibinfo {author} {\bibfnamefont {A.}~\bibnamefont {Ker\"anen}}, \bibinfo {author} {\bibfnamefont {A.}~\bibnamefont {Alizadeh}}, \bibinfo {author} {\bibfnamefont {T.}~\bibnamefont {M\"orstedt}}, \bibinfo {author} {\bibfnamefont {A.}~\bibnamefont {Sah}}, \bibinfo {author} {\bibfnamefont {A.}~\bibnamefont {Gunyh\'o}}, \bibinfo {author} {\bibfnamefont {W.}~\bibnamefont {Liu}},\ and\ \bibinfo {author} {\bibfnamefont {M.}~\bibnamefont {M\"ott\"onen}},\ }\bibfield  {title} {\bibinfo {title} {Compact inductor-capacitor resonators at sub-gigahertz frequencies},\ }\href {https://doi.org/10.1103/PhysRevResearch.5.043126} {\bibfield  {journal} {\bibinfo  {journal} {Phys. Rev. Res.}\ }\textbf {\bibinfo {volume} {5}},\ \bibinfo {pages} {043126} (\bibinfo
  {year} {2023}{\natexlab{b}})}\BibitemShut {NoStop}%
\bibitem [{\citenamefont {McKenzie-Sell}\ \emph {et~al.}(2019)\citenamefont {McKenzie-Sell}, \citenamefont {Xie}, \citenamefont {Lee}, \citenamefont {Robinson}, \citenamefont {Ciccarelli},\ and\ \citenamefont {Haigh}}]{McKenzie2019}%
  \BibitemOpen
  \bibfield  {author} {\bibinfo {author} {\bibfnamefont {L.}~\bibnamefont {McKenzie-Sell}}, \bibinfo {author} {\bibfnamefont {J.}~\bibnamefont {Xie}}, \bibinfo {author} {\bibfnamefont {C.-M.}\ \bibnamefont {Lee}}, \bibinfo {author} {\bibfnamefont {J.~W.~A.}\ \bibnamefont {Robinson}}, \bibinfo {author} {\bibfnamefont {C.}~\bibnamefont {Ciccarelli}},\ and\ \bibinfo {author} {\bibfnamefont {J.~A.}\ \bibnamefont {Haigh}},\ }\bibfield  {title} {\bibinfo {title} {Low-impedance superconducting microwave resonators for strong coupling to small magnetic mode volumes},\ }\href {https://doi.org/10.1103/PhysRevB.99.140414} {\bibfield  {journal} {\bibinfo  {journal} {Phys. Rev. B}\ }\textbf {\bibinfo {volume} {99}},\ \bibinfo {pages} {140414} (\bibinfo {year} {2019})}\BibitemShut {NoStop}%
\bibitem [{\citenamefont {Geerlings}\ \emph {et~al.}(2012)\citenamefont {Geerlings}, \citenamefont {Shankar}, \citenamefont {Edwards}, \citenamefont {Frunzio}, \citenamefont {Schoelkopf},\ and\ \citenamefont {Devoret}}]{Geerlings2012}%
  \BibitemOpen
  \bibfield  {author} {\bibinfo {author} {\bibfnamefont {K.}~\bibnamefont {Geerlings}}, \bibinfo {author} {\bibfnamefont {S.}~\bibnamefont {Shankar}}, \bibinfo {author} {\bibfnamefont {E.}~\bibnamefont {Edwards}}, \bibinfo {author} {\bibfnamefont {L.}~\bibnamefont {Frunzio}}, \bibinfo {author} {\bibfnamefont {R.~J.}\ \bibnamefont {Schoelkopf}},\ and\ \bibinfo {author} {\bibfnamefont {M.~H.}\ \bibnamefont {Devoret}},\ }\bibfield  {title} {\bibinfo {title} {{Improving the quality factor of microwave compact resonators by optimizing their geometrical parameters}},\ }\href {https://doi.org/10.1063/1.4710520} {\bibfield  {journal} {\bibinfo  {journal} {Applied Physics Letters}\ }\textbf {\bibinfo {volume} {100}},\ \bibinfo {pages} {192601} (\bibinfo {year} {2012})},\ \Eprint {https://arxiv.org/abs/https://pubs.aip.org/aip/apl/article-pdf/doi/10.1063/1.4710520/13984673/192601\_1\_online.pdf} {https://pubs.aip.org/aip/apl/article-pdf/doi/10.1063/1.4710520/13984673/192601\_1\_online.pdf} \BibitemShut {NoStop}%
\bibitem [{\citenamefont {Zotova}\ \emph {et~al.}(2023)\citenamefont {Zotova}, \citenamefont {Wang}, \citenamefont {Semenov}, \citenamefont {Zhou}, \citenamefont {Khrapach}, \citenamefont {Tomonaga}, \citenamefont {Astafiev},\ and\ \citenamefont {Tsai}}]{Zotova2023}%
  \BibitemOpen
  \bibfield  {author} {\bibinfo {author} {\bibfnamefont {J.}~\bibnamefont {Zotova}}, \bibinfo {author} {\bibfnamefont {R.}~\bibnamefont {Wang}}, \bibinfo {author} {\bibfnamefont {A.}~\bibnamefont {Semenov}}, \bibinfo {author} {\bibfnamefont {Y.}~\bibnamefont {Zhou}}, \bibinfo {author} {\bibfnamefont {I.}~\bibnamefont {Khrapach}}, \bibinfo {author} {\bibfnamefont {A.}~\bibnamefont {Tomonaga}}, \bibinfo {author} {\bibfnamefont {O.}~\bibnamefont {Astafiev}},\ and\ \bibinfo {author} {\bibfnamefont {J.-S.}\ \bibnamefont {Tsai}},\ }\bibfield  {title} {\bibinfo {title} {Compact superconducting microwave resonators based on ${\text{al-alo}}_{x}$-al capacitors},\ }\href {https://doi.org/10.1103/PhysRevApplied.19.044067} {\bibfield  {journal} {\bibinfo  {journal} {Phys. Rev. Appl.}\ }\textbf {\bibinfo {volume} {19}},\ \bibinfo {pages} {044067} (\bibinfo {year} {2023})}\BibitemShut {NoStop}%
\bibitem [{\citenamefont {Cho}\ \emph {et~al.}(2013)\citenamefont {Cho}, \citenamefont {Patel}, \citenamefont {Podkaminer}, \citenamefont {Gao}, \citenamefont {Folkman}, \citenamefont {Bark}, \citenamefont {Lee}, \citenamefont {Zhang}, \citenamefont {Pan}, \citenamefont {McDermott},\ and\ \citenamefont {Eom}}]{Cho2013}%
  \BibitemOpen
  \bibfield  {author} {\bibinfo {author} {\bibfnamefont {K.-H.}\ \bibnamefont {Cho}}, \bibinfo {author} {\bibfnamefont {U.}~\bibnamefont {Patel}}, \bibinfo {author} {\bibfnamefont {J.}~\bibnamefont {Podkaminer}}, \bibinfo {author} {\bibfnamefont {Y.}~\bibnamefont {Gao}}, \bibinfo {author} {\bibfnamefont {C.~M.}\ \bibnamefont {Folkman}}, \bibinfo {author} {\bibfnamefont {C.~W.}\ \bibnamefont {Bark}}, \bibinfo {author} {\bibfnamefont {S.}~\bibnamefont {Lee}}, \bibinfo {author} {\bibfnamefont {Y.}~\bibnamefont {Zhang}}, \bibinfo {author} {\bibfnamefont {X.~Q.}\ \bibnamefont {Pan}}, \bibinfo {author} {\bibfnamefont {R.}~\bibnamefont {McDermott}},\ and\ \bibinfo {author} {\bibfnamefont {C.~B.}\ \bibnamefont {Eom}},\ }\bibfield  {title} {\bibinfo {title} {{Epitaxial Al2O3 capacitors for low microwave loss superconducting quantum circuits}},\ }\href {https://doi.org/10.1063/1.4822436} {\bibfield  {journal} {\bibinfo  {journal} {APL Materials}\ }\textbf {\bibinfo {volume} {1}},\ \bibinfo {pages} {042115} (\bibinfo
  {year} {2013})},\ \Eprint {https://arxiv.org/abs/https://pubs.aip.org/aip/apm/article-pdf/doi/10.1063/1.4822436/14555064/042115\_1\_online.pdf} {https://pubs.aip.org/aip/apm/article-pdf/doi/10.1063/1.4822436/14555064/042115\_1\_online.pdf} \BibitemShut {NoStop}%
\bibitem [{\citenamefont {Paik}\ and\ \citenamefont {Osborn}(2010)}]{Paik2010}%
  \BibitemOpen
  \bibfield  {author} {\bibinfo {author} {\bibfnamefont {H.}~\bibnamefont {Paik}}\ and\ \bibinfo {author} {\bibfnamefont {K.~D.}\ \bibnamefont {Osborn}},\ }\bibfield  {title} {\bibinfo {title} {{Reducing quantum-regime dielectric loss of silicon nitride for superconducting quantum circuits}},\ }\href {https://doi.org/10.1063/1.3309703} {\bibfield  {journal} {\bibinfo  {journal} {Applied Physics Letters}\ }\textbf {\bibinfo {volume} {96}},\ \bibinfo {pages} {072505} (\bibinfo {year} {2010})},\ \Eprint {https://arxiv.org/abs/https://pubs.aip.org/aip/apl/article-pdf/doi/10.1063/1.3309703/14426272/072505\_1\_online.pdf} {https://pubs.aip.org/aip/apl/article-pdf/doi/10.1063/1.3309703/14426272/072505\_1\_online.pdf} \BibitemShut {NoStop}%
\bibitem [{\citenamefont {Deng}\ \emph {et~al.}(2014)\citenamefont {Deng}, \citenamefont {Otto},\ and\ \citenamefont {Lupascu}}]{Deng2014}%
  \BibitemOpen
  \bibfield  {author} {\bibinfo {author} {\bibfnamefont {C.}~\bibnamefont {Deng}}, \bibinfo {author} {\bibfnamefont {M.}~\bibnamefont {Otto}},\ and\ \bibinfo {author} {\bibfnamefont {A.}~\bibnamefont {Lupascu}},\ }\bibfield  {title} {\bibinfo {title} {{Characterization of low-temperature microwave loss of thin aluminum oxide formed by plasma oxidation}},\ }\href {https://doi.org/10.1063/1.4863686} {\bibfield  {journal} {\bibinfo  {journal} {Applied Physics Letters}\ }\textbf {\bibinfo {volume} {104}},\ \bibinfo {pages} {043506} (\bibinfo {year} {2014})},\ \Eprint {https://arxiv.org/abs/https://pubs.aip.org/aip/apl/article-pdf/doi/10.1063/1.4863686/14302986/043506\_1\_online.pdf} {https://pubs.aip.org/aip/apl/article-pdf/doi/10.1063/1.4863686/14302986/043506\_1\_online.pdf} \BibitemShut {NoStop}%
\bibitem [{\citenamefont {Kouwenhoven}\ \emph {et~al.}(2024)\citenamefont {Kouwenhoven}, \citenamefont {van Doorn}, \citenamefont {Buijtendorp}, \citenamefont {de~Rooij}, \citenamefont {Lamers}, \citenamefont {Thoen}, \citenamefont {Murugesan}, \citenamefont {Baselmans},\ and\ \citenamefont {de~Visser}}]{Kouwenhouven2024}%
  \BibitemOpen
  \bibfield  {author} {\bibinfo {author} {\bibfnamefont {K.}~\bibnamefont {Kouwenhoven}}, \bibinfo {author} {\bibfnamefont {G.}~\bibnamefont {van Doorn}}, \bibinfo {author} {\bibfnamefont {B.}~\bibnamefont {Buijtendorp}}, \bibinfo {author} {\bibfnamefont {S.}~\bibnamefont {de~Rooij}}, \bibinfo {author} {\bibfnamefont {D.}~\bibnamefont {Lamers}}, \bibinfo {author} {\bibfnamefont {D.}~\bibnamefont {Thoen}}, \bibinfo {author} {\bibfnamefont {V.}~\bibnamefont {Murugesan}}, \bibinfo {author} {\bibfnamefont {J.}~\bibnamefont {Baselmans}},\ and\ \bibinfo {author} {\bibfnamefont {P.}~\bibnamefont {de~Visser}},\ }\bibfield  {title} {\bibinfo {title} {Geometry dependence of two-level-system noise and loss in $a$-$\mathrm{Si}\mathrm{C}$:$\mathrm{H}$ parallel-plate capacitors for superconducting microwave resonators},\ }\href {https://doi.org/10.1103/PhysRevApplied.21.044036} {\bibfield  {journal} {\bibinfo  {journal} {Phys. Rev. Appl.}\ }\textbf {\bibinfo {volume} {21}},\ \bibinfo {pages} {044036} (\bibinfo {year}
  {2024})}\BibitemShut {NoStop}%
\bibitem [{\citenamefont {Johansson}\ \emph {et~al.}(2014)\citenamefont {Johansson}, \citenamefont {Johansson},\ and\ \citenamefont {Nori}}]{Johansson}%
  \BibitemOpen
  \bibfield  {author} {\bibinfo {author} {\bibfnamefont {J.~R.}\ \bibnamefont {Johansson}}, \bibinfo {author} {\bibfnamefont {G.}~\bibnamefont {Johansson}},\ and\ \bibinfo {author} {\bibfnamefont {F.}~\bibnamefont {Nori}},\ }\bibfield  {title} {\bibinfo {title} {Optomechanical-like coupling between superconducting resonators},\ }\href {https://doi.org/10.1103/PhysRevA.90.053833} {\bibfield  {journal} {\bibinfo  {journal} {Phys. Rev. A}\ }\textbf {\bibinfo {volume} {90}},\ \bibinfo {pages} {053833} (\bibinfo {year} {2014})}\BibitemShut {NoStop}%
\bibitem [{\citenamefont {Kim}\ \emph {et~al.}(2015)\citenamefont {Kim}, \citenamefont {Johansson},\ and\ \citenamefont {Nori}}]{Kim}%
  \BibitemOpen
  \bibfield  {author} {\bibinfo {author} {\bibfnamefont {E.-j.}\ \bibnamefont {Kim}}, \bibinfo {author} {\bibfnamefont {J.~R.}\ \bibnamefont {Johansson}},\ and\ \bibinfo {author} {\bibfnamefont {F.}~\bibnamefont {Nori}},\ }\bibfield  {title} {\bibinfo {title} {Circuit analog of quadratic optomechanics},\ }\href {https://doi.org/10.1103/PhysRevA.91.033835} {\bibfield  {journal} {\bibinfo  {journal} {Phys. Rev. A}\ }\textbf {\bibinfo {volume} {91}},\ \bibinfo {pages} {033835} (\bibinfo {year} {2015})}\BibitemShut {NoStop}%
\bibitem [{\citenamefont {Hardal}\ \emph {et~al.}(2017)\citenamefont {Hardal}, \citenamefont {Aslan}, \citenamefont {Wilson},\ and\ \citenamefont {M\"ustecapl\ifmmode \imath \else \i \fi{}o\ifmmode~\breve{g}\else \u{g}\fi{}lu}}]{Hardal2017}%
  \BibitemOpen
  \bibfield  {author} {\bibinfo {author} {\bibfnamefont {A.~U.~C.}\ \bibnamefont {Hardal}}, \bibinfo {author} {\bibfnamefont {N.}~\bibnamefont {Aslan}}, \bibinfo {author} {\bibfnamefont {C.~M.}\ \bibnamefont {Wilson}},\ and\ \bibinfo {author} {\bibfnamefont {O.~E.}\ \bibnamefont {M\"ustecapl\ifmmode \imath \else \i \fi{}o\ifmmode~\breve{g}\else \u{g}\fi{}lu}},\ }\bibfield  {title} {\bibinfo {title} {Quantum heat engine with coupled superconducting resonators},\ }\href {https://doi.org/10.1103/PhysRevE.96.062120} {\bibfield  {journal} {\bibinfo  {journal} {Phys. Rev. E}\ }\textbf {\bibinfo {volume} {96}},\ \bibinfo {pages} {062120} (\bibinfo {year} {2017})}\BibitemShut {NoStop}%
\bibitem [{\citenamefont {Rasola}\ and\ \citenamefont {M{\"o}tt{\"o}nen}(2024)}]{Rasola2024}%
  \BibitemOpen
  \bibfield  {author} {\bibinfo {author} {\bibfnamefont {M.}~\bibnamefont {Rasola}}\ and\ \bibinfo {author} {\bibfnamefont {M.}~\bibnamefont {M{\"o}tt{\"o}nen}},\ }\bibfield  {title} {\bibinfo {title} {Autonomous quantum heat engine based on non-markovian dynamics of an optomechanical hamiltonian},\ }\href {https://doi.org/10.1038/s41598-024-59881-z} {\bibfield  {journal} {\bibinfo  {journal} {Scientific Reports}\ }\textbf {\bibinfo {volume} {14}},\ \bibinfo {pages} {9448} (\bibinfo {year} {2024})}\BibitemShut {NoStop}%
\bibitem [{\citenamefont {Gevorgian}\ \emph {et~al.}(1995)\citenamefont {Gevorgian}, \citenamefont {Linner},\ and\ \citenamefont {Kollberg}}]{Gevorgian95}%
  \BibitemOpen
  \bibfield  {author} {\bibinfo {author} {\bibfnamefont {S.}~\bibnamefont {Gevorgian}}, \bibinfo {author} {\bibfnamefont {L.}~\bibnamefont {Linner}},\ and\ \bibinfo {author} {\bibfnamefont {E.}~\bibnamefont {Kollberg}},\ }\bibfield  {title} {\bibinfo {title} {Cad models for shielded multilayered cpw},\ }\href {https://doi.org/10.1109/22.375223} {\bibfield  {journal} {\bibinfo  {journal} {IEEE Transactions on Microwave Theory and Techniques}\ }\textbf {\bibinfo {volume} {43}},\ \bibinfo {pages} {772} (\bibinfo {year} {1995})}\BibitemShut {NoStop}%
\bibitem [{\citenamefont {{Watanabe}}\ \emph {et~al.}(1994)\citenamefont {{Watanabe}}, \citenamefont {{Yoshida}}, \citenamefont {{Aoki}},\ and\ \citenamefont {{Kohjiro}}}]{Watanabe94}%
  \BibitemOpen
  \bibfield  {author} {\bibinfo {author} {\bibfnamefont {K.}~\bibnamefont {{Watanabe}}}, \bibinfo {author} {\bibfnamefont {K.}~\bibnamefont {{Yoshida}}}, \bibinfo {author} {\bibfnamefont {T.}~\bibnamefont {{Aoki}}},\ and\ \bibinfo {author} {\bibfnamefont {S.}~\bibnamefont {{Kohjiro}}},\ }\bibfield  {title} {\bibinfo {title} {{Kinetic inductance of superconducting coplanar waveguides}},\ }\href@noop {} {\bibfield  {journal} {\bibinfo  {journal} {Japanese Journal of Applied Physics Regular Papers Short Notes and Review Papers}\ }\textbf {\bibinfo {volume} {33}},\ \bibinfo {pages} {5708} (\bibinfo {year} {1994})}\BibitemShut {NoStop}%
\bibitem [{\citenamefont {Maxfield}\ and\ \citenamefont {McLean}(1965)}]{Maxfield1965}%
  \BibitemOpen
  \bibfield  {author} {\bibinfo {author} {\bibfnamefont {B.~W.}\ \bibnamefont {Maxfield}}\ and\ \bibinfo {author} {\bibfnamefont {W.~L.}\ \bibnamefont {McLean}},\ }\bibfield  {title} {\bibinfo {title} {Superconducting penetration depth of niobium},\ }\href {https://doi.org/10.1103/PhysRev.139.A1515} {\bibfield  {journal} {\bibinfo  {journal} {Phys. Rev.}\ }\textbf {\bibinfo {volume} {139}},\ \bibinfo {pages} {A1515} (\bibinfo {year} {1965})}\BibitemShut {NoStop}%
\bibitem [{\citenamefont {Sage}\ \emph {et~al.}(2011)\citenamefont {Sage}, \citenamefont {Bolkhovsky}, \citenamefont {Oliver}, \citenamefont {Turek},\ and\ \citenamefont {Welander}}]{Sage2011}%
  \BibitemOpen
  \bibfield  {author} {\bibinfo {author} {\bibfnamefont {J.~M.}\ \bibnamefont {Sage}}, \bibinfo {author} {\bibfnamefont {V.}~\bibnamefont {Bolkhovsky}}, \bibinfo {author} {\bibfnamefont {W.~D.}\ \bibnamefont {Oliver}}, \bibinfo {author} {\bibfnamefont {B.}~\bibnamefont {Turek}},\ and\ \bibinfo {author} {\bibfnamefont {P.~B.}\ \bibnamefont {Welander}},\ }\bibfield  {title} {\bibinfo {title} {{Study of loss in superconducting coplanar waveguide resonators}},\ }\href {https://doi.org/10.1063/1.3552890} {\bibfield  {journal} {\bibinfo  {journal} {Journal of Applied Physics}\ }\textbf {\bibinfo {volume} {109}},\ \bibinfo {pages} {063915} (\bibinfo {year} {2011})},\ \Eprint {https://arxiv.org/abs/https://pubs.aip.org/aip/jap/article-pdf/doi/10.1063/1.3552890/15071073/063915\_1\_online.pdf} {https://pubs.aip.org/aip/jap/article-pdf/doi/10.1063/1.3552890/15071073/063915\_1\_online.pdf} \BibitemShut {NoStop}%
\bibitem [{\citenamefont {O’Connell}\ \emph {et~al.}(2008)\citenamefont {O’Connell}, \citenamefont {Ansmann}, \citenamefont {Bialczak}, \citenamefont {Hofheinz}, \citenamefont {Katz}, \citenamefont {Lucero}, \citenamefont {McKenney}, \citenamefont {Neeley}, \citenamefont {Wang}, \citenamefont {Weig}, \citenamefont {Cleland},\ and\ \citenamefont {Martinis}}]{OConnell2008}%
  \BibitemOpen
  \bibfield  {author} {\bibinfo {author} {\bibfnamefont {A.~D.}\ \bibnamefont {O’Connell}}, \bibinfo {author} {\bibfnamefont {M.}~\bibnamefont {Ansmann}}, \bibinfo {author} {\bibfnamefont {R.~C.}\ \bibnamefont {Bialczak}}, \bibinfo {author} {\bibfnamefont {M.}~\bibnamefont {Hofheinz}}, \bibinfo {author} {\bibfnamefont {N.}~\bibnamefont {Katz}}, \bibinfo {author} {\bibfnamefont {E.}~\bibnamefont {Lucero}}, \bibinfo {author} {\bibfnamefont {C.}~\bibnamefont {McKenney}}, \bibinfo {author} {\bibfnamefont {M.}~\bibnamefont {Neeley}}, \bibinfo {author} {\bibfnamefont {H.}~\bibnamefont {Wang}}, \bibinfo {author} {\bibfnamefont {E.~M.}\ \bibnamefont {Weig}}, \bibinfo {author} {\bibfnamefont {A.~N.}\ \bibnamefont {Cleland}},\ and\ \bibinfo {author} {\bibfnamefont {J.~M.}\ \bibnamefont {Martinis}},\ }\bibfield  {title} {\bibinfo {title} {{Microwave dielectric loss at single photon energies and millikelvin temperatures}},\ }\href {https://doi.org/10.1063/1.2898887} {\bibfield  {journal} {\bibinfo  {journal} {Applied
  Physics Letters}\ }\textbf {\bibinfo {volume} {92}},\ \bibinfo {pages} {112903} (\bibinfo {year} {2008})},\ \Eprint {https://arxiv.org/abs/https://pubs.aip.org/aip/apl/article-pdf/doi/10.1063/1.2898887/14389401/112903\_1\_online.pdf} {https://pubs.aip.org/aip/apl/article-pdf/doi/10.1063/1.2898887/14389401/112903\_1\_online.pdf} \BibitemShut {NoStop}%
\bibitem [{\citenamefont {Martinis}\ \emph {et~al.}(2005)\citenamefont {Martinis}, \citenamefont {Cooper}, \citenamefont {McDermott}, \citenamefont {Steffen}, \citenamefont {Ansmann}, \citenamefont {Osborn}, \citenamefont {Cicak}, \citenamefont {Oh}, \citenamefont {Pappas}, \citenamefont {Simmonds},\ and\ \citenamefont {Yu}}]{Martinis2005}%
  \BibitemOpen
  \bibfield  {author} {\bibinfo {author} {\bibfnamefont {J.~M.}\ \bibnamefont {Martinis}}, \bibinfo {author} {\bibfnamefont {K.~B.}\ \bibnamefont {Cooper}}, \bibinfo {author} {\bibfnamefont {R.}~\bibnamefont {McDermott}}, \bibinfo {author} {\bibfnamefont {M.}~\bibnamefont {Steffen}}, \bibinfo {author} {\bibfnamefont {M.}~\bibnamefont {Ansmann}}, \bibinfo {author} {\bibfnamefont {K.~D.}\ \bibnamefont {Osborn}}, \bibinfo {author} {\bibfnamefont {K.}~\bibnamefont {Cicak}}, \bibinfo {author} {\bibfnamefont {S.}~\bibnamefont {Oh}}, \bibinfo {author} {\bibfnamefont {D.~P.}\ \bibnamefont {Pappas}}, \bibinfo {author} {\bibfnamefont {R.~W.}\ \bibnamefont {Simmonds}},\ and\ \bibinfo {author} {\bibfnamefont {C.~C.}\ \bibnamefont {Yu}},\ }\bibfield  {title} {\bibinfo {title} {Decoherence in josephson qubits from dielectric loss},\ }\href {https://doi.org/10.1103/PhysRevLett.95.210503} {\bibfield  {journal} {\bibinfo  {journal} {Phys. Rev. Lett.}\ }\textbf {\bibinfo {volume} {95}},\ \bibinfo {pages} {210503} (\bibinfo
  {year} {2005})}\BibitemShut {NoStop}%
\bibitem [{\citenamefont {Lindstr\"om}\ \emph {et~al.}(2009)\citenamefont {Lindstr\"om}, \citenamefont {Healey}, \citenamefont {Colclough}, \citenamefont {Muirhead},\ and\ \citenamefont {Tzalenchuk}}]{Lindstrom2009}%
  \BibitemOpen
  \bibfield  {author} {\bibinfo {author} {\bibfnamefont {T.}~\bibnamefont {Lindstr\"om}}, \bibinfo {author} {\bibfnamefont {J.~E.}\ \bibnamefont {Healey}}, \bibinfo {author} {\bibfnamefont {M.~S.}\ \bibnamefont {Colclough}}, \bibinfo {author} {\bibfnamefont {C.~M.}\ \bibnamefont {Muirhead}},\ and\ \bibinfo {author} {\bibfnamefont {A.~Y.}\ \bibnamefont {Tzalenchuk}},\ }\bibfield  {title} {\bibinfo {title} {Properties of superconducting planar resonators at millikelvin temperatures},\ }\href {https://doi.org/10.1103/PhysRevB.80.132501} {\bibfield  {journal} {\bibinfo  {journal} {Phys. Rev. B}\ }\textbf {\bibinfo {volume} {80}},\ \bibinfo {pages} {132501} (\bibinfo {year} {2009})}\BibitemShut {NoStop}%
\bibitem [{\citenamefont {Wang}\ \emph {et~al.}(2009)\citenamefont {Wang}, \citenamefont {Hofheinz}, \citenamefont {Wenner}, \citenamefont {Ansmann}, \citenamefont {Bialczak}, \citenamefont {Lenander}, \citenamefont {Lucero}, \citenamefont {Neeley}, \citenamefont {O’Connell}, \citenamefont {Sank}, \citenamefont {Weides}, \citenamefont {Cleland},\ and\ \citenamefont {Martinis}}]{Wang2009}%
  \BibitemOpen
  \bibfield  {author} {\bibinfo {author} {\bibfnamefont {H.}~\bibnamefont {Wang}}, \bibinfo {author} {\bibfnamefont {M.}~\bibnamefont {Hofheinz}}, \bibinfo {author} {\bibfnamefont {J.}~\bibnamefont {Wenner}}, \bibinfo {author} {\bibfnamefont {M.}~\bibnamefont {Ansmann}}, \bibinfo {author} {\bibfnamefont {R.~C.}\ \bibnamefont {Bialczak}}, \bibinfo {author} {\bibfnamefont {M.}~\bibnamefont {Lenander}}, \bibinfo {author} {\bibfnamefont {E.}~\bibnamefont {Lucero}}, \bibinfo {author} {\bibfnamefont {M.}~\bibnamefont {Neeley}}, \bibinfo {author} {\bibfnamefont {A.~D.}\ \bibnamefont {O’Connell}}, \bibinfo {author} {\bibfnamefont {D.}~\bibnamefont {Sank}}, \bibinfo {author} {\bibfnamefont {M.}~\bibnamefont {Weides}}, \bibinfo {author} {\bibfnamefont {A.~N.}\ \bibnamefont {Cleland}},\ and\ \bibinfo {author} {\bibfnamefont {J.~M.}\ \bibnamefont {Martinis}},\ }\bibfield  {title} {\bibinfo {title} {{Improving the coherence time of superconducting coplanar resonators}},\ }\href {https://doi.org/10.1063/1.3273372}
  {\bibfield  {journal} {\bibinfo  {journal} {Applied Physics Letters}\ }\textbf {\bibinfo {volume} {95}},\ \bibinfo {pages} {233508} (\bibinfo {year} {2009})},\ \Eprint {https://arxiv.org/abs/https://pubs.aip.org/aip/apl/article-pdf/doi/10.1063/1.3273372/13129793/233508\_1\_online.pdf} {https://pubs.aip.org/aip/apl/article-pdf/doi/10.1063/1.3273372/13129793/233508\_1\_online.pdf} \BibitemShut {NoStop}%
\bibitem [{\citenamefont {Chen}\ \emph {et~al.}(2022)\citenamefont {Chen}, \citenamefont {Pfeiffer}, \citenamefont {Partanen}, \citenamefont {Fesquet}, \citenamefont {Honasoge}, \citenamefont {Kronowetter}, \citenamefont {Nojiri}, \citenamefont {Renger}, \citenamefont {Fedorov}, \citenamefont {Marx}, \citenamefont {Deppe},\ and\ \citenamefont {Gross}}]{Chen2022}%
  \BibitemOpen
  \bibfield  {author} {\bibinfo {author} {\bibfnamefont {Q.-M.}\ \bibnamefont {Chen}}, \bibinfo {author} {\bibfnamefont {M.}~\bibnamefont {Pfeiffer}}, \bibinfo {author} {\bibfnamefont {M.}~\bibnamefont {Partanen}}, \bibinfo {author} {\bibfnamefont {F.}~\bibnamefont {Fesquet}}, \bibinfo {author} {\bibfnamefont {K.~E.}\ \bibnamefont {Honasoge}}, \bibinfo {author} {\bibfnamefont {F.}~\bibnamefont {Kronowetter}}, \bibinfo {author} {\bibfnamefont {Y.}~\bibnamefont {Nojiri}}, \bibinfo {author} {\bibfnamefont {M.}~\bibnamefont {Renger}}, \bibinfo {author} {\bibfnamefont {K.~G.}\ \bibnamefont {Fedorov}}, \bibinfo {author} {\bibfnamefont {A.}~\bibnamefont {Marx}}, \bibinfo {author} {\bibfnamefont {F.}~\bibnamefont {Deppe}},\ and\ \bibinfo {author} {\bibfnamefont {R.}~\bibnamefont {Gross}},\ }\bibfield  {title} {\bibinfo {title} {Scattering coefficients of superconducting microwave resonators. i. transfer matrix approach},\ }\href {https://doi.org/10.1103/PhysRevB.106.214505} {\bibfield  {journal} {\bibinfo  {journal}
  {Phys. Rev. B}\ }\textbf {\bibinfo {volume} {106}},\ \bibinfo {pages} {214505} (\bibinfo {year} {2022})}\BibitemShut {NoStop}%
\bibitem [{\citenamefont {Lake}\ \emph {et~al.}(2017)\citenamefont {Lake}, \citenamefont {Govenius}, \citenamefont {Kokkoniemi}, \citenamefont {Tan}, \citenamefont {Partanen}, \citenamefont {Virtanen},\ and\ \citenamefont {Möttönen}}]{Lake2017}%
  \BibitemOpen
  \bibfield  {author} {\bibinfo {author} {\bibfnamefont {R.~E.}\ \bibnamefont {Lake}}, \bibinfo {author} {\bibfnamefont {J.}~\bibnamefont {Govenius}}, \bibinfo {author} {\bibfnamefont {R.}~\bibnamefont {Kokkoniemi}}, \bibinfo {author} {\bibfnamefont {K.~Y.}\ \bibnamefont {Tan}}, \bibinfo {author} {\bibfnamefont {M.}~\bibnamefont {Partanen}}, \bibinfo {author} {\bibfnamefont {P.}~\bibnamefont {Virtanen}},\ and\ \bibinfo {author} {\bibfnamefont {M.}~\bibnamefont {Möttönen}},\ }\bibfield  {title} {\bibinfo {title} {Microwave admittance of gold-palladium nanowires with proximity-induced superconductivity},\ }\href {https://doi.org/https://doi.org/10.1002/aelm.201600227} {\bibfield  {journal} {\bibinfo  {journal} {Advanced Electronic Materials}\ }\textbf {\bibinfo {volume} {3}},\ \bibinfo {pages} {1600227} (\bibinfo {year} {2017})},\ \Eprint {https://arxiv.org/abs/https://onlinelibrary.wiley.com/doi/pdf/10.1002/aelm.201600227} {https://onlinelibrary.wiley.com/doi/pdf/10.1002/aelm.201600227} \BibitemShut
  {NoStop}%
\bibitem [{\citenamefont {Probst}\ \emph {et~al.}(2015)\citenamefont {Probst}, \citenamefont {Song}, \citenamefont {Bushev}, \citenamefont {Ustinov},\ and\ \citenamefont {Weides}}]{Probst2015}%
  \BibitemOpen
  \bibfield  {author} {\bibinfo {author} {\bibfnamefont {S.}~\bibnamefont {Probst}}, \bibinfo {author} {\bibfnamefont {F.~B.}\ \bibnamefont {Song}}, \bibinfo {author} {\bibfnamefont {P.~A.}\ \bibnamefont {Bushev}}, \bibinfo {author} {\bibfnamefont {A.~V.}\ \bibnamefont {Ustinov}},\ and\ \bibinfo {author} {\bibfnamefont {M.}~\bibnamefont {Weides}},\ }\bibfield  {title} {\bibinfo {title} {{Efficient and robust analysis of complex scattering data under noise in microwave resonators}},\ }\href {https://doi.org/10.1063/1.4907935} {\bibfield  {journal} {\bibinfo  {journal} {Review of Scientific Instruments}\ }\textbf {\bibinfo {volume} {86}},\ \bibinfo {pages} {024706} (\bibinfo {year} {2015})},\ \Eprint {https://arxiv.org/abs/https://pubs.aip.org/aip/rsi/article-pdf/doi/10.1063/1.4907935/15732678/024706\_1\_online.pdf} {https://pubs.aip.org/aip/rsi/article-pdf/doi/10.1063/1.4907935/15732678/024706\_1\_online.pdf} \BibitemShut {NoStop}%
\bibitem [{\citenamefont {Pappas}\ \emph {et~al.}(2011)\citenamefont {Pappas}, \citenamefont {Vissers}, \citenamefont {Wisbey}, \citenamefont {Kline},\ and\ \citenamefont {Gao}}]{Pappas2011}%
  \BibitemOpen
  \bibfield  {author} {\bibinfo {author} {\bibfnamefont {D.~P.}\ \bibnamefont {Pappas}}, \bibinfo {author} {\bibfnamefont {M.~R.}\ \bibnamefont {Vissers}}, \bibinfo {author} {\bibfnamefont {D.~S.}\ \bibnamefont {Wisbey}}, \bibinfo {author} {\bibfnamefont {J.~S.}\ \bibnamefont {Kline}},\ and\ \bibinfo {author} {\bibfnamefont {J.}~\bibnamefont {Gao}},\ }\bibfield  {title} {\bibinfo {title} {Two level system loss in superconducting microwave resonators},\ }\href {https://doi.org/10.1109/TASC.2010.2097578} {\bibfield  {journal} {\bibinfo  {journal} {IEEE Transactions on Applied Superconductivity}\ }\textbf {\bibinfo {volume} {21}},\ \bibinfo {pages} {871} (\bibinfo {year} {2011})}\BibitemShut {NoStop}%
\bibitem [{\citenamefont {Gao}\ \emph {et~al.}(2008)\citenamefont {Gao}, \citenamefont {Daal}, \citenamefont {Vayonakis}, \citenamefont {Kumar}, \citenamefont {Zmuidzinas}, \citenamefont {Sadoulet}, \citenamefont {Mazin}, \citenamefont {Day},\ and\ \citenamefont {Leduc}}]{Gao2008}%
  \BibitemOpen
  \bibfield  {author} {\bibinfo {author} {\bibfnamefont {J.}~\bibnamefont {Gao}}, \bibinfo {author} {\bibfnamefont {M.}~\bibnamefont {Daal}}, \bibinfo {author} {\bibfnamefont {A.}~\bibnamefont {Vayonakis}}, \bibinfo {author} {\bibfnamefont {S.}~\bibnamefont {Kumar}}, \bibinfo {author} {\bibfnamefont {J.}~\bibnamefont {Zmuidzinas}}, \bibinfo {author} {\bibfnamefont {B.}~\bibnamefont {Sadoulet}}, \bibinfo {author} {\bibfnamefont {B.~A.}\ \bibnamefont {Mazin}}, \bibinfo {author} {\bibfnamefont {P.~K.}\ \bibnamefont {Day}},\ and\ \bibinfo {author} {\bibfnamefont {H.~G.}\ \bibnamefont {Leduc}},\ }\bibfield  {title} {\bibinfo {title} {{Experimental evidence for a surface distribution of two-level systems in superconducting lithographed microwave resonators}},\ }\href {https://doi.org/10.1063/1.2906373} {\bibfield  {journal} {\bibinfo  {journal} {Applied Physics Letters}\ }\textbf {\bibinfo {volume} {92}},\ \bibinfo {pages} {152505} (\bibinfo {year} {2008})},\ \Eprint
  {https://arxiv.org/abs/https://pubs.aip.org/aip/apl/article-pdf/doi/10.1063/1.2906373/14393210/152505\_1\_online.pdf} {https://pubs.aip.org/aip/apl/article-pdf/doi/10.1063/1.2906373/14393210/152505\_1\_online.pdf} \BibitemShut {NoStop}%
\bibitem [{\citenamefont {McRae}\ \emph {et~al.}(2020)\citenamefont {McRae}, \citenamefont {Wang}, \citenamefont {Gao}, \citenamefont {Vissers}, \citenamefont {Brecht}, \citenamefont {Dunsworth}, \citenamefont {Pappas},\ and\ \citenamefont {Mutus}}]{McRae2020}%
  \BibitemOpen
  \bibfield  {author} {\bibinfo {author} {\bibfnamefont {C.~R.~H.}\ \bibnamefont {McRae}}, \bibinfo {author} {\bibfnamefont {H.}~\bibnamefont {Wang}}, \bibinfo {author} {\bibfnamefont {J.}~\bibnamefont {Gao}}, \bibinfo {author} {\bibfnamefont {M.~R.}\ \bibnamefont {Vissers}}, \bibinfo {author} {\bibfnamefont {T.}~\bibnamefont {Brecht}}, \bibinfo {author} {\bibfnamefont {A.}~\bibnamefont {Dunsworth}}, \bibinfo {author} {\bibfnamefont {D.~P.}\ \bibnamefont {Pappas}},\ and\ \bibinfo {author} {\bibfnamefont {J.}~\bibnamefont {Mutus}},\ }\bibfield  {title} {\bibinfo {title} {{Materials loss measurements using superconducting microwave resonators}},\ }\href {https://doi.org/10.1063/5.0017378} {\bibfield  {journal} {\bibinfo  {journal} {Review of Scientific Instruments}\ }\textbf {\bibinfo {volume} {91}},\ \bibinfo {pages} {091101} (\bibinfo {year} {2020})},\ \Eprint {https://arxiv.org/abs/https://pubs.aip.org/aip/rsi/article-pdf/doi/10.1063/5.0017378/19786221/091101\_1\_online.pdf}
  {https://pubs.aip.org/aip/rsi/article-pdf/doi/10.1063/5.0017378/19786221/091101\_1\_online.pdf} \BibitemShut {NoStop}%
\bibitem [{\citenamefont {Beldi}\ \emph {et~al.}(2019)\citenamefont {Beldi}, \citenamefont {Boussaha}, \citenamefont {Hu}, \citenamefont {Monfardini}, \citenamefont {Traini}, \citenamefont {Levy-Bertrand}, \citenamefont {Chaumont}, \citenamefont {Gonzales}, \citenamefont {Firminy}, \citenamefont {Reix}, \citenamefont {Rosticher}, \citenamefont {Mignot}, \citenamefont {Piat},\ and\ \citenamefont {Bonifacio}}]{Beldi2019}%
  \BibitemOpen
  \bibfield  {author} {\bibinfo {author} {\bibfnamefont {S.}~\bibnamefont {Beldi}}, \bibinfo {author} {\bibfnamefont {F.}~\bibnamefont {Boussaha}}, \bibinfo {author} {\bibfnamefont {J.}~\bibnamefont {Hu}}, \bibinfo {author} {\bibfnamefont {A.}~\bibnamefont {Monfardini}}, \bibinfo {author} {\bibfnamefont {A.}~\bibnamefont {Traini}}, \bibinfo {author} {\bibfnamefont {F.}~\bibnamefont {Levy-Bertrand}}, \bibinfo {author} {\bibfnamefont {C.}~\bibnamefont {Chaumont}}, \bibinfo {author} {\bibfnamefont {M.}~\bibnamefont {Gonzales}}, \bibinfo {author} {\bibfnamefont {J.}~\bibnamefont {Firminy}}, \bibinfo {author} {\bibfnamefont {F.}~\bibnamefont {Reix}}, \bibinfo {author} {\bibfnamefont {M.}~\bibnamefont {Rosticher}}, \bibinfo {author} {\bibfnamefont {S.}~\bibnamefont {Mignot}}, \bibinfo {author} {\bibfnamefont {M.}~\bibnamefont {Piat}},\ and\ \bibinfo {author} {\bibfnamefont {P.}~\bibnamefont {Bonifacio}},\ }\bibfield  {title} {\bibinfo {title} {High q-factor near infrared and visible al2o3-based parallel-plate
  capacitor kinetic inductance detectors},\ }\href {https://doi.org/10.1364/OE.27.013319} {\bibfield  {journal} {\bibinfo  {journal} {Opt. Express}\ }\textbf {\bibinfo {volume} {27}},\ \bibinfo {pages} {13319} (\bibinfo {year} {2019})}\BibitemShut {NoStop}%
\bibitem [{\citenamefont {Mamin}\ \emph {et~al.}(2021)\citenamefont {Mamin}, \citenamefont {Huang}, \citenamefont {Carnevale}, \citenamefont {Rettner}, \citenamefont {Arellano}, \citenamefont {Sherwood}, \citenamefont {Kurter}, \citenamefont {Trimm}, \citenamefont {Sandberg}, \citenamefont {Shelby}, \citenamefont {Mueed}, \citenamefont {Madon}, \citenamefont {Pushp}, \citenamefont {Steffen},\ and\ \citenamefont {Rugar}}]{Mamin2021}%
  \BibitemOpen
  \bibfield  {author} {\bibinfo {author} {\bibfnamefont {H.}~\bibnamefont {Mamin}}, \bibinfo {author} {\bibfnamefont {E.}~\bibnamefont {Huang}}, \bibinfo {author} {\bibfnamefont {S.}~\bibnamefont {Carnevale}}, \bibinfo {author} {\bibfnamefont {C.}~\bibnamefont {Rettner}}, \bibinfo {author} {\bibfnamefont {N.}~\bibnamefont {Arellano}}, \bibinfo {author} {\bibfnamefont {M.}~\bibnamefont {Sherwood}}, \bibinfo {author} {\bibfnamefont {C.}~\bibnamefont {Kurter}}, \bibinfo {author} {\bibfnamefont {B.}~\bibnamefont {Trimm}}, \bibinfo {author} {\bibfnamefont {M.}~\bibnamefont {Sandberg}}, \bibinfo {author} {\bibfnamefont {R.}~\bibnamefont {Shelby}}, \bibinfo {author} {\bibfnamefont {M.}~\bibnamefont {Mueed}}, \bibinfo {author} {\bibfnamefont {B.}~\bibnamefont {Madon}}, \bibinfo {author} {\bibfnamefont {A.}~\bibnamefont {Pushp}}, \bibinfo {author} {\bibfnamefont {M.}~\bibnamefont {Steffen}},\ and\ \bibinfo {author} {\bibfnamefont {D.}~\bibnamefont {Rugar}},\ }\bibfield  {title} {\bibinfo {title} {Merged-element
  transmons: Design and qubit performance},\ }\href {https://doi.org/10.1103/PhysRevApplied.16.024023} {\bibfield  {journal} {\bibinfo  {journal} {Phys. Rev. Appl.}\ }\textbf {\bibinfo {volume} {16}},\ \bibinfo {pages} {024023} (\bibinfo {year} {2021})}\BibitemShut {NoStop}%
\bibitem [{son(2024)}]{sonnet}%
  \BibitemOpen
  \href@noop {} {\bibinfo {title} {Sonnet software official website}},\ \bibinfo {howpublished} {\url{https://www.sonnetsoftware.com/}} (\bibinfo {year} {2024}),\ \bibinfo {note} {accessed: 2024-11-13}\BibitemShut {NoStop}%
\end{thebibliography}%

\end{document}